\documentstyle[prd,aps,eqsecnum,epsfig,twocolumn]{revtex}
\begin{document}


\onecolumn

\title{Geometry Eigenvalues and Scalar Product \\
from Recoupling Theory in Loop Quantum Gravity}

\author{Roberto De Pietri\footnote[1]{
E-mail address: depietri@vaxpr.pr.infn.it}}
\address{Dipartimento Di Fisica, Universit\`a di Parma and \\
        I.N.F.N. Sezione di Milano, Gruppo Collegato di Parma,\\
        I-43100 Parma (PR), Italy}
\author{Carlo Rovelli\footnote[2]{
E-mail address: rovelli@pitt.edu}}
\address{Department of Physics and Astronomy, \\
         University of Pittsburgh, Pittsburgh Pa 15260, USA.}

\date{\today}
\maketitle


\vspace{.2cm}
\begin{abstract}
\noindent
We summarize the basics of the loop representation of quantum
gravity and describe the main aspects of the formalism, including
its latest developments, in a reorganized and consistent form.
Recoupling theory, in its graphical tangle-theoretic 
Temperley-Lieb formulation, provides a powerful calculation tool
in this context.  We describe its application to the loop
representation in detail.  Using recoupling theory, we derive
general expressions for the spectrum of the quantum area and the
quantum volume operators.  We compute several volume eigenvalues
explicitly. We introduce a scalar product with respect to which
area and volume are symmetric operators, and (the trivalent
expansions of) the spin network states are orthonormal.
\end{abstract}

\pacs{04.60.-m, 02.70.-c, 04.60.Ds, 03.70+k}




\section{Introduction}

We start with a citation from R.\ Penrose\cite{Penrose72}:
``My own view is that ultimately physical laws should find
their most natural expression in terms of essentially
combinatorial principles, that is to say, in terms of finite
processes such as counting or other basically simple
manipulation procedures. Thus, in accordance with such a
view, some form of discrete or combinatorial space time
should emerge.''  The loop approach to quantum general
relativity \cite{Rovelli88,Rovelli90} seems to be leading
precisely to a realization of such a vision of a
combinatorial space-time, deriving it solely from a strict
application of conventional quantum ideas to standard
general relativity.\footnote{For an overview of current
ideas on quantum geometry, see \cite{Isham96}, \cite{jmp95}
and \cite{Baez94b}; for a recent overview of canonical
gravity, see \cite{Ehlers94}; for introductions to
loop quantum gravity, see 
\cite{Rovelli91,Ashtekar92,Smolin93,Gambini,Bruegmann,Ezawa}.}

A number of recent advances in this direction have
strengthened this hope. First of all, there is the
mathematically rigorous development of the connection
representation \cite{Ashtekar92,Isham,Ashtekar95,AshtekarLewand}
which has lead to recovering the 
loop representation formalism from a general quantization 
program.  This approach has sharpened various loop
representation results using rigorous $C^*$ algebraic and 
measure theoretical techniques, and has put them on a 
solid mathematical footing. For a discussion
of the precise relation between the two formulations 
of loop quantum gravity, see Refs.~\cite{Lewandowski96,DePietri96}.
Furthermore: a simplification of the formalism
due to the introduction in quantum gravity of the spin network basis
\cite{Rovelli95a} (see also \cite{Baez95a,Foxon}); the result that area
\cite{Ashtekar92a} and volume operators 
\cite{Rovelli95,Loll95} have discrete eigenvalues; the idea
that in the presence of matter these eigenvalues might be
taken as physical predictions on quantum geometry
\cite{Rovelli93a}; a Hamiltonian generating
clock time evolution \cite{Rovelli94b} and a tentative
perturbation scheme for computing diffeomorphism invariant
transition amplitudes \cite{Rovelli95b}; the extension of
the theory to fermions \cite{Rovelli94} to the
electromagnetic field \cite{Krasnov95,Pullin}.  
This rapid 
development has produced a certain amount of confusion in
the notation and the basics of the theory.  A first aim of
this paper is to bring some order in the kinematics of the
loop representation 
formalism, by presenting the basics formulas, notations and
results in a consistent and self-contained form.  This
allows us to insert a novel sign factor into the very
definition of the loop representation, short-cutting sign
complications of the previous formulation.  In a sense, we
bring to full maturity the insights of reference
\cite{Rovelli95a}.

With the new sign factor, loop states of the loop
representation satisfy the axioms of Penrose's ``binor
calculus'' \cite{Penrose71a}, or, equivalently, the axioms
of the tangle-theoretic formulation of recoupling theory
\cite{Kauffman94} for the special ``classical'' value $A=-1$
of the deformation parameter.  This fact brings a powerful
set of computational techniques at the service of quantum
gravity.  We describe here in detail how this calculus can
be used.  Calculations in loop quantum gravity were first
performed using the grasping operation on single loops
\cite{Rovelli90}.  It was then realized, mainly in
\cite{Ashtekar95}, that such combinatorial techniques
admitted a group theoretical interpretation ($su(2)$
representation theory admits a fully combinatorial
description).  Recoupling theory is a further --and far more
powerful-- level of sophistication for the same calculus.

The idea that recoupling theory plays a role in loop quantum
gravity has been advocated by Reisenberger
\cite{Reisenberger96} and by Smolin.  Motivated by certain
physical and mathematical considerations, Borissov, Major
and Smolin \cite{Smolin95a,Borissov} have considered
deformations of the standard loop representation theory.
Recoupling theory with general values of the deformation
parameter $A$ plays a key role in the definition of these
deformations.  What we do here is very different in spirit:
we remain within the framework of the standard
loop-representation quantum GR, and use recoupling theory
merely as a computational tool.

Using recoupling theory, we derive general formulas for area
and volume in quantum gravity.  The spectrum of the area
agrees with previously published results \cite{Rovelli95}.
The derivation presented here is simpler and more elegant
than the one in Ref.~\cite{Rovelli95}.  The first of our
main new results is a general formula for the volume.  We
present it here expressed in terms of $su(2)$ 6-j symbols
(and related quantities).  We confirm the fact that
trivalent vertices have zero volume, first pointed out by
Loll \cite{Loll95}. We explicitly compute many eigenstates
for four- and five-valent vertices.  Loll has computed a few
of these eigenvalues in \cite{Loll95a} using a different
technique.  We find agreement with the numbers published by
Loll (see also \cite{Borissov}).  We show that the absolute
value and the square root that appear in the definition of
the volume operator are well defined.  Indeed, we show that
the arguments of the absolute value are finite dimensional
matrices diagonalizable and with real eigenvalues; and that
the arguments of the square root are finite dimensional
matrices diagonalizable and with real non-negative
eigenvalues. We show in general that the eigenvalues of the
volume are real and non-negative.

Finally, the technique introduced allows us to define a
scalar product in the loop representation, by requiring that
area and volume be symmetric, and that spin network states
be orthogonal to each other -- whatever the trivalent
decomposition of high valence vertices we use. This is our
second main new result.

The structure of this paper is the following. In Sec.~II, we
review the basics of the Ashtekar formulation of general
relativity and we define the loop variables (in the new form
that leads directly to recoupling theory).  In Sec.~III, we
derive the basic equations of the loop representation.  In
Sec.~IV we discuss the role of recoupling theory. In Sec.~V
we define the spin network basis. In Sec.~VI, we discuss the
area operator, and in Sec.~VII the volume operator. In
Sec.~VIII we define the scalar product. Sec.~IX contains our
conclusions.


\section{Loop variables in classical GR}

We begin by reviewing the canonical formulation of general
relativity in the real Ashtekar formalism~\cite{Barbero,Ashtekar91,Thiemann}.
This is given as follows. We fix a three-dimensional
manifold $M$ and consider two real (smooth) 
$SO(3)$ fields $A_a^i(x)$ and $\tilde{E}^a_i(x)$ on~$M$.  We use
$a,b,\ldots=1,2,3$ for (abstract) spatial indices and
$i,j,\ldots=1,2,3$ for internal $SO(3)$ indices.  We
indicate coordinates on $M$ with~$x$.  
The relation between these fields and conventional metric 
gravitational variables is as follows: 
$\tilde{E}^a_i(x)$ is the (densitized) inverse triad, related to the 
three-dimensional metric $g_{ab}(x)$ of the constant-time surface by
\begin{eqnarray}
g\ g^{ab} = \tilde{E}^a_i\tilde{E}^b_i, 
\end{eqnarray}
where $g$ is the determinant of $g_{ab}$; and 
\begin{equation}
A_a^i(x)=\Gamma_a^i(x)-k_a^i(x), 
\label{real}
\end{equation}
where $\Gamma_a^i(x)$ is the $SU(2)$ spin connection 
associated to the triad and $k_a^i(x)$ is the 
extrinsic curvature of the three surface (up to indices' position).  
Notice the absence of the $i$ in (\ref{real}), which yields the 
{\it real \/} Ashtekar connection. The real Ashtekar 
connection (\ref{real}) is the natural variable for the Riemannian theory,
but it can be used as the basic field for the Lorentzian
theory as well, at the price of a more complicated form of
the Hamiltonian constraint. Recently, Thiemann and Ashtekar \cite{Thiemann}
have argued that the most promising strategy for implementing
the quantum reality conditions is to start from the real 
Ashtekar connection, and circumvent the difficulties due to the 
complicate form of the Lorentzian Hamiltonian constraint by expressing 
it in terms of the Riemannian Hamiltonian constraint via a  
generalized Wick transform (see also \cite{Loll96,Immirzi}). 
Here, we will not discuss the dynamics. 
Therefore our results can be significative for the Riemannian theory as 
well as for the Lorentzian theory.  We will discuss the necessary
modifications of the formalism for applying it to the complex Ashtekar 
connection at the end of section VII. 

It is useful for
what follows to consider the dimensional character of the
field with care. We set the dimension of the fields as
follow:
\begin{eqnarray}
   ~[ g_{ab} ] &=& L^2 , \ \ \   
   ~[ \tilde{E}^a_i ] = L^2 , \nonumber\\
   ~[ A_a^i] &=& \mbox{dimensionless}.
\end{eqnarray}
The popular choice of taking the metric dimensionless is
not very sensible in GR. It forces coordinates to have
dimensions of a length; but the freedom of arbitrary
transformations on the coordinates is hardly compatible
with dimensional coordinates.  Coordinates, for instance,
can be angles, and assigning angles dimension of a length
makes no sense.  The Einstein action can be rewritten
(see for example \cite{Rovelli91}) as
\begin{eqnarray}
&&  S = \frac{1}{G} \int\!d^4\rlap x\  ~\sqrt{g}\ R 
=  \frac{1}{G} \int  dx^0 \int\!d^3 x ~\bigg[
  - \dot{A}^i_a \tilde{E}^a_i 
  + \dot{A}^i_0 \tilde{C}_i
  + N^a \tilde{C}_a
  + \rlap{\lower1ex\hbox{$\sim$}}{N} \tilde{\tilde{W}} \bigg],  
\end{eqnarray}
where we have set
\begin{equation}
   G = \frac{16 \pi\ G_{{\rm Newton}} }{c^3}
\end{equation}
$G_{{\rm Newton}}$ being Newton's gravitational constant, 
and $\tilde{C}_i$, $\tilde{C}_a$, $\tilde{W}$ 
the diffeomorphism, Gauss and Hamiltonian constraints.
It follows that the momentum canonically conjugate to 
$A_a^i$ is 
\begin{equation}
  p_i^a (x) = \frac{\delta S}{\delta \dot{A}_a^i(x) }
            = -  \frac{1}{G}\ \tilde{E}^a_i. 
\end{equation}  
and therefore the fundamental Poisson bracket
of the Hamiltonian theory is
\begin{equation}
 \{ A_a^i (x) , \tilde{E}^b_j (y) \} 
        =  G~ \delta_a^b \delta_j^i \delta^3 (x,y)
\end{equation}

The spinorial version of
the Ashtekar variables is given in terms of the Pauli matrices
${\sigma_i}, i=1,2,3$, or the $su(2)$ generators
${\tau_i} = -  \frac{{\rm i}}{2} \ {\sigma_i}$, by
\begin{eqnarray}
\tilde{E}^{a}(x) 
        &=& - {\rm i}\ \tilde{E}^a_i(x) \ {\sigma_i}
         =  2 \tilde{E}^a_i(x) \ {\tau_i} \\
A_{a}(x)&=& - \frac{{\rm i}}{{2}}\ {A}_a^i(x)\ {\sigma_i}
         = {A}_a^i(x)\ {\tau_i}\,.\,.
\end{eqnarray}
$A_a(x)$ and $\tilde{E}^a(x)$ are $2\times 2$
complex matrices.   We use upper case
indices $A, B ... = 1,2$ for the spinor space on which the
Pauli matrices act. Thus, the components of the
gravitational fields are $A_a{}_A{}^B(x)$ and
$\tilde{E}^a{}_A{}^B(x)$.

In order to construct the loop variables, we start from some
definitions.
\begin{description}
\item[Segment. ]  A segment $\gamma$ is a continuous and
         piecewise smooth map from the closed interval $[0,1]$
         into $M$.  We write: $\gamma: s \longmapsto
         \gamma^a(s)$.
\item[Loop. ]  A loop $\alpha$ is a segment such that
	$\alpha^a(0) = \alpha^a(1)$. Equivalently, it is a
	continuous, piecewise smooth, map from the circle $S_1$
	into $M^3$.
\item[Free Loop Algebra. ] We consider (formal) linear
	combinations $\Phi$ of (formal) products of loops, as
	in:
\begin{equation}
\Phi = c_0 + \sum_i c_i ~[\alpha_i] + \sum_{jk} c_{jk}
                 ~[\alpha_j][\alpha_k]~ + \ldots ~~,
\label{fla}
\end{equation}
	where the $c$'s are arbitrary complex number and the
	$\alpha$'s are loops; we denote the space of such
	objects as the Free Loop Algebra ${\cal A}^f[{\cal
	L}]$. (See also \cite{Ashtekar92})
\item[Multiloop. ] We denote the monomials in ${\cal A}^f[{\cal
	L}]$, namely the elements of the form $\Phi =
	[\alpha_1] ... [\alpha_n]$ as multiloops.  We indicate
	multiloops by a Greek letter, in the same manner as
	(single) loops: $[\alpha] = [\alpha_1] ...  [\alpha_n]$
	.
\end{description}

Given a segment $\gamma$, we consider the parallel
propagator of $A_a$ along $\gamma$, This is defined by the equation
\begin{equation}
  \frac{d}{d\tau} U_\gamma (\tau,\tau_0)
       + \frac{d \gamma^a(\tau)}{d\tau} 
       A_a(\gamma(\tau)) U_\gamma (\tau,\tau_0) = 0, 
\label{p}
\end{equation}
with the boundary condition 
$U_\gamma(\tau_0,\tau_0)_A^{~B} = \delta_A^{~B}$. 
The formal solution is 
\begin{equation}
U_\gamma (\tau,\tau_0) = {\cal P} e^{-  
\int_{\tau_0}^\tau d\tau \ 
                         \dot{\gamma}^a A_a(\gamma(\tau))},
\label{pp}
\end{equation}
where ${\cal P}$ indicates the path ordering of the
exponential. We also write --in a somewhat imprecise
notation-- $U_\gamma = U_\gamma(0,1)$ and
$U_\gamma(s_2,s_1) = U_\gamma(\tau_2,\tau_1)$ if $s_2 =
\gamma(\tau_2)$ and $s_1 = \gamma(\tau_1)$

We can now define the fundamental loop variables. Given a
loop $\alpha$ and the points $s_1,s_2,\ldots,s_n\in\alpha$
we define:
\begin{eqnarray}
  {\cal T}[\alpha]          &=& - {\rm Tr} [U_\alpha] ,
\label{t0}  \\
  {\cal T}^a[\alpha](s)     
      &=& - {\rm Tr} [U_\alpha(s,s) \tilde{E}^a(s)] 
\end{eqnarray}
and, in general
\begin{eqnarray}
{\cal T}^{a_1a_2}[\alpha](s_1,s_2) &=& 
  - {\rm Tr} [U_\alpha(s_1,s_2) \tilde{E}^{a_2}(s_2) 
                U_\alpha(s_2,s_1) \tilde{E}^{a_1}(s_1) ] ,
\\
{\cal T}^{a_1\ldots a_N}[\alpha](s_1,\ldots,s_N) 
&=&  - {\rm Tr} [U_\alpha(s_1,s_N) \tilde{E}^{a_N}(s_N)
                U_\alpha(s_N,s_{N-1}) \ldots 
~~\ldots U_\alpha(s_{2},s_1) \tilde{E}^{a_1}(s_1) ] . 
\nonumber
\end{eqnarray}
The function ${\cal T}[\alpha]$ defined in (\ref{t0}) for
a single loop, can be defined over the whole
free loop algebra ${\cal A}^f[{\cal L}]$: given
the generic element $\Phi \in {\cal A}^f[{\cal L}]$ in
(\ref{fla}), we pose
\begin{equation}
{\cal T}[\Phi] 
    = -2 c_0  
     + \sum_i c_i ~{\cal T}[\alpha_i]~
     + \sum_{ij} c_{ij} ~{\cal T}[\alpha_i]\,{\cal T}[\alpha_j]~
     + \ldots
\end{equation}
The reason for the $-2$ in the first term is the
following.  We may think of the first term of the sum as
corresponding to the ``point loop'', or a loop whose
image is a point. For this loop, the exponent in
(\ref{pp}) is zero, the holonomy is the identity (in
$sl(2,C)$, namely in 2d) and $\cal T$ is therefore $-2$.

Notice that there is a sign difference between the usual 
loops observables \cite{Rovelli88,Rovelli90} (denoted
$T$-variables) and these new loop observables, denoted
${\cal T}$-variables.  This is a key technicality at the
origin of the simplification of the formalism presented
here.  The new sign takes care immediately of the sign
complications extensively discussed in reference
\cite{Rovelli95a}.  The suggestion that those sign
complications could be avoided by inserting a minus sign
in front of the trace was considered by S.~Major as
well\cite{major}. Let us illustrate the consequences of having this
sign. Consider an $N$ component multiloop $\alpha =
\alpha_1 \alpha_2 \ldots \alpha_N$. We have:
\begin{eqnarray*}
 {\cal T}[ ~[\alpha_1]\cdots[\alpha_N]~] 
    &=& {\cal T}[\alpha_1]\cdots{\cal T}[\alpha_1]
 \\ &=& (-{\rm Tr}[U_{\alpha_1}]) \cdots (-{\rm Tr}[U_{\alpha_N}])
 \\ &=& (-1)^N~ {\rm Tr}[U_{\alpha_1}] \cdots {\rm 
Tr}[U_{\alpha_N}]
 \\ &=& (-1)^N ~T[ ~\{ \alpha\}~].
\end{eqnarray*}
This shows that the new sign choice implements in the
formalism the sign factor for the number of loops that
was recognized in \cite{Rovelli95a} as the key to
transform the spinor relation into a local relation. In
fact, we have 
\begin{eqnarray}
   {\rm Tr}[U_\alpha]{\rm Tr}[U_\beta] 
 - {\rm Tr}[U_\alpha U_\beta] 
 - {\rm Tr}[U_\alpha U_{\beta^{-1}}] = 0  ,
\\
    {\cal T}[\alpha] ~ {\cal T}[\beta]
  + {\cal T}[ \alpha \#_s \beta ]
  + {\cal T}[ \alpha \#_s \beta^{-1} ] = 0 ,
\\  
    {\cal T}[ ~[\alpha][\beta]~]
  + {\cal T}[ ~[\alpha \#_s \beta]~ ]
  + {\cal T}[ ~[\alpha \#_s \beta^{-1}]~ ] = 0 ; 
\label{spinor}
\end{eqnarray}
namely the spinor identity (one $+$ and two $-$) has become
a binor identity (all $+$) (see Penrose
\cite{Penrose71}).  While the first is non local, the
second is local, and is the basic identity at the roots
of binor calculus and $A=-1$ recoupling theory. For the 
notations $\circ$ and $\#$ (to be used in a moment), see for 
instance \cite{Rovelli91}.

We recall here, for later use, the retracing identity.
For all loops $\alpha$ and segments $\gamma$, we have
\cite{Rovelli90}
\begin{equation}
 {\cal T}[\alpha]
  = {\cal T}[\alpha\circ\gamma\circ\gamma^{-1}].
\label{retracing}
\end{equation}

The Poisson bracket algebra of these loop variables is
easily computed.  For a rigorous way of performing these
computations, see \cite{Rovelli93}. We give here the
Poisson bracket of the ${\cal T}$ variables of order $0$
and $1$.
\begin{eqnarray}
 \{ {\cal T}[\alpha] , {\cal T}[\beta] \} &=& 0 , \\
 \{ {\cal T}^a[\alpha](s) , {\cal T}[\beta] \} 
     &=& - G~ \Delta^a[\beta,s] ~\cdot
        ~\frac{1}{2} \left\{ {\cal T}[\alpha \#_s \beta]  
                           -{\cal T}[\alpha \#_s \beta^{-1}]  
        \right\} ,
\end{eqnarray}
where we have defined:
\begin{equation}
\Delta^a[\beta,s] = \int_\beta d\tau ~\dot{\beta}^a(\tau) 
                   \delta^3 [\beta(\tau),s].
\label{delta}
\end{equation} 
The factor $-\frac{1}{2}$, different than in previous
papers, is due to the new conventions.


\section{The loop representation of quantum gravity}

We now define the loop representation\cite{Gambini81} of
quantum gravity as a linear representation of the Poisson
algebra of the ${\cal T}$ variables.  First, we define the
carrier space of the representation.  To this aim, we
consider the linear subspace $\cal K$ of the free loop
algebra defined by
\begin{equation}
{\cal K} = \{\Phi\in {\cal A}^f[{\cal L}]\ \ |\ \ {\cal T}[\Phi] 
=0\},
\end{equation}
and we define the carrier space $\cal V\ $ of the
representation by
\begin{equation}
{\cal V}= {\cal A}^f[{\cal L}]/{\cal K}.
\end{equation}
In other words, the state space of the loop representation
is defined as the space of the equivalence classes of
linear combinations of multiloops, under the equivalence
defined by the Mandelstam relations
\begin{equation}
  \Phi\sim\Psi$\ \ {\rm if} \ \ ${\cal T}[\Phi] ={\cal
T}[\Psi],
\label{eq:Mandelstam}
\end{equation}
namely by the equality of the corresponding holonomies
\cite{Ashtekar92} .\footnote{${\cal T}[\Phi]$ is a function
on configuration space, namely a function over the space of
smooth connections. Equality between functions means of
course having the same value for any value of the
independent variable; here, for all (smooth) connections.}
We denote the equivalence classes defined in his way, namely
the elements of the quantum state space of the theory as
Mandelstam classes, and we indicate them in Dirac notation
as ${\langle \Phi |}$.  Clearly, the multiloop states
${\langle \alpha |}$ span (actually, overspan) the state
space $\cal V$. Later we will define a scalar product on
$\cal V$, and promote it to a Hilbert space.  The reason for
preferring a bra notation over a ket notation is just
historical at this point. We recall that the loop
representation was originally defined in terms of kets
$|{\psi}\rangle$ in the dual of $\cal V$.  These are
represented on the (overcomplete) basis ${\langle \alpha |}$
by loop functionals
\begin{equation}
    \psi(\alpha)=\langle\alpha | \psi\rangle.
\end{equation}

The principal consequences of the Mandelstam relations are
the following.
\begin{enumerate}
\item The element ${\langle \alpha |}$ does not depends on the
	orientation of $\alpha$: $[\alpha] \sim
	[\alpha^{-1}]$.
\item The element ${\langle \alpha |}$ does not depend on the
      parameterization of $\alpha$: $[\alpha] \sim [\beta]\
      \ \ {\rm if}\ \ \beta^a(\tau) = \alpha^a(f(\tau))$.
\item Retracing: if $\gamma$ is a {\it segment} starting
      in a point of $\alpha$ then.  \begin{equation}
      [\alpha\circ\gamma\circ\gamma^{-1}] \sim [\alpha]
      \label{ri}. \end{equation}
\item {Binor identity:} \begin{equation} [\alpha]
      \cdot [\beta] \sim - [\alpha \#_s \beta] - [\alpha
      \#_s \beta^{-1}]. \label{binor} \end{equation}
\end{enumerate}
It has been conjectured that all Mandelstam relations can
be derived by repeated use of these identities. We expect
that the methods described below may allow to prove this
conjecture, but we do not discuss this issue here.

Next, we define the quantum operators corresponding to the
$\cal T$-variables as linear operators on $\cal V$. These
form a representation of the loop variables Poisson algebra.
We define the loop operators as acting on the bra states
${\langle \Phi |}$ from the right.  (Since they act on the
right, they define, more precisely, an {\it
anti}-representation of the Poisson algebra.)  We define the
${{\hat{\cal T}}}[\alpha] $ operator by
\begin{eqnarray}
\bigg\langle {c_0 + \sum_i c_i ~[\alpha_i]
  + \sum_{ij} c_{ij} ~[\alpha_i][\alpha_j]~
  + \ldots}\bigg| {\hat{\cal T}}[\alpha] 
\label{3.7} 
&=&\bigg\langle {c_0 [\alpha]+ \sum_i c_i ~[\alpha_i] [\alpha]
  + \sum_{ij} c_{ij} ~[\alpha_i][\alpha_j] [\alpha]
  + \ldots}\bigg| \ \ .
\end{eqnarray}
Next, we define the ${\hat{\cal T}}^a[\alpha](s)$
operator. This is a derivative operator (i.e. it satisfies
Leibniz rule) over the free loop algebra such that
\begin{equation}
   {\langle [\beta] |} {{\hat{\cal T}}}^a[\alpha](s) 
    = -{\rm i} l^2_0 \Delta^a[\beta,s] 
    {1\over 2}
   \left( {\langle \![\alpha\#_s\beta] |}
  - {\langle [\alpha\#_s\beta^{-1}] |}\right),
\label{t1}\label{3.8} 
\end{equation}
where we have introduced the elementary length $l_0$ by
\begin{equation}
 l^2_0 = \hbar G = \frac{16 \pi \hbar G_{{\rm Newton}}}{c^3} 
 = 16 \pi\ l^2_{Planck}.
\end{equation}
The definition extends on the entire free loop algebra by
Leibniz rule and linearity.  The two operators commute with
the Mandelstam relations and are therefore well defined on
$\cal V$.

Notice that the factor $\Delta^a[\beta,s]$ in (\ref{t1})
depends on the orientation of the loop $\beta$: it changes
sign if the orientation of $\beta$ is reversed.  So does the
difference in the parentheses, therefore the r.h.s of
(\ref{t1}) is independent from the orientation of $\beta$,
as the l.h.s..  On the other hand, both the r.h.s and the
l.h.s of (\ref{t1}) change sign if we reverse the
orientation of $\alpha$.

The action of the ${{\hat{\cal T}}}^a[\alpha](s)$ operator
on a state ${\langle [\beta] |}$ can be visualized
graphically.  The graphical action is denoted a ``grasp'',
and it can be described as follows: i. Disjoin the two edges
of the loop $\beta$ and the two edges of the loop $\alpha$,
that enter the intersection point $s$.  ii.  Pairwise join
the four open ends of $\alpha$ and $\beta$ in the two
possible alternative ways. This defines two new
states. Consider the difference between these two states
(arbitrarily choosing one of the two as
positive). iii. Multiply this difference by the factor
$-{\rm i} l_0^2~\Delta^a[\beta,s]$, where the direction of
$\beta$ (which determines the sign of $\Delta^a[\beta,s]$)
is determined as follows: it is the direction induced on
$\beta$ by $\alpha$ (which {\it is} oriented) in the term
chosen as positive.  A moment of reflection shows that the
definition is consistent, and independent from the choice of
the positive term.  An explicit computation shows that the
operators defined realize a linear representation of the
Poisson algebra of the corresponding classical observables.

The grasping rule generalizes to higher order ${\cal
T}$-variables.  The action of ${\hat{\cal T}}^{a_1\ldots
a_n}[\alpha](s_1,\ldots,s_n)$, over a single loop-state
$[\beta]$ is given as follows. First the result vanishes
unless $\beta$ crosses all the $n$ points $s_i$. If it does,
the action of ${\hat{\cal T}}^{a_1\ldots
a_n}[\alpha](s_1,\ldots,s_n)$ is given by the simultaneous
grasp on all intersection points.  This action produces
$2^n$ terms.  These terms are summed algebraically with
alternate signs, and the result is multiplied by a factor
$-{\rm i} l_0^2 ~\Delta^a[\beta,s_1]$ for each grasp, where
the sign of each coefficient $\Delta^a[\beta,s_i]$ is
determined assuming that $\beta$ is oriented consistently
with $\alpha$ in the term chosen as positive.  Again, a
moment of reflection shows that the definition is
consistent, and independent from the choice of the positive
terms.  The generalization to arbitrary states, using
linearity and the Leibnitz rule, is straightforward.  This
concludes the construction of the linear ingredients of the
loop representation.


\section{Loop states and recoupling theory}

A quantum state ${\langle \Phi |}$ in the state space $\cal V$
is a Mandelstam equivalence class of elements of the form
(\ref{fla}).  We now show that because of the equivalence
relation, these states are related to tangles --in the
sense of Kauffman \cite{Kauffman91}-- and they obey the
formal identities that define the Temperley-Lieb-Kauffman
recoupling theory described in Ref.\cite{Kauffman94}.
This fact yields two results. First, we can write a basis
in $\cal V$. This basis is constructed in the next
section. Second, recoupling theory becomes a powerful
calculus in loop quantum gravity.

Consider the element $\Phi$, given in (\ref{fla}), of the
vector space ${\cal A}^f[{\cal L}]$.  We need some
definitions.
\begin{description}
\item[Graph of a state. ]  We denote the union in ${M}$ of
the images of all the loops in the r.h.s of (\ref{fla})
as the ``graph of $\Phi$'', and we indicate it as
$\Gamma_\Phi$.  Notice that $\Gamma_\Phi$ is a graph in
the sense of graph theory \cite{Graph}, embedded in ${M}$.
\item[Vertex. ] We denote the points $i$ where $\Gamma_\Phi$
	fails to be a smooth submanifold of ${M}$ as
	``vertices''.
\item[Edge. ] We denote the lines $e$ of the graph connecting
	the vertices as ``edges''.
\item[Valence. ] We say that a vertex $i$ has valence $n$, or
is $n$-valent, if $n$ edges are adjacent to it.  A vertex
can have any positive integer valence, including 1 and 2.
\end{description}
Clearly, ${\Phi}$ is not uniquely determined by its graph 
$\Gamma_\Phi$.  
If our only information about a state is its graph, 
then we do not know how the state is decomposed into 
multiloops, nor how many single loops run along each edge, 
nor how the single loops are rooted through the vertices. We
now introduce a graphical technique to represent this missing 
information.  The technique is based on the idea of 
``blowing up'' the graph -as if viewed through 
an infinite magnifying glass- and representing the 
additional information in terms of planar tangles on the 
blown up graph.  As we will see, these tangles obey
recoupling theory.

First, draw a graph isomorphic to $\Gamma_\Phi$
in the sense of graph theory (that is, the isomorphism
preserves only adjacency relations between vertices and
edges), on a two dimensional surface.  As usual in graph
theory, we must distinguish points representing vertices
from accidental intersections between edges generated by
the fact that we are representing a non-planar graph on a
plane.  Denote these accidental intersections as ``false
intersections''.  Next, replace each vertex (not the false
intersections) by (the interior of) a circle in the plane,
and each edge by a ribbon connecting two circles. (At false
intersections, ribbons bridge each other without merging.)
In this way, we construct a ``thickened out'' graph: a
two-dimensional oriented surface which (loosely speaking)
has the topology of the graph $\Gamma_\Phi$ times the
$[0,1]$ interval.
\begin{description}
\item[Ribbon-net. ] We call this two-dimensional surface
	the ``ribbon-net'' (or simply the ribbon) of the
	graph $\Gamma_\Phi$, and we denote it as $R_\Phi$.
	Notice that the graph $\Gamma_\Phi$ is embedded in
	$M$, while its ribbon-net $R_\Phi$ is not.
\end{description}
Now we can represent the missing information needed to
reconstruct $\Phi$ from $\Gamma_\Phi$ as (a formal linear
combination of) tangles drawn on the surface $R_\Phi $.
First, we represent each multiloop in (\ref{fla}) by means
of a closed line over $R_\Phi $:
\begin{description}
\item[Planar (representation of a) multiloop. ]  For each
	loop $\alpha_i$ in a given multiloop $\alpha$ we
	draw a loop $\alpha_i$ over the ribbon-net $R_\Phi
	$, wrapping around $R_\Phi $ in the same way in
	which $\alpha_i$ wraps around $\Gamma_\Phi$.  We
	denote the drawing (over $R_\Phi $) of all the
	loops of a multiloop as ``the planar
	representation'' of the multiloop $\alpha$, or
	simply as the ``planar multiloop''.  We indicate
	it as $P_\alpha$.
\end{description}
For technical reasons, we allow edges and vertices of the
ribbon-net to be empty of loops as well.  Thus, we
identify a ribbon-net containing a planar multiloop, with
a second one obtained from the first by adding edges and
vertices empty of loops.  Finally:
\begin{description}
\item[Planar (representation of a) state. ] Every state
   ${\langle \Phi |}$ is a formal linear combination of
   multiloops: ${\langle \Phi |}=\sum_j c_j~ [\alpha_j]$ (up to
   equivalence).  We denote the corresponding formal
   linear combination $P_\Phi=\sum_j c_j~P_{\alpha_j}$ of
   planar multiloops on the ribbon-net $R_\Phi $ (up to
   equivalence), as a planar representation of
   ${\langle \Phi |}$.
\end{description}

We have split the information contained in $\Phi$ in two
parts: $\Phi$ determines a graph $\Gamma_\Phi$ embedded in
${M}$ and a planar state $P_\Phi$. $P_\Phi$ is a linear
combinations of drawings of loops over a surface (the
ribbon-net $R_\Phi $) and codes the information on which
loops are present and how they are rooted through
intersections.  This information is {\it purely
combinatorial\ }. On the other hand, $\Gamma_\Phi$ contains
the information on how the loops are embedded into ${M}$.

Notice that a multiloop determines its planar representation
only up to smooth planar deformations of the lines within
the circles and the ribbons of the ribbon-net. In other
words, we can arbitrarily deform the lines within each
circle and within each ribbon, without changing $\Phi$.  In
particular, the lines of the planar representation will
intersect in points of $R_\Phi $, and we can apply
Reidemeister\cite{Reidemeister} moves \cite{Kauffman91} to
such intersections (that is, disentangle them). Under- and
over-crossings of loops within $R_\Phi $ are not
distinguished.

\begin{figure}[t]
$$
\begin{array}{c}\mbox{\epsfig{file=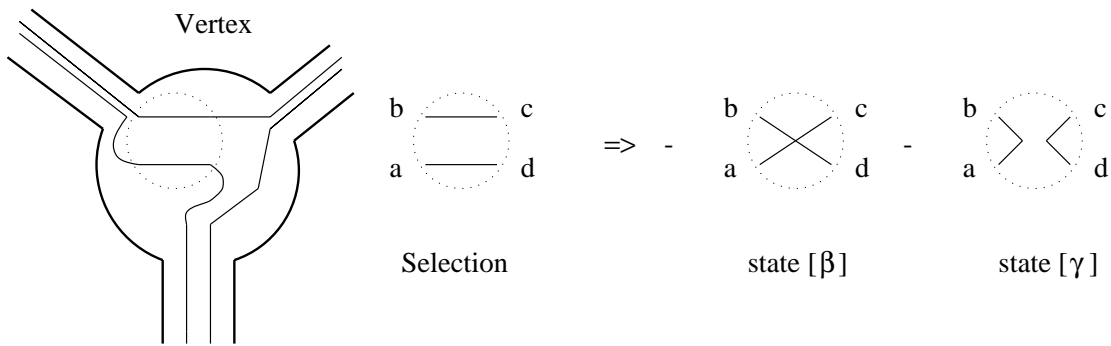}}\end{array}
$$
\caption{The binor identity.}
\label{FigBinorID}
\end{figure}

Let us come to the key observation on which the possibility of
using recoupling theory relies.  Consider an element $\Phi$ of
the free vector algebra.  For simplicity, let us momentarily
assume that $\Phi$ is formed by a single loop $\Phi = [\alpha]$
(which may self-intersect and run over itself).  Thus $\Phi =
(\Gamma_\alpha, P_\alpha)$.  Consider an intersection of two
lines (two segments of $P_\alpha$) in $R_\Phi $.  Break the two
lines meeting at this intersection, and pairwise rejoin the
four legs, in the two alternative possible ways, as in Figure
\ref{FigBinorID}.

We obtain two new loops on $R_\Phi $, which
we denote as $P_{[\beta]}$ and $P_{[\gamma]}$.  Consider the
element $\Psi$ of the free vector algebra uniquely determined
by the graph $\Gamma_\Psi=\Gamma_\Phi$, and by the linear
combination of planar representations
$P_\Psi=-P_\beta-P_\gamma$. Notice that $\Psi$ is different 
than $\Phi$ as an element of the free vector
algebra; however, the two are
in the same Mandelstam equivalence class because of the binor
relation (\ref{binor}), and therefore they define the same
element of the quantum state space $\cal V$.  Namely
${\langle \Psi |}= {\langle \Phi |}$.  We say that two planar
representations $P_\Phi$ and $P_\Psi$ are ``equivalent'' if
${\langle \Psi |}= {\langle \Phi |}$.  Thus, in dealing with planar
representations of a quantum state ${\langle \Phi |}$, we can freely
use the identity
\begin{equation}
\begin{array}{c}{\setlength{\unitlength}{1 pt}
\begin{picture}(10,15) 
\put( 0,0){\line( 2,3){10}}
\put(10,0){\line(-2,3){10}}
\end{picture}}\end{array}
= - \begin{array}{c}{\setlength{\unitlength}{1 pt}
\begin{picture}(10,15) 
\put( 0,0){\line(0,1){15}}
\put(10,0){\line(0,1){15}}
\end{picture}}\end{array}
 - \begin{array}{c}{\setlength{\unitlength}{1 pt}
\begin{picture}(10,15) 
\put(5, 0){\oval(10,10)[t]}
\put(5,15){\oval(10,10)[b]} 
\end{picture}}\end{array}
\label{gbi} 
\end{equation}
on $P_\Phi$ without changing the quantum state. This
identity is the identity (i) in page 7 of reference
\cite{Kauffman94}, (equation (\ref{eq:KB1}) in Appendix B) 
which is the key axiom of recoupling
theory -- with the value of the $A$ parameter set to $-1$.

An easy to derive consequence is that every closed line
entirely contained within a circle, or within a ribbon,
can be replaced by a factor $d=-2$.  Furthermore, it is
easy to see that the retracing identity (\ref{ri}) implies
that the loops of $P_\Phi$ can be arbitrarily deformed
within the {\it entire\ } ribbon-net, without changing the
state ${\langle \Phi |}$. In particular, every loop contractible
in $R_\Phi $ can be replaced by a factor $d=-2$.  This is
the second axiom of recoupling theory (equation 
(\ref{eq:KB2}) in Appendix B) in the $A=-1$
case. The value $A=-1$, correspond to the case in which
the distinction between over- and under-crossings can be
neglected, consistently with the fact such distinction is
irrelevant for the planar representation of a loop.

Thus $P_\Phi$ can be interpreted as a linear combination
of tangles in the sense of reference
\cite{Kauffman94}. The tangles obey the axioms of
recoupling theory. They are confined inside the oriented
surface $R_\Phi$ with has a highly nontrivial topology.
This is the key result of this section.

The relation between loop states and recoupling theory is
subtle, and may generate confusion.  A source of confusion
is given by the fact that the relation between recouplings
and knots in knot theory \cite{Kauffman94} is different
from the relation between recouplings and knots in quantum
gravity.  In both cases recouplings enters as a
consequence of a skein (or binor) equation --as equation
(\ref{gbi})-- holding at intersections.  But in knot
theory this equation is satisfied by the Kauffman brackets
at the ``false intersections'' of the planar projection of
a loop.  Contrary to this, in quantum gravity equation
(\ref{gbi}) {\it does not hold} for the false
intersections. It holds for the intersections of lines
{\it within} $R_\Phi$.

On the other hand, knots play a role in quantum gravity as well
\cite{Rovelli88}, because of the diffeomorphism
constraint. GR's diffeomorphism invariance identifies states
that have equivalent $P_\Phi$, and whose {\it graphs} can be
deformed into each other by 3-d diffeomorphism of $M$ in the
connected component of the identity. To clarify this point, let
us require that the ribbon-net $R_\Phi$ is generated by a two
dimensional projection of $\Gamma_\Phi$, and let us keep track
of the resulting over- and under- crossings at false
intersections. Then diffeomorphism invariance identifies all
states that have equivalent $P_\Phi$, and whose ribbon-nets can
be transformed into each other by Reidemeister moves
{\it at the false intersections }. Thus, as
far as diff-invariant states are concerned, Reidemeister moves
can be used at the tangles' intersections {\it within the
ribbon-net\ } as well as at the false intersections.  But in
the first case a skein equation (equation (\ref{gbi})) holds,
in the second it doesn't.\footnote{This is true in
general. One may wonder if there is any special quantum state
${\langle \Phi_0 |}$ for which the relation (\ref{gbi}) holds at
false intersections as well.  The possibility that such a
special state could exist in quantum gravity has been explored,
with various motivations, by various authors
\cite{Smolin95a,FalseIntersec}.
} Mixing up the two cases has generated a certain confusion in
the past.

An immediate consequence of the result is that we can write a
basis in $\cal V$ following \cite{Kauffman94}. Given a state
${\langle \Psi |}$, and its ribbon-net $R_\Phi $, we can use
(\ref{gbi}) to eliminate all intersections from the $P_\alpha$
of each multiloop.  Next, we can retrace each single line that
returns over itself, and eliminate every loop contractible in
$R_\Phi$.  We obtain parallel lines without intersections along
each ribbon and routings without intersections at each vertex.
No further use of the retracing or binor identity is then
possible without altering this form.  This procedure defines a
basis of independent states, labeled by the graph, the number
of lines along each edge, and elementary routings at each node.
An elementary routing is a planar rooting of loops through the
vertex of the ribbon-net, having no intersections.  This basis
is not very practical for calculations. In the next section, we
use the technology of \cite{Kauffman94} to define a more useful
basis.\footnote{A basis in a linear space is a set of linearly
independent vectors that span the linear space.  The fact that for 
the moment we are still
working in linear spaces without fixing a scalar product (we will
fix a scalar product only later, in section VIII) has raised
some confusion in the past. It is perhaps worthwhile recalling
that the notions of basis, eigenvalues and eigenvectors are
well defined notions for linear spaces, not just for Hilbert
spaces.  (They do not require a scalar product to be defined in
order to make sense.)  Similarly, the fact that a linear
operator is diagonalizable, or has real eigenvalues does not
depend on the presence of a scalar product. Given an arbitrary
linear basis $v_i$ in a finite dimensional linear space, a
linear operator $A$ is hermitian in this basis if its matrix
elements (defined by $(Av)_i = A_i^j v_j$) satisfy
$A_i^j=\bar{A}_j^i$. If $A$ is hermitian in a basis, then $A$
is diagonalizable and has real eigenvalues. This is true
independently from any scalar product.}

\section{The spin network basis}

The representation $(\Gamma_\Phi, P_\Phi)$ of a state
${\left\langle{{\Phi}}\right|}$ can be expanded in terms of a ``virtual''
trivalent representation as follows.

\begin{figure}[t]
$$
\begin{array}{c}\mbox{\epsfig{file=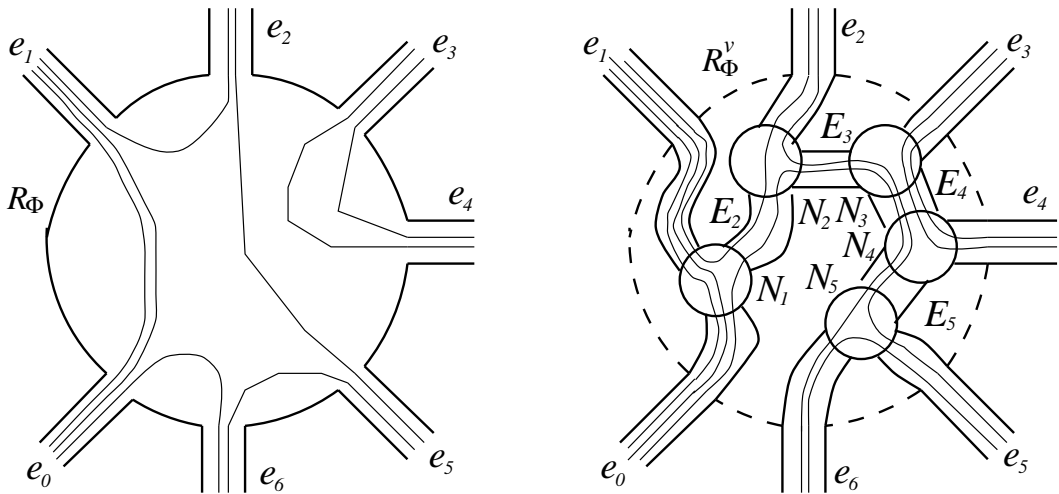}}\end{array}
$$
\caption{Construction of ``virtual'' vertices and ``virtual''
strips over an $n$-valent vertex.}
\label{TheVirtualNode}
\end{figure}

\begin{description}
\item[Virtual graph. ]  To every graph $\Gamma$, we can
	associate a trivalent graph $\Gamma^v$ as follows.
	For each $n$-valent vertex $v$ of $\Gamma$,
	(arbitrarily) label the adjacent edges as $e_0
	... e_{(n-1)}$, and disjoin them from $v$.  Then,
	replace $v$ with $n-2$ trivalent vertices $N_1
	... N_{n-2}$, denoted ``virtual'' vertices.  Join
	the virtual vertices with $n-3$ ``virtual'' edges
	$E_2...E_{n-2}$, where $E_i$ joins $N_{i-1}$ and
	$N_i$.  Prolong the edges $e_2 ... e_{(n-1)}$ to
	reach the corresponding virtual vertices $N_1
	... N_{n-2}$, and the edges $e_1$ and $e_{(n-1)}$
	to reach the virtual vertices $N_1$ and $
	N_{n-2}$.  Denote the resulting trivalent graph
	$\Gamma^v$ as the virtual graph associated to
	$\Gamma$ (for the chosen ordering of edges).
\item[Virtual ribbon-net. ] We denote the ribbon-net of
	$\Gamma^v_\Phi$ as the virtual ribbon-net
	$R^v_\Phi$ of $\Phi$. We view it as a subset of
	$R_\Phi$, namely we view the virtual circles $N_1
	... N_{n-2}$ and the virtual ribbons
	$E_2...E_{n-2}$ as drawn inside the circle $c$
	representing $v$.  This circle $c$ indicates that
	the virtual vertices $N_1 ... N_{n- 2}$ correspond
	all to the same point of $M$.
	(Thus, a virtual ribbon-net is a trivalent
	ribbon-net with strings of adjacent intersections
	specified.)
\item[Virtual representation. ] Finally, deform $P_\Phi$
   	so that it lies entirely inside $R^v_\Phi$.  We
   	indicate the deformed $P_\Phi$ as $P^v_\Phi$, and
   	call it the ``virtual'' planar representation of
   	$\Phi$.  The virtual representation $P^v_\Phi$ of
   	a state is not unique, due to the arbitrariness of
   	assigning the ordering $e_0 \ldots e_{(n-1)}$ to
   	the edges of $n$-valent intersections.
\end{description}
The above construction is more difficult to describe in
words than to visualize, and is illustrated in 
Figure \ref{TheVirtualNode}.

Consider now deformations of the tangle $P^v_\Phi$ within
$R^v_\Phi$ --a subset of the deformations within the full
$R_\Phi $.  We can move all intersections of deform
$P^v_\Phi$ away from the vertices (to the virtual or real
ribbons), leaving trivalent vertices free from
intersections.  Next, we can use the binor relation to
remove all intersections from the ribbons, leaving
non-intersecting tangles with $n$ inputs and $n$ outputs
along each single ribbon $e$.  As described in Sec 2.2 of
\cite{Kauffman94}, tangles of this kind can be described
as elements of the (tangle-theoretic) Temperley-Lieb
algebras $T^{(e)}_n$.  A basis of this algebra is obtained
by using the Jones-Wenzl projectors $\Pi^{(e)}_n$.  Since
we are here in the case $A=-1$, the $\Pi^{(e)}_n$ are just
normalized antisymmetrizers.  More precisely, given the
multiloop $P_\alpha$ with $n$ lines along the ribbon $e$,
call $P^{(p)}_{\alpha}, p=1...n!$ the multiloops obtained
by all possible permutations $p$ in the way the $n$ lines
entering $e$ are connected to the $n$ outgoing lines, and
$|p|$ the parity of the permutation, then
\begin{equation}
	\Pi^{(e)}_n P_\alpha = \frac{1}{n!} \sum_p
(-1)^{|p|}\ P^{(p)}_{\alpha}.
\end{equation}
Notice the ${1}/{n!}$ factor, which was not present in
previous conventions \cite{Rovelli95}.  It follows from
the completeness of the Jones-Wenzel projectors that a
basis for all planar loops over a given $R^v_\Phi$
is given by the linear combination of 
loops in which the lines along each
(virtual and real) edge are fully antisymmetrized.  We can
therefore expand every state in states in which lines are
fully antisymmetrized along each ribbon.  A state in which
the lines along each (virtual or real) ribbon are fully
antisymmetrized is a spin network state.  Thus, we recover
the result of reference \cite{Rovelli95a}, to which we
refer for details.

A spin network state is characterized by a graph $\Gamma$
in $M$, by the assignment of an ordering to the edges
adjacent to each vertex, and by the number $p_e$ of
(antisymmetrized) lines in each virtual or real edge 
$e$. We denote the integer $p_e$ as the ``color'' of the
corresponding edge $e$ of $\Gamma^v$.  We will use also
the ``spin'' $j_e$ of the edge, defined as half its color:
$j_e =\frac{1}{2} p_e$ .\footnote{The oscillation between
the historically motivated half integer terminology  ``spin''
and the rationally motivated integer terminology ``color''
goes back to Penrose's papers on spin networks
\cite{Penrose71}.}  At each vertex, the colors $p_1$,
$p_2$ and $p_3$ of the three adjacent edges satisfy a
compatibility condition: there must exist three positive
integers $a$, $b$ and $c$ (the number of lines rooted
through each pair of edges) such that
\begin{equation}
p_1 = a+b, \ \ \ p_2=b+c, \ \ \ p_3=c+a,
\label{cond}
\end{equation}
It is easy to see that this condition is equivalent to the
Clebsh-Gordon condition that each of the three $su(2)$
representations of spin $j_i={1}/{2}~p_i$ is contained in
the tensor product of the other two \cite{Penrose71}.

The spin network states form a basis in $\cal V$.  The basis
elements are given as follows. For every graph $\Gamma$
embedded in $M$, choose an ordering of the edges at each
node.  This choice associates an oriented trivalent virtual
graph $\Gamma^v$ (non-embedded) to every $\Gamma$.
\begin{description}
\item[Spin network. ] A spin network $S$ is given by a graph
$\Gamma_S$ in $M$, and by a compatible coloring $\{p_e\}$ of
the associated oriented trivalent virtual graph
$\Gamma^v$. Thus $S=(\Gamma_S, \{p_e\})$.
\item[Spin network state. ] For every spin network $S$, the
	spin network quantum state
	${\left\langle{{S}}\right|}=(\Gamma_S,P_S)$ is the
	element of $\cal V$ determined by the graph
	$\Gamma_S$ and by the linear combination $P_S$ of
	planar multiloops obtained as follows. Draw $p_e$
	lines on each ribbon $e$ of the ribbon-net $R^v_S$;
	connect lines at intersections without crossings;
	this gives a planar multiloop $P^{(0)}_S$; then
\begin{equation} 
 P_S = \prod_{e\in\Gamma} \Pi^{(e)}_{p_e} P^{(0)}_S.
\end{equation}
\end{description}
We can represent a spin network state as a colored trivalent
graph over the ribbon-net $R^v_S$ (with a single edge along
each ribbon).  This representation satisfies the identities
of recoupling theory.  We describe the main ones of these
identities in Appendix E.  As an example, we give here the
formula that allows one to express the basis elements of a
4-valent intersection in terms of the basis elements of a
different trivalent expansion. Using the recoupling theorem
of \cite{Kauffman94} (pg.\ 60), we have immediately
\begin{equation}
\begin{array}{c}\setlength{\unitlength}{1 pt}
\begin{picture}(50,40)
          \put( 0,0){$a$}\put( 0,30){$b$}
          \put(45,0){$d$}\put(45,30){$c$}
          \put(10,10){\line(1,1){10}}\put(10,30){\line(1,-1){10}}
          \put(30,20){\line(1,1){10}}\put(30,20){\line(1,-1){10}}
          \put(20,20){\line(1,0){10}}\put(22,25){$j$}
          \put(20,20){\circle*{3}}\put(30,20){\circle*{3}}
\end{picture}\end{array}
    = \sum_i  \left\{\begin{array}{ccc}
                      a  & b & i \\
                      c  & d & j  
              \end{array}\right\}
\begin{array}{c}\setlength{\unitlength}{1 pt}
\begin{picture}(40,40)
      \put( 0,0){$a$}\put( 0,40){$b$}
      \put(35,0){$d$}\put(35,40){$c$}
      \put(10,10){\line(1,1){10}}\put(10,40){\line(1,-1){10}}
      \put(20,30){\line(1,1){10}}\put(20,20){\line(1,-1){10}}
      \put(20,20){\line(0,1){10}}\put(22,22){$i$}
      \put(20,20){\circle*{3}}\put(20,30){\circle*{3}}
\end{picture}\end{array}
\end{equation}
where the quantities $\left\{\begin{array}{ccc}
a & b & i \\ c & d & j  \end{array}\right\}$ are $su(2)$
six-j symbols (normalized as in \cite{Kauffman94}; see
Appendices).

A side remark should be added. An embedded colored
trivalent graph specify a state $\Phi$ only up to a global
sign, because it does not fix the overall sign of the
antisymmetrized linear combination of multiloops.  To keep
track of this overall sign, one needs {\it oriented}
trivalent graphs, as in reference \cite{Penrose71} where
Penrose considered oriented spin networks and in
\cite{Rovelli95a}.  An orientation of a trivalent graph is
an assignment of a cyclic order to the edges of each node,
modulo $Z_2$ (that is, identifying two orientations if
they differ in an even number of intersections).
$\Gamma^v$ is oriented by the order assigned to the edges
entering each vertex, and ribbon-nets are oriented
(consistently, we assume) as graphs because they are
oriented as two-surfaces: edges can be ordered --say--
clockwise.

\subsection{The Action of the operators in the
 spin-network basis}

We now describe how the ${{\hat{\cal T}}}$ operators act on
the spin network states.  From Eq.\ (\ref{3.7}), the
operator ${\hat{\cal T}}[\alpha]$, acting on a state
${\left\langle{\Phi}\right|}$ simply adds a loop to
${\left\langle{{\Phi}}\right|}$. Consider the graph $\Gamma$
formed by the union (in ${M}$) of the graphs of $\Phi$ and
$\alpha$. Since we admit empty edges, we can represent
$\Phi$ over the ribbon-net $R$ associated to $\Gamma$.  In
this representation, the action of ${\hat{\cal T}}[\alpha]$
consists in adding the draw of $\alpha$ over $R$.  Using the
expression for the Jones-Wenzl projectors in
\cite{Kauffman94} (pg.\ 96), one can expand the
non-antisymmetrized lines, if any, in combinations of
antisymmetrized ones.

\begin{figure}[t]
$$
{{\begin{array}{c}\mbox{
\epsfig{file=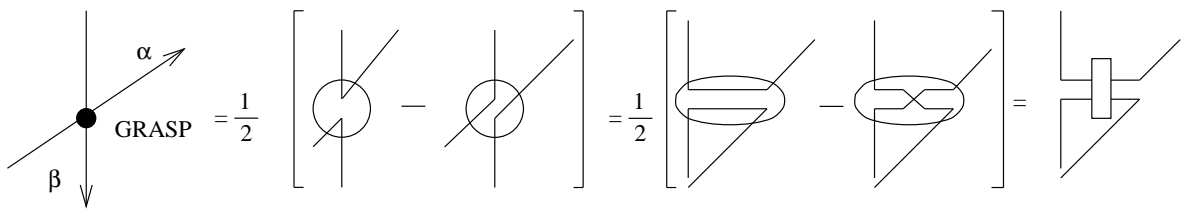}} 
\end{array}}}
$$
\caption{Action of the grasp.}
\label{FigGrasp}
\end{figure}

Higher order loop operators are expressed in terms of the
elementary grasp operation, Eq.\ (\ref{3.8}).  The ribbon
construction allows us to represent the grasp operation in a
simpler form.  Indeed, one easily sees that Eq.\ (\ref{3.8})
is equivalent to the following: acting on an edge with color
1, the grasp creates two virtual trivalent vertices (inside
the same circle, corresponding to the intersection point) --
one on the spin-network state and one the loop of the
operator.  The two vertices are joined by a virtual strip of
color 2, and the overall multiplicative factor is determined
as follows.  The sign of the tangent of $\beta$ in
$\Delta^a[\beta,s]$ is determined by the orientation of
$\beta$ consistent with the positive-terms of the loop
expansion of the spin network.  The equivalence between the
old definition of the grasp and the new one is illustrated
in Figure \ref{FigGrasp}.

A straightforward computation, using Leibnitz rule, shows
that acting on an edge with color $p$, the grasp has the
very same action, with the multiplicative factor multiplied
by $p$. Finally, notice that the two antisymmetrized loops
form a (virtual) spin network edge of color 2. Therefore, we
can express the action of the grasp in the spin network
basis by the following equation
\begin{equation}
  \begin{array}{c}\mbox{\epsfig{file=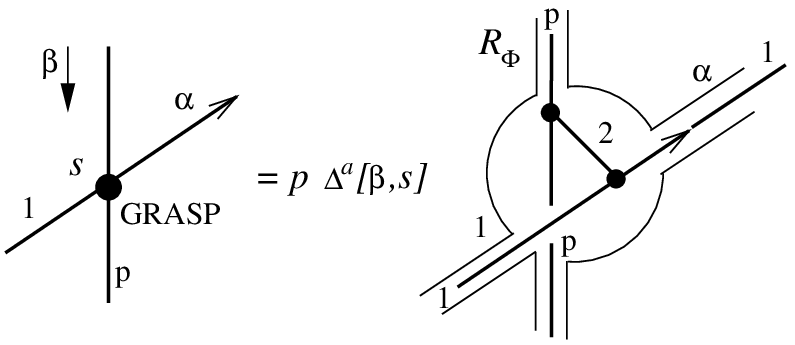}}\end{array}
\end{equation}
This simple form of the action of the loop operators on the
spin-network basis is the reason that enables us to use
recoupling-theory in actual calculations involving quantum
gravity operators.  Notice that it is the ribbon-net
construction that allows us to ``open up'' the intersection
point and represent it by means of two vertices (one over
$\alpha$ and one over $\beta$) and a (``zero length'') edge
connecting the two vertices.  These two vertices and this
edge are all in the same point of the three-manifold M.

\begin{figure}[t]
$$
\begin{array}{c}\mbox{\epsfig{file=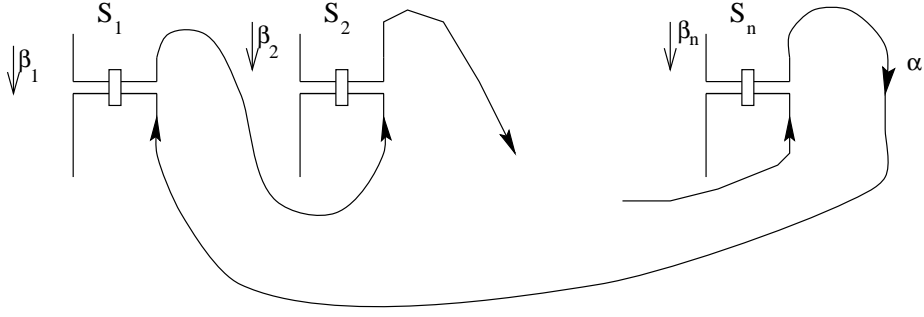}}\end{array}
$$
\caption{Representation of the $n$ grasp
of the ${\cal T}^{a_1\ldots a_n}[\alpha](s_1,\ldots,s_n)$ 
operator.} 
\label{TheTNoperatore}
\end{figure}

Higher order loop operators act similarly, as sketched in
Figure \ref{TheTNoperatore}.

\section{The area operator}

A surface $\Sigma$ in $M$ is an embedding of a
2-dimensional manifold $\Sigma$, with coordinates
$\sigma^u=(\sigma^1,\sigma^2),~~u,v=1,2$, into $M$.  We
write $S:\Sigma \longrightarrow M^3, \sigma^u
\longrightarrow ~x^a(\sigma)$. The metric and the normal
one form on $\Sigma$ are given by
\begin{eqnarray}
g^{\Sigma}   &=& S^\star~g,     
\qquad\qquad  g^{\Sigma}_{uv}= 
\frac{\partial x^a}{\partial \sigma^u}
\frac{\partial x^b}{\partial \sigma^v}  g_{ab};  \\
n_a  &=& \frac{1}{2}\epsilon^{uv} \epsilon_{abc} 
\frac{\partial x^b}{\partial \sigma^u}
\frac{\partial x^c}{\partial \sigma^v}.
\end{eqnarray}
The area of $\Sigma$ is 
\begin{eqnarray}
A[\Sigma] &=& \int_\Sigma\!d^2\sigma~\sqrt{\det g^{\Sigma}} 
= \int_\Sigma\!d^2\sigma~\sqrt{\frac{1}{2} 
\epsilon^{u\bar{u}}\epsilon^{\bar{v}\bar{v}} 
g^{\Sigma}_{uv}g^{\Sigma}_{\bar{u}\bar{v}}} 
= \int_\Sigma\! d^2\sigma
~\sqrt{n_a n_b \tilde{E}^{ai} \tilde{E}^{b}_{i}}, 
\label{area}
\end{eqnarray}
where we have used
\begin{eqnarray*}
\epsilon^{u\bar{u}}\epsilon^{\bar{v}\bar{v}} 
g^{\Sigma}_{uv} g^{\Sigma}_{\bar{u}\bar{v}}
&=& \epsilon^{u\bar{u}}\epsilon^{\bar{v}\bar{v}} 
    \frac{\partial x^a}{\partial \sigma^u}
    \frac{\partial x^b}{\partial \sigma^v}  
         g_{ab} 
    \frac{\partial x^{\bar{a}}}{\partial \sigma^{\bar{u}}}
    \frac{\partial x^{\bar{b}}}{\partial \sigma^{\bar{v}}}  
         g_{\bar{a}\bar{b}}  , 
\\
\epsilon^{u\bar{u}} 
\frac{\partial x^a}{\partial \sigma^u}
\frac{\partial x^{\bar{a}}}{\partial \sigma^{\bar{u}}}
      &=& \frac{1}{2} \epsilon^{u\bar{u}} 
        \frac{\partial x^{a'}}{\partial \sigma^u}
        \frac{\partial x^{\bar{a}'}}{\partial \sigma^{\bar{u}}}
        \epsilon_{a'\bar{a}'c} \epsilon^{a\bar{a}c}
       = n_c  \epsilon^{a\bar{a}c},   \\
g g^{c\bar{c}} 
      &=& \frac{1}{2}
          \epsilon^{a\bar{a}c} \epsilon^{b\bar{b}\bar{c}}
          g_{ab} g_{\bar{a}\bar{b}}. 
\end{eqnarray*}
(On the role of played by surface area in the Ashtekar's
formulation of GR, see \cite{Rovelli93b}.)  We want to
construct the quantum area operator $\hat A[\Sigma]$,
namely a function of the loop representation operators whose
classical limit is $A[\Sigma]$.  Following conventional
quantum field theoretical techniques, we deal with
operator products by defining $\hat A[\Sigma]$ as a limit
of regularized operators $\hat A_{\epsilon}[\Sigma]$ that
do not contain operator products.  The difficulty in the
present context is to find a regularization that does not
break general covariance.  This can be achieved by a
geometrical regularization \cite{Ashtekar92a,Smolin93}.

Following \cite{Rovelli95}, we begin by constructing a
classical regularized expression for the area, namely a
one parameter family of classical functions of the loop
variables $A_\epsilon[\Sigma]$ which converges to the area
as $\epsilon$ approaches zero.  Consider a small region
$\Sigma_\epsilon$ of the surface $\Sigma$, whose
coordinate area goes to zero with $\epsilon^2$.  For every
$s$ in $\Sigma$, the smoothness of the classical fields
implies that $\tilde{E}^a(s) = \tilde{E}^a(x_I) +
O(\epsilon)$, where $x_I$ is an arbitrary fixed point in
$\Sigma_\epsilon$. Also, $U_\alpha(s,t)_A^{~B} =
\openone_A^{~B} + O(\epsilon)$ for any $s,t\in \Sigma_I$
and $\alpha$ a (coordinate straight) segment joining $s$
and $t$.  It follows that (because of (\ref{tr2tau})) to
zeroth order in $\epsilon$
\begin{eqnarray}
 {\cal T}^{ab}[\alpha_{st}](s,t) 
     &=& - {\rm Tr}\left[
             \tilde{E}^a(s) U_\alpha(s,t)      
             \tilde{E}^b(t) U_\alpha(t,s)      
           \right]
= 2 \tilde{E}^{ai}(x_I)\tilde{E}^{b}_i(x_I) .
\label{2Tlimit}
\end{eqnarray}
Using this, we can write  
\begin{eqnarray} 
&& \epsilon^4 \tilde{E}^{ai}(x_I)\tilde{E}^{b}_i(x_I) 
  =\frac{1}{2}\int_{\Sigma_{\epsilon}} d^2\!\sigma~ n_a(\sigma) 
              \int_{\Sigma_{\epsilon}} d^2\!\tau ~  n_b(\tau)
         ~{\cal T}^{ab}[\alpha_{\sigma\tau}](\sigma,\tau) 
	 + O(\epsilon),
\label{ee}
\end{eqnarray}
where $\alpha_{\sigma\tau}$ is, say, a (coordinate) circular
loop with the two points $\sigma$ and $\tau$ on antipodal
points.  Next, consider the area of the full surface
$\Sigma$.  By the very definition of Riemann integral,
(\ref{area}) can be written as
\begin{eqnarray}
A[\Sigma] &=& \int_\Sigma\! d^2\sigma
~\sqrt{n_a n_b \tilde{E}^{ai} \tilde{E}^{b}_{i}} 
= \lim_{ \stackrel{{\scriptstyle  N\to\infty}}{
         {\scriptstyle  \epsilon\to 0}}} 
    \sum_{I_\epsilon}  \epsilon^2 
~\sqrt{n_a(x_I) n_b(x_I) 
  \tilde{E}^{ai}(x_I) \tilde{E}^{b}_{i}(x_I)}
\label{riemann} 
\end{eqnarray}
where, following Riemann, we have partitioned the surface
$\Sigma$ in $N$ small surfaces $\Sigma_{I_\epsilon}$ of
coordinate area $\epsilon^2$ and $x_I$ is an arbitrary
point in $\Sigma_{I_\epsilon}$. The convergence of the
limit to the integral, and its independence from the
details of the construction, are assured by the Riemann
theorem for all bounded smooth fields.  Inserting
(\ref{ee}) in (\ref{riemann}), we obtain the desired
regularized expression for the classical area, suitable to
be promoted to a quantum loop operator
\begin{eqnarray}
A[\Sigma] &=& \lim_{\epsilon\to 0}\ A_\epsilon[\Sigma] ~~,  
\label{limitc}
\\
A_\epsilon[\Sigma] &=& \sum_{I_\epsilon} \sqrt{A^2_{I_\epsilon}}
~~,\\
A^2_{I_\epsilon} &=& \frac{1}{2}
       \int_{\Sigma_{I_\epsilon}\otimes\Sigma_{I_\epsilon}} 
       \!\!\!\!\!\!\!\!\! d^2\!\sigma  d^2\!\tau ~ 
        n_a(\sigma) n_b(\tau) 
          ~{\cal T}^{ab}[\alpha_{\sigma\tau}](\sigma,\tau). 
\end{eqnarray}
Notice that the powers of the regulator $\epsilon$ in
(\ref{ee}) and (\ref{riemann}) combine nicely, so that
$\epsilon$ appears in (\ref{limitc}) only in the
integration domains.

We are now ready to define the area operator:
\begin{eqnarray}
\hat A[\Sigma] &=& 
\lim_{\epsilon\to 0}\ \hat A_\epsilon[\Sigma],  
\label{limit}
\\
 A_\epsilon[\Sigma] &=&  \sum_{I_\epsilon} 
\sqrt{\hat{A}^2_{I_\epsilon}}, 
\label{root}
\\
\hat{A^2_{I_\epsilon}} &=& \frac{1}{2}
        \int_{\Sigma_{I_\epsilon}\otimes\Sigma_{I_\epsilon}} 
           \!\!\!\!\!\!\!\!\!\!\! d^2\!\sigma  d^2\!\tau ~ 
           n_a(\sigma) n_b(\tau) 
           ~{\hat{\cal T}}^{ab}[\alpha_{\sigma\tau}](\sigma,\tau). 
\label{sigmatau}
\end{eqnarray}
The meaning of the limit in (\ref{limit}) needs to be
specified. The specification of the topology in which the
limit is taken is an integral part of the definition of the
operator.  As it is usual for limits involved in the
regularization of quantum field theoretical operators, the
limit cannot be taken in the Hilbert space topology where,
in general, it does not exist.  The limit must be taken in a
topology that ``remembers'' the topology in which the
corresponding classical limit (\ref{limitc}) is taken.  This
is easy to do in the present context.  We say that a
sequence of (multi) loops $\alpha_\epsilon$ converges to
$\alpha$ if ${\alpha_\epsilon}$ converges pointwise to
$\alpha$; we say that a sequence of quantum states
${\left\langle{{\alpha_\epsilon}}\right|}$ converges to the
state ${\left\langle{{\alpha}}\right|}$ if
$\alpha_\epsilon\to\alpha$ for at least one
$\alpha_\epsilon\in{\left\langle{{\alpha_\epsilon}}\right|}$
($\forall\epsilon$) and one
$\alpha\in{\left\langle{{\alpha}}\right|}$.  This definition
extends immediately to general states
${\left\langle{{\Phi}}\right|}$ by linearity, and defines a
topology on the state space, and the corresponding operator
topology: $\hat O_\epsilon\to\hat O$ iff
${\left\langle{{\Phi}}\right|}\!\hat O_\epsilon\to{\langle
\Phi |}\!\hat O,~\forall{\left\langle{{\Phi}}\right|}$.
Notice that the above is equivalent to say that
${\left\langle{{\Phi_\epsilon}}\right|}$ converges to
${\left\langle{{\Phi}}\right|}$ if ${\cal T}[\Phi_\epsilon]$
converges pointwise to ${\cal T}[\Phi]$, which is the
topology implicitly used in \cite{Ashtekar95} to regularize
the area operator.

An important consequence of the use of this topology is the
following.  Let ${\left\langle{{\Phi_\epsilon}}\right|}$
converge to ${\left\langle{{\Phi}}\right|}$.  Then the
graphs $\Gamma_{\Phi_\epsilon}$ converge to $\Gamma_{\Phi}$
in the topology of $M$.  In other words, given a
$\delta$-neighborhood of $\Gamma_{\Phi}$, there exists an
$\epsilon$ such that $\Gamma_{\Phi_\epsilon'}$ is included
in the $\delta$-neighborhood for all $\epsilon'<\epsilon$.
Visually, we can imagine that the ribbon-nets
$R_{\Phi_\epsilon}$ ``merge'' into the ribbon-net
$\Gamma^{ex}_{\Phi}$ as $\epsilon$ approaches zero.  In
addition, the representations $P_{\Phi_\epsilon}$ go to
$P_{\Phi_\epsilon}$, up to equivalence. This fact allows us
to separate the study of a limit in two steps. First, we
study of the graph of the limit state. In this process, the
representations $P_{\Phi_\epsilon}$ are merged into the
ribbon-net $R$ of the limit state.  Second, we can use
recoupling theory on $R$, in order to express the limit
representation in terms of the spin network basis.

We now study the action of the area operator $\hat
A[\Sigma]$ given in (\ref{limit}) on a spin network state
${\left\langle{{S}}\right|}$. Namely, we compute
${\left\langle{{S}}\right|}\hat A[\Sigma]$.  Let
$S\cap\Sigma$ be the set of the points $i$ in the
intersection of $\Gamma_S$ and $\Sigma$.  In other words, we
label by an index $i$ the points where the spin network
graph $\Gamma_S$ and the surface $\Sigma$
intersect. Generically $S\cap\Sigma$ is numerable, and does
not include vertices of $S$.  Here we disregard spin
networks that have a vertex lying on $\Sigma$ or a
continuous number of intersection points with $\Sigma$.  It
was pointed out by A.\ Ashtekar that spin networks with a
vertex {\it and\ } one -or more- of its adjacent edges lying
on $\Sigma$ are eigenstates of the area with eigenvalues
that are not included in the spectrum of the operator
computed in \cite{Rovelli95} -and derived again below.
Therefore the spectrum of the area given in \cite{Rovelli95}
is not complete.  The physical relevance of these
``degenerate'' cases is unclear to us.\footnote{
	Note added: the complete spectrum of the area has been 
	obtained in the meanwhile in \cite{AshtekarLewand}, and then 
	reobtained in \cite{Frittelli} using the methods developed in this 	paper.}

For small enough $\epsilon$, each intersection $i$ will lie
inside a distinct $\Sigma_{I_\epsilon}$
surface.\footnote{The (perhaps cavilling) issue that an
intersection may fall on the {\it boundary\ } between two
$I_\epsilon$ surfaces has been raised.  This eventuality,
however, does not generate difficulties for the following
reason.  The integrals we are using are not Lebesgue
integrals, because, due to the presence of the $\delta$'s,
regions of zero measure of the integration domain cannot be
neglected -- nor doubly counted.  Therefore in selecting the
partition of $\Sigma$ in the $I_\epsilon$ surfaces one must
include each boundary in one and only one of the two
surfaces (which are therefore partially open and partially
closed). Boundary points are then normal points that fall
inside one and only one integration domain.}  Let us call
$\Sigma_{i_\epsilon}$ the surface containing the
intersection $i$ (at every fixed $\epsilon$), and $e_i$ the
edge through the intersection $i$.  Notice that
${\left\langle{{S}}\right|}\hat A^2_{\Sigma_{I_\epsilon}}$
vanishes for all surfaces $I_\epsilon$ except the ones
containing intersections. Thus the sum over surfaces
$\sum_{I_\epsilon}$ reduces to a sum over intersections.
Bringing the limit inside the sum and the square root, we
can write
\begin{eqnarray}
  {\left\langle{{S}}\right|} \hat A[\Sigma] &=& 
  \sum_{i\in\{S\cap\Sigma\}}\ {\left\langle{{S}}\right|} \sqrt{\hat A^2_i} 
\label{root2}\label{sumarea}
\\
\hat A^2_i &=& \lim_{\epsilon\to 0} \hat A^2_{i_\epsilon} 
\label{limit2}
\end{eqnarray}

For finite $\epsilon$, the state ${\left\langle{{S}}\right|}\hat
A^2_{i_\epsilon}$ has support on the union of the graphs
of $S$ and the graph of the loop $\alpha_{\sigma\tau}$ in
the argument of the operator (\ref{sigmatau}). But the
last converges to a point on $\Gamma_S$ as $\epsilon$ goes
to zero.  Therefore
\begin{equation}
	\lim_{\epsilon\to 0} \Gamma_{{\langle S |} 
           \hat A^2_{i_\epsilon}} 
	= \Gamma_S. 
\end{equation} 
The operator $\hat A[\Sigma]$ does not affect the graph of
${\langle S |}$.  Next, we have to compute the planar
representation of $ \Gamma_{{\langle S |} \hat A[\Sigma]}$,
which is a tangle on $R_{{\langle S |}\hat A[\Sigma]}$, namely a
tangle on $R_{S}$.  By equation (\ref{sumarea}), this is
given by a sum of terms, one for each
$i\in\{S\cap\Sigma\}$. Consider one of these terms.  By
definition of the ${\hat{\cal T}}$ loop operators and of the grasp
operation (Section 3), this is obtained by inserting two
trivalent intersections on the spin network edge $e_i$
(inside its ribbon), connected by a new edge of color 2.
This is because the circle $\Gamma_{\alpha_{\sigma\tau}}$
has converged to a point on $e_i$; in turn, this point is
then expanded inside the ribbon as a degenerate loop
following back and forward a segment connecting the two
intersections.  By indicating the representation of the
spin network simply by means of its $e_i$ edge, we thus
have
\begin{eqnarray}
~{\left\langle{{\big|}^{p_e}~}\right|}~ 
   \hat A^2_{i_\epsilon} &=&\frac{1}{2}
          \int_{\Sigma_{i_\epsilon}\otimes\Sigma_{i_\epsilon}} 
          \!\!\! d^2\!\sigma  d^2\!\tau ~ 
          n_a(\sigma) n_b(\tau) 
          ~{\left\langle{{\big|}^{p_e}~}\right|}~
         {\hat{\cal T}}^{ab}[\alpha_{\sigma\tau}](\sigma,\tau) 
\\
&=& - \frac{l_0^4}{2} 
        \int_{\Sigma_{i_\epsilon}\otimes\Sigma_{i_\epsilon}} 
        \!\!\! d^2\!\sigma  d^2\!\tau ~ 
          n_a(\sigma)  \Delta^a[\beta_e,\sigma] 
          n_b(\tau)    \Delta^b[\beta_e,\tau]\  
          p_e^2 
{\left\langle{ \begin{array}{c}
  \setlength{\unitlength}{1 pt}
  \begin{picture}(30,40)
  \put(12, 0){\line(0,1){40}}
  \put(12,10){\circle*{4}}\put(12,30){\circle*{4}}
  \put( 0,4){${\scriptstyle  p_e}$}
  \put( 0,20){${\scriptstyle  p_e}$}
  \put( 0,34){${\scriptstyle  p_e}$}
  \put(12,20){\oval(20,20)[r]}
  \put(24,20){${}^2$}
\end{picture}\end{array} }\right|}  ,
\nonumber
\end{eqnarray}
where we have already taken the limit (inside the
integral) in the state enclosed in the brackets
${\langle {~~} |}$. Notice that this does not depend on the
integration variables anymore, because the loop it
contains does not represent the grasped loop for a finite
$\epsilon$, but the a ribbon expansion of the limit state.
Notice also that the two integrals are independent, and
equal. Thus, we can write
\begin{equation}
~{\left\langle{{\big|}^{p_e}~}\right|}~ 
      \hat A^2_{i_\epsilon} = - \frac{l_0^4}{2} \left(
          \int_{\Sigma_{I_\epsilon}} 
          \!\!\! d^2\!\sigma   ~ 
          n_a(\sigma)  \Delta^a[\beta_e,\sigma] \right)^2
          p_e^2 
{\left\langle{ \begin{array}{c}
  \setlength{\unitlength}{1 pt}
  \begin{picture}(30,40)
  \put(12, 0){\line(0,1){40}}
  \put(12,10){\circle*{4}}\put(12,30){\circle*{4}}
  \put( 0,4){${\scriptstyle  p_e}$}
  \put( 0,20){${\scriptstyle  p_e}$}
  \put( 0,34){${\scriptstyle  p_e}$}
  \put(12,20){\oval(20,20)[r]}
  \put(24,20){${}^2$}
\end{picture}\end{array} }\right|} 
\end{equation}
The parenthesis is easy to compute. Using (\ref{delta}),
it becomes the analytic form of the intersection number
between the edge and the surface
\begin{eqnarray}
\int_{\Sigma_{i_\epsilon}} 
      \!\!\! d^2\!\sigma   ~ 
      n_a(\sigma)  \Delta^a[\beta_e,\sigma] 
&=& \int_{\Sigma_{i_\epsilon}} 
         \!\!\! d^2\!\sigma   ~ n_a(\sigma)  
         \int_{\beta_e} d\tau ~\dot{\beta}^a_e(\tau) 
         \delta^3 [\beta_e(\tau),s]
~=~ \pm 1,
\label{eq:IntersectionN}
\end{eqnarray}
where the sign, which depends on the relative orientation of
the loop and the surface, becomes then irrelevant because of
the square.  Thus
\begin{equation}
~{\left\langle{{\big|}^{p_e}~}\right|}~ \hat A^2_{i} 
= - \frac{l_0^4}{2}\  p_e^2 
{\left\langle{ \begin{array}{c}
  \setlength{\unitlength}{1 pt}
  \begin{picture}(30,40)
  \put(12, 0){\line(0,1){40}}
  \put(12,10){\circle*{4}}\put(12,30){\circle*{4}}
  \put( 0,4){${\scriptstyle  p_e}$}
  \put( 0,20){${\scriptstyle  p_e}$}
  \put( 0,34){${\scriptstyle  p_e}$}
  \put(12,20){\oval(20,20)[r]}
  \put(24,20){${}^2$}
\end{picture}\end{array} }\right|}  ,
\end{equation}
where we have trivially taken the limit (\ref{limit2}),
since there is no residual dependence on $\epsilon$. We
have now to express the tangle inside the bracket in terms
of (an edge of) a spin network state.  But tangles inside
ribbons satisfy recoupling theory, and we can therefore
use the formula (\ref{ThetaInALine}) in the appendix,
obtaining
\begin{eqnarray}
~{\left\langle{{\big|}^{p_e}~}\right|}~ \hat A^2_{i_\epsilon} &=& 
 - l_0^4~ p_e^2~ \frac{\theta(p_e,p_e,2)}{2 
                       \Delta_{p_e}}~{\left\langle{{\big|}^{p_e}~}\right|}~ 
    =  l_0^4~ \frac{p_e(p_e+2)}{4} ~{\left\langle{{\big|}^{p_e}~}\right|}~
~=~ l_0^4~ \frac{p_e}{2}\left(\frac{p_e}{2}+1\right)
     ~{\left\langle{{\big|}^{p_e}~}\right|}. 
\nonumber
\end{eqnarray}
The square root in (\ref{root2}) is now easy to take
because the operator $\hat A^2_i$ is diagonal.
\begin{equation}
~{\left\langle{{\big|}^{p_e}~}\right|}~ \hat A_i 
  =~{\left\langle{{\big|}^{p_e}~}\right|}~ \sqrt{\hat A^2_i} = 
~=~ \sqrt{l_0^4~ \frac{p_e}{2}\left(\frac{p_e}{2}+1\right)}
     ~{\left\langle{{\big|}^{p_e}~}\right|}. 
\end{equation}
Inserting in the sum (\ref{root2}), and shifting from
color to spin notation, we obtain the final result
\begin{equation}
{\langle S |}\ \hat{A}[\Sigma] = 
\left(l^2_0 \sum_{i\in\{S\cap\Sigma\}}  
\sqrt{j_i(j_i+1)}\right) \  {\langle S |}
\end{equation}
where $j_i$ is the spin of the edge crossing $\Sigma$ in
$i$.  This result shows that the spin network states (with
a finite number of intersection points with the surface
and no vertices on the surface) are eigenstates of the
area operator. The corresponding spectrum is labeled by
multiplets $\vec j = (j_1, ..., j_n)$ of positive half
integers, with arbitrary $n$, and given by
\begin{equation}
   A_{\vec j}\,[\Sigma] = l^2_0 \, \sum_i  \sqrt{j_i(j_i+1)}.
\end{equation}
The spectral values of the degenerate cases in which
$\Gamma_S\cap\Sigma$ includes vertices or a continuous number
of points, and a discussion on the relevance of these cases, 
will be given elsewhere.


\section{The volume operator}

\subsection{The volume in terms of loop variables}

Consider a three dimensional region ${\cal R}$.  The volume
of ${\cal R}$ is given by
\begin{eqnarray}
V[{\cal R}] &=& \int_{\cal R} d^3\!x \sqrt{\det{g}}
= \int_{\cal R} d^3\!x \sqrt{
             \frac{1}{3!}\bigg|
             \epsilon_{abc}\epsilon_{ijk}
             \tilde{E}^{ai} \tilde{E}^{bj} \tilde{E}^{ck} 
             \bigg|}
~~, 
\end{eqnarray}
In order to construct a regularized form of this
expression, consider the three index (three hands) loop
variable:
\begin{eqnarray}
&& {{\cal T}}^{abc}[\alpha](s,t,r) =  
- {\rm Tr}[
     \tilde{E}^a(s) U_\alpha(s,t)\
     \tilde{E}^b(t) U_\alpha(t,r)
     \tilde{E}^c(r) U_\alpha(r,s)].  
\end{eqnarray}
Because of (\ref{tr3tau}), in the limit of the loop
$[\alpha]$ shrinking to a point $x$ we have:
\begin{equation}
  {{\cal T}}^{abc}[\alpha](s,t,r) \rightarrow
  2 \epsilon_{ijk} \tilde{E}^{ai} \tilde{E}^{bj} \tilde{E}^{ck}
  = 2\ \epsilon^{abc} {\rm det}(\tilde{E})  .
\label{The3Tvalue}
\end{equation} 
Following \cite{Rovelli95}, fix an arbitrary chart of 
$M$, and consider a small cubic region
${\cal R}_I$ of coordinate volume $\epsilon^3$. Let $x_I$
be an arbitrary but fixed point in ${\cal R}_I$.  Since
classical fields are smooth we have $\tilde{E}(s) =
\tilde{E}(x_I) + O(\epsilon) $ for every $s \in {\cal
R}_I$, and $U_\alpha(s,t)_A^{~B} = \openone_A^{~B} +
O(\epsilon)$ for any $s,t\in {\cal R}_I$ and straight
segment $\alpha$ joining $s$ and $t$. Consider the quantity
\begin{eqnarray}
W_I &=& \frac{1}{16\ 3! \epsilon^6} 
     \int_{\partial {\cal R}_I}\!\!\!\! d^2\!\sigma
     \int_{\partial {\cal R}_I}\!\!\!\! d^2\!\tau
     \int_{\partial {\cal R}_I}\!\!\!\! d^2\!\rho  \cdot 
\label{WclasDEF} 
\cdot \big| 
      ~n_a(\sigma)  ~n_b(\tau) ~n_c(\rho)
      {{\cal T}}^{abc}[\alpha_{\sigma\tau\rho}](\sigma,\tau,\rho) 
     \big|
~~,
\end{eqnarray}
where $\alpha_{\sigma\tau\rho}$ is a triangular loop
joining the points $\sigma$, $\tau$ and $\rho$. Because of
(\ref{The3Tvalue}), we have, to lowest order in $\epsilon$
\begin{eqnarray}
   W_I &=&  \frac{1}{8\ 3! \epsilon^6} \  
   \big| {\rm det}(\tilde{E}(x_I) \big|  
          \int_{\partial {\cal R}_I}\! d^2\!\sigma
          \int_{\partial {\cal R}_I}\! d^2\!\tau
          \int_{\partial {\cal R}_I}\! d^2\!\rho  \cdot
          \big|  n_a(\sigma)  n_b(\tau) n_c(\rho)
               \epsilon^{abc}
          \big|
= \big| {\rm det}\tilde{E}(x_I) \big| 
~~,
\label{WclasAPROX} 
\end{eqnarray}
Thus, $W_I$ is a non-local quantity that approximates
${\rm det}g(x_I)$ for small $\epsilon$.  Using the
Riemann theorem as in the case of the area, we can then 
write the volume $V[{\cal R}]$ of the region $\cal R$ as
follows. For every $\epsilon$, we 
partition of $\cal R$ in cubes ${\cal R}_{I_\epsilon}$ of
coordinate volume $\epsilon^3$. Then
\begin{eqnarray}
V[{\cal R}] &=& \lim_{\epsilon\to 0} V_\epsilon[{\cal R}];
\\
V_\epsilon[{\cal R}] &=& 
   \sum_{I_\epsilon} \epsilon^3 W^{1/2}_{I_\epsilon}. 
\end{eqnarray}


\subsection{Quantum volume operator}

We have then immediately a definition of the quantum volume
operator \cite{Rovelli95}
\begin{eqnarray}
\hat V[{\cal R}] &=&
 \lim_{\epsilon\to 0} \hat V_\epsilon[{\cal R}];
\\
\hat  V_\epsilon[{\cal R}] &=& 
    \sum_{I_\epsilon} \epsilon^3 \hat W^{1/2}_{I_\epsilon};  
\label{sumvol}
\\
\hat W_{I_\epsilon} &=&  \frac{1}{16\ 3! \epsilon^6} 
	\int_{\partial {\cal R}_I}\!\!\! d^2\!\sigma
          \int_{\partial {\cal R}_I}\!\!\! d^2\!\tau
          \int_{\partial {\cal R}_I}\!\!\! d^2\!\rho   
\cdot \big| 
        \, n_a(\sigma)  n_b(\tau) n_c(\rho)
       {{\hat{\cal T}}}^{abc}[\alpha_{\sigma\tau\rho}](\sigma,\tau,\rho) 
         \big| .
      \label{opWI}
\end{eqnarray}
Notice the crucial cancellation of the $\epsilon^6$ factor. We
refer to the previous section on the area operator for the
discussion on the meaning of the limit and the split of the
action of the operator in the computation of the graph and the
representation.  We will discuss the meaning of the square root
later.  For alternative definitions of the volume operators, and a 
discussion on the relation between these, see \cite{Loll95,Loll95a} and \cite{Lewandowski96}. 
 
Let us now begin to compute the action of this operator on a
spin network state. The three surface integrals on the surface
of the cube and the line integrals along the loops combine --as
in the case of the area-- to give three intersection numbers,
which select three intersection points between the spin network
and the boundary of the cube. In these three points, which we
denote as $r$, $s$ and $t$, the loop $\alpha_{\sigma\tau\rho}$
of the operator grasps the spin network.

Notice that the integration domain of the (three) surface
integrals is a six dimensional space --the space of the
possible positions of three points on the surface of a cube.
Let us denote this integration domain as $D^6$.  The absolute
value in (\ref{opWI}) plays a crucial role here: contributions
from different points of $D^6$ have to be taken in their
absolute value, while contributions from the same point of
$D^6$ have to be summed algebraically before taking the
absolute value.  The position of each hand of the operator is
integrated over the surface, and therefore each hand grasps
each of the three points $r$, $s$ and $t$, producing $3^3$
distinct terms.  However, because of the absolute value, a term
in which two hands grasp the same point, say $r$, vanishes.
This happens because the result of the grasp is symmetric, but
the operator is antisymmetric, in the two hands -- as follows
from the antisymmetry of the trace of three sigma matrices.
Thus, only terms in which each hand grasps a distinct point
give non vanishing contributions. For each triple of points of
intersection between spin network and cube's surface $r$, $s$
and $t$, there are $3!$ ways in which the three hands can grasp
the three points. These $3!$ terms have alternating signs
because of the antisymmetry of the operator, but the absolute
value prevents the sum from vanishing, and yields the same
contribution for each of the $3!$ terms.

If there are only two intersection points between the boundary
of the cube and the spin network, then there are always two
hands grasping in the same point; contributions have to be
summed before taking the absolute value, and thus they cancel.
Thus, the sum in (\ref{sumvol}) reduces to a sum over the cubes
$I_\epsilon^i$ whose boundary has at least three distinct
intersections with the spin network, and the surface
integration reduces to a sum over the triple-graspings in {\it
distinct\ } points.  For $\epsilon$ small enough, the only
cubes whose surface has at least three intersections with the
spin network are the cubes containing a vertex $i$ of the spin
network . Therefore, the sum over cubes reduces to a sum over
the vertices $i\in\{{S\cap{\cal R}}\}$ of the spin network,
contained inside $\cal R$.  Let us denote by $I_{i_\epsilon}$
the cube containing the vertex $i$. We then have
\begin{eqnarray}
{\langle S |} \hat V[{\cal R}] &=&  \lim_{\epsilon\to 0}
 \sum_{i\in\{{S\cap{\cal V}}\}}  \epsilon^3 
{\langle S |} ~\sqrt{ \left| \hat W_{I^i_\epsilon} \right|} 
\nonumber \\
{\langle S |} \hat W_{I^i_\epsilon} 
  &=&
  \frac{{\rm i} l_0^6}{16\ 3! \epsilon^6} 
\sum_{s,t,r} {\left\langle{ S \tilde{\#}_{s, t, r}\alpha_{s, t, r}
              }\right|}, 
\end{eqnarray}
where $s, t$ and $r$ are three {\it distinct\ }
intersections between the spin network and the boundary
of the box, and we have indicated by 
$\left\langle {S\tilde{\#}_s\tilde{\#}_t\tilde{\#}_r\alpha_{str}}
 \right|$
the result of the triple grasp of the three hands operator 
with loop $\alpha_{str}$ on $S$.

Let us compute one of the terms above, corresponding to a given
triple of grasps, over an $n$-valent intersections.  First of
all, in the limit $\epsilon\to 0$ the operator does not change
the graph of the quantum state, for the same reason the area
operator doesn't.  Thus, the computation reduces to a
combinatorial computation of the action of the operator on the
representation of the planar state, involving recoupling theory.

Let us represent a spin network state simply by means of the
portion of its virtual net containing the vertex on which the
operator is acting.  We have
\begin{eqnarray}
&&\Bigg\langle{ \begin{array}{c}
   \setlength{\unitlength}{1 pt}
   \begin{picture}(120,55)
  \put( 0, 0){$P_0$}\put(15, 0){\line(1,1){10}}
  \put( 0,20){$P_1$}\put(15,20){\line(1,0){10}}
  \put(10,40){$P_2$}\put(25,40){\line(1,-1){10}}
  \put(25,10){\line(0,1){10}} \put(27,10){${\scriptstyle  i_1}$}
  \put(25,20){\line(1,1){10}} \put(32,18){${\scriptstyle  i_2}$}
  \put(35,30){\line(1,0){10}} \put(40,22){${\scriptstyle  i_3}$}
  \put(25,20){\circle*{3}}\put(35,30){\circle*{3}}
  \put(50,40){$\cdots$}        \put(50,30){$\ldots$} 
  \put(70,30){\line(1, 0){10}} \put(58,20){$ $} 
  \put(80,30){\line(1,-1){10}} \put(65,20){${\scriptstyle  i_{n-2}}$}
  \put(90,20){\line(0,-1){10}} \put(72, 8){${\scriptstyle  i_{n-1}}$}
  \put(90,20){\circle*{3}}\put(80,30){\circle*{3}}
  \put(95,40){$P_{n-3}$}\put(80,30){\line(1,1){10}}
  \put(105,20){$P_{n-2}$}\put(90,20){\line(1,0){10}}
  \put(105, 0){$P_{n-1}$}\put(90,10){\line(1,-1){10}} 
\end{picture}\end{array} 
~~}\Bigg| 
  ~~{\hat W_{I_\epsilon^i}}~ = 
\label{eq:AzWe}\\
&&~~=  \frac{{\rm i} l_0^6}{16~3!} 
 \sum_{\begin{array}{l} {\scriptstyle  r=0,\ldots,n-1} \\[-1mm]
                        {\scriptstyle  t=0,\ldots,n-1} \\[-1mm]
                        {\scriptstyle  s=0,\ldots,n-1} 
       \end{array}}
    \int_{{\partial {\cal V}_I} 
          \otimes{\partial {\cal V}_I} 
          \otimes{\partial {\cal V}_I}}
    \!\!\!\!\! \!\!\!\!\! \!\!\!\!\! \!\!\!\!\! 
    d^2\!\sigma d^2\!\tau d^2\!\rho
  ~\Big|
    ~n_a(\sigma)\Delta^a[\gamma,\sigma]
    ~n_b(\tau)\Delta^b[\gamma,\tau]
    ~n_c(\rho)\Delta^c[\gamma,\rho]
  \Big|
\nonumber\\ &&\qquad\qquad\qquad\qquad\qquad 
\cdot
  ~~{\bigg\langle{ \begin{array}{c}
  \setlength{\unitlength}{1 pt}
  \begin{picture}(120,55)
  \put( 0, 0){$P_0$}\put(15, 0){\line(1,1){10}}
  \put( 0,20){$P_1$}\put(15,20){\line(1,0){10}}
  \put(10,40){$P_2$}\put(25,40){\line(1,-1){10}}
  \put(25,10){\line(0,1){10}} \put(27,10){${\scriptstyle  i_1}$}
  \put(25,20){\line(1,1){10}} \put(32,18){${\scriptstyle  i_2}$}
  \put(35,30){\line(1,0){10}} \put(40,22){${\scriptstyle  i_3}$}
  \put(25,20){\circle*{3}}\put(35,30){\circle*{3}}
  \put(50,40){$\cdots$}        \put(50,30){$\ldots$} 
  \put(70,30){\line(1, 0){10}} \put(58,20){$ $} 
  \put(80,30){\line(1,-1){10}} \put(65,20){${\scriptstyle  i_{n-2}}$}
  \put(90,20){\line(0,-1){10}} \put(72, 8){${\scriptstyle  i_{n-1}}$}
  \put(90,20){\circle*{3}}\put(80,30){\circle*{3}}
  \put(95,40){$P_{n-3}$}\put(80,30){\line(1,1){10}}
  \put(105,20){$P_{n-2}$}\put(90,20){\line(1,0){10}}
  \put(105, 0){$P_{n-1}$}\put(90,10){\line(1,-1){10}} 
\end{picture}\end{array} ~~}\bigg|}
  ~~   \Big|  \hat W_{[rts]}^{(n)} ~  \Big|
\nonumber
\end{eqnarray}
\noindent
where $\hat W_{[rts]}^{(n)}$ is the 
operator that grasp the $r$, $t$ and $s$ edge of the 
the $n$-valent vertex as follow:
\begin{eqnarray}
&&
{\bigg\langle{ \begin{array}{c}
  \setlength{\unitlength}{1 pt}
  \begin{picture}(120,55)
  \put( 0, 0){$P_0$}\put(15, 0){\line(1,1){10}}
  \put( 0,20){$P_1$}\put(15,20){\line(1,0){10}}
  \put(10,40){$P_2$}\put(25,40){\line(1,-1){10}}
  \put(25,10){\line(0,1){10}} \put(27,10){${\scriptstyle  i_1}$}
  \put(25,20){\line(1,1){10}} \put(32,18){${\scriptstyle  i_2}$}
  \put(35,30){\line(1,0){10}} \put(40,22){${\scriptstyle  i_3}$}
  \put(25,20){\circle*{3}}\put(35,30){\circle*{3}}
  \put(50,40){$\cdots$}        \put(50,30){$\ldots$} 
  \put(70,30){\line(1, 0){10}} \put(58,20){$ $} 
  \put(80,30){\line(1,-1){10}} \put(65,20){${\scriptstyle  i_{n-2}}$}
  \put(90,20){\line(0,-1){10}} \put(72, 8){${\scriptstyle  i_{n-1}}$}
  \put(90,20){\circle*{3}}\put(80,30){\circle*{3}}
  \put(95,40){$P_{n-3}$}\put(80,30){\line(1,1){10}}
  \put(105,20){$P_{n-2}$}\put(90,20){\line(1,0){10}}
  \put(105, 0){$P_{n-1}$}\put(90,10){\line(1,-1){10}} 
\end{picture}\end{array} ~~}\bigg|}
  ~~\hat W_{[rts]}^{(n)} ~ = 
   P_r P_t P_s 
 \Bigg\langle{ 
\begin{array}{c}
  \setlength{\unitlength}{1 pt}
  \begin{picture}(155,60)
  \put( 0, 0){$P_0$}\put(15, 0){\line(1,1){10}}
  \put( 0,20){$ $}\put(15,20){\line(1,0){10}}
  \put(10,40){$ $}\put(25,40){\line(1,-1){10}}
  \put(25,10){\line(0,1){10}} \put(27,10){${\scriptstyle  i_1}$}
  \put(25,20){\line(1,1){10}} \put(32,18){${\scriptstyle  i_2}$}
  \put(35,30){\line(1,0){10}} \put(40,22){${\scriptstyle   }$}
  \put(25,20){\circle*{3}}\put(35,30){\circle*{3}}
  \put(50,30){\line(1,0){20}}   \put(57,22){${\scriptstyle  i_r}$}
  \put(65,30){\line(0,1){18}}   \put(62,50){$P_r$}
  \put(65,32){\oval(20,20)[tl]} 
  \put(65,42){\circle*{3}}      \put(65,30){\circle*{3}}
  \put(75,30){\line(1,0){20}}   \put(82,22){${\scriptstyle  i_t}$}
  \put(90,30){\line(0,1){18}}   \put(90,50){$P_t$}
  \put(90,32){\oval(20,20)[tl]} 
  \put(90,42){\circle*{3}}      \put(90,30){\circle*{3}}
  \put(115,32){\oval(20,20)[tl]}
  \put(115,42){\line(1,0){12}}  \put(127,42){\circle*{3}}
  \put(135,45){$P_{s}$}         \put(115,30){\line(1,1){15}}
  \put(115,30){\line(1,-1){10}} \put(107,22){${\scriptstyle  i_{s}}$}
  \put( 55,28){\line(0,-1){ 8}} \put(65, 2){${\scriptstyle  2}$}
  \put( 80,28){\line(0,-1){18}} \put(72,15){${\scriptstyle  2}$}
  \put(105,28){\line(0,-1){ 8}} \put(90, 2){${\scriptstyle  2}$}
  \put( 80,20){\oval(50,20)[b]} \put(80,10){\circle*{3}}
  \put(105,30){\line(1,0){10}}  \put( 93,20){$ $}
  \put(125,20){\line(0,-1){10}} \put(107,8){${\scriptstyle   }$}
  \put(125,20){\circle*{3}}     \put(115,30){\circle*{3}}
  \put(140,20){$ $}             \put(125,20){\line(1,0){10}}
  \put(140, 0){$P_{n-1}$}       \put(125,10){\line(1,-1){10}}
\end{picture}\end{array} ~~}\Bigg|~ 
\label{eq:recVol}
\\[1mm] 
&& \qquad \qquad
= \sum_{k_2,\ldots,k_{N-2}} 
 W^{(n)}_{[rts]}{}_{i_{2}\ldots i_{n-2}}^{k_{2}\ldots k_{n-2}}
         (P_{0},\ldots,P_{n-1}) \cdot
{\bigg\langle{ \begin{array}{c}
  \setlength{\unitlength}{1 pt}
  \begin{picture}(120,55)
  \put( 0, 0){$P_0$}\put(15, 0){\line(1,1){10}}
  \put( 0,20){$P_1$}\put(15,20){\line(1,0){10}}
  \put(10,40){$P_2$}\put(25,40){\line(1,-1){10}}
  \put(25,10){\line(0,1){10}} \put(27,10){${\scriptstyle  k_1}$}
  \put(25,20){\line(1,1){10}} \put(32,18){${\scriptstyle  k_2}$}
  \put(35,30){\line(1,0){10}} \put(40,22){${\scriptstyle  k_3}$}
  \put(25,20){\circle*{3}}\put(35,30){\circle*{3}}
  \put(50,40){$\cdots$}\put(50,30){$\ldots$}
  \put(70,30){\line(1,0){10}}  \put(58,20){$ $}
  \put(80,30){\line(1,-1){10}} \put(65,20){${\scriptstyle  k_{n-2}}$}
  \put(90,20){\line(0,-1){10}}  \put(72,8){${\scriptstyle  k_{n-1}}$}
  \put(90,20){\circle*{3}}\put(80,30){\circle*{3}}
  \put( 95,40){$P_{n-3}$}\put(80,30){\line(1,1){10}}
  \put(105,20){$P_{n-2}$}\put(90,20){\line(1,0){10}}
  \put(105, 0){$P_{n-1}$}\put(90,10){\line(1,-1){10}}
\end{picture}\end{array} ~~}\bigg|}
~~,
\nonumber
\end{eqnarray}
Notice that we have replaced the triangular loop with
vertices $r$, $s$ and $t$ by three edges of color 2 joining
the three points $r$, $s$ and $t$ to a trivalent vertex.
This can be done as follows. First we deform the triangle
over the ribbon-net.  Indeed, as remarked for the case of
the area, the tangle above does not represent a tangle
extended in $M$, but just the expansion over the ribbon net
of a rooting of lines in a single point of $M$.  Second, we
notice that we can antisymmetrize the two lines that exit
from the hand of an operator by using the binor identity,
because tracing a hand with a zero length loop gives a
vanishing quantity.

The last equality in the last equation follows from the fact
that trivalent spin network form a basis (see Sec.~V).  From
eq.\ (\ref{eq:AzWe}) we see that the action of ${\hat
W_{I_\epsilon^i}}$ splits into a multiplication by a
numerical prefactor and a recoupling part given by
eq.~(\ref{eq:recVol}), which does not depend on the
integration variables.  Using eq.~(\ref{eq:IntersectionN})
we can perform the integration in eq.~(\ref{eq:AzWe}). This
yields the intersection number between the edges $r$, $s$
and $t$ and the surface of the cube ${\cal V}_I$.  The sign
of the intersection number, coming from the relative
orientation of the loop and the surface, is irrelevant,
because of the presence of the absolute value.

Because of the symmetry properties of the $3$-valent node
$(222)$, the $3!$ terms in eq.\ (\ref{eq:recVol}) are
related by:
\begin{equation}
\hat W_{[i_1 i_2 i_3]}^{(n)}
 = (-1)^p \hat W_{[i_{p_1} i_{p_2} i_{p_3}]}^{(n)}
\end{equation}
where $p_i$ it is a permutation of $123$, and $p$ it
is the order of the permutation.
Thus, the action the volume operator
on a generic spin network state ${\langle S |}$ is given by:
\begin{eqnarray}
 \hat V[{\cal V}] &=& {l_0^3}
 \sum_{i\in\{{S\cap{\cal V}}\}}
~\sqrt{  \sum_{\begin{array}{l} 
	{\scriptstyle  r=0,\ldots,n-3}  \\[-1mm]
        {\scriptstyle  t=r+1,\ldots,n-2} \\[-1mm]
        {\scriptstyle  s=t+1,\ldots,n-1} 
       		\end{array}}
  \left| \frac{{\rm i}}{16}~ \hat W_{[rts]}^{(n_i)} ~\right|
	} 
\label{eq:VolOpSN}
\end{eqnarray}
where $n_i$ is the valence of the $i$-th intersection.
Equations (\ref{eq:recVol}) and (\ref{eq:VolOpSN}) completely
define the volume operator.  There are two remaining tasks: to
find the explicit expression for the matrix ${\rm i}
W^{(n)}_{[rst]}{}^{i_{n-2}\ldots i_3i_2}_{k_{n-2}\ldots k_3k_2}
(P_{n-1},\ldots,P_0)$, which is defined in eq.\
(\ref{eq:recVol}) only implicitly; and to show that the
absolute value and the square root in equation
(\ref{eq:VolOpSN}) are well defined.  Below, we complete both
tasks: we provide an explicit expression for ${\rm i}
W^{(n)}_{[rst]}{}^{i_{n-2}\ldots i_3i_2}_{k_{n-2}\ldots k_3k_2}
(P_{n-1},\ldots,P_0)$, and we prove that the argument of the
absolute value is a diagonalizable finite dimensional matrix
with real eigenvalues, and the argument of the square root is a
finite dimensional diagonalizable matrix with positive real
eigenvalues.

\subsection{Trivalent vertices}

We begin studying the case $n=3$.  
It is easy to see that  $W^{(3)}_{[012]} = 0$ 
from the relation
\begin{equation}
\begin{array}{c}
  \setlength{\unitlength}{1 pt}
  \begin{picture}(90,60)
  \put(15,40){\line(0,1){8}}   \put(20,50){$P_0$}
  \put(15,32){\oval(20,20)[tl]} 
  \put(15,42){\circle*{3}}     
  \put(40,30){\line(0,1){18}}   \put(45,50){$P_1$}
  \put(40,32){\oval(20,20)[tl]} 
  \put(40,42){\circle*{3}}      \put(40,30){\circle*{3}}
  \put(65,40){\line(0,1){8}}    \put(70,50){$P_2$}
  \put(65,32){\oval(20,20)[tl]} 
  \put(65,42){\circle*{3}}     
  \put( 5,32){\line(0,-1){12}} \put(15, 2){${\scriptstyle  2}$}
  \put(30,28){\line(0,-1){18}} \put(22,15){${\scriptstyle  2}$}
  \put(55,28){\line(0,-1){ 8}} \put(40, 2){${\scriptstyle  2}$}
  \put(30,20){\oval(50,20)[b]} \put(30,10){\circle*{3}}
  \put(40,40){\oval(50,20)[b]}
\end{picture}\end{array} 
=  W^{(3)}_{[012]} 
\begin{array}{c}\setlength{\unitlength}{1 pt}
\begin{picture}(40,40)
       \put(15,15){\line(-1, 1){10}} \put( 4,27){$P_0$}
       \put(15,15){\line( 1, 1){10}} \put(22,27){$P_1$}
       \put(15, 5){\line(0,1){10}}   \put(17,1){$P_2$}
       \put(15,15){\circle*{3}}
\end{picture}\end{array}
~.
\end{equation}
In fact, by closing the generic $3$-valent node with itself we
have
\begin{equation}
 P_0 P_1 P_2 
\begin{array}{c}
  \setlength{\unitlength}{1 pt}
  \begin{picture}(90,60)
  \put(15,40){\line(0,1){10}}   \put(20,50){$P_0$}
  \put(15,32){\oval(20,20)[tl]} 
  \put(15,42){\circle*{3}}     
  \put(40,30){\line(0,1){30}}   \put(45,50){$P_1$}
  \put(40,32){\oval(20,20)[tl]} 
  \put(40,42){\circle*{3}}     
  \put(65,40){\line(0,1){10}}    \put(70,50){$P_2$}
  \put(65,32){\oval(20,20)[tl]} 
  \put(65,42){\circle*{3}}     
  \put( 5,32){\line(0,-1){12}} \put(15, 2){${\scriptstyle  2}$}
  \put(30,28){\line(0,-1){18}} \put(22,15){${\scriptstyle  2}$}
  \put(55,28){\line(0,-1){ 8}} \put(40, 2){${\scriptstyle  2}$}
  \put(30,20){\oval(50,20)[b]} \put(30,10){\circle*{3}}
  \put(40,40){\oval(50,20)[b]}  \put(40,30){\circle*{3}}
  \put(40,50){\oval(50,20)[t]}  \put(40,60){\circle*{3}}
\end{picture}\end{array}
 =  W^{(3)}_{[012]} 
\begin{array}{c}\setlength{\unitlength}{1 pt}\begin{picture}(40,40)
        \put(18,32){$P_0$}
        \put(18,17){$P_1$} 
        \put(18, 2){$P_2$} 
        \put(20,15){\oval(40,30)} \put( 0,15){\line(1,0){40}} 
        \put( 0,15){\circle*{3}}  \put(40,15){\circle*{3}}
\end{picture}\end{array} ~.   
\end{equation}
Thus $W^{(3)}_{[012]}$ its determined
by the Wigner $9J$-symbol (the evaluation of the hexagonal net)
as:
\begin{equation}
W^{(3)}_{[012]} = 
  \frac{P_0 P_1 P_2
  \left\{\begin{array}{ccc} 
         P_0 & P_1 & P_2 \\
         P_0 & P_1 & P_2 \\
         2   &  2  & 2  
  \end{array}\right\}
  }{\theta(P_0,P_1,P_2)}.
\end{equation} 
But the hexagonal net (in the case of $A=\pm 1$) it is
antisymmetric for the exchange of two columns or of two
rows. Therefore the matrix $W^3$ vanishes, and the trivalent
vertices give no contribution to the volume. We have re-derived
the result that the volume of a $3$-valent vertex is zero,
first obtained by Loll \cite{Loll95}.

\subsection{Four-valent vertices}

Next, we study the $n=4$ case. 
\begin{equation}
\hat W^{(4)}_{[012]}{}
\begin{array}{c}\setlength{\unitlength}{1 pt}
\begin{picture}(50,40)
          \put( 0,0){$P_0$}\put( 0,30){$P_1$}
          \put(45,0){$P_3$}\put(45,30){$P_2$}
          \put(10,10){\line(1,1){10}}\put(10,30){\line(1,-1){10}}
          \put(30,20){\line(1,1){10}}\put(30,20){\line(1,-1){10}}
          \put(20,20){\line(1,0){10}}\put(22,25){$i$}
          \put(20,20){\circle*{3}}\put(30,20){\circle*{3}}
\end{picture}\end{array}
= \sum_j~
W^{(4)}_{[012]}{}_i^j
\begin{array}{c}\begin{picture}(50,40)
          \put( 0,0){$P_0$}\put( 0,30){$P_1$}
          \put(45,0){$P_3$}\put(45,30){$P_2$}
          \put(10,10){\line(1,1){10}}\put(10,30){\line(1,-1){10}}
          \put(30,20){\line(1,1){10}}\put(30,20){\line(1,-1){10}}
          \put(20,20){\line(1,0){10}}\put(22,25){$j$}
          \put(20,20){\circle*{3}}\put(30,20){\circle*{3}}
\end{picture}\end{array}
\label{eq:ActOver4}
\end{equation}

Using the same technique of the $3$-valent node we can
compute the matrix $W^{(4)}_{[012]}{}_i^j$ for a  
$4$-valent node as follows
\begin{eqnarray}
&&\FL  P_0 P_1 P_2 
\begin{array}{c}
  \setlength{\unitlength}{1 pt}
  \begin{picture}(110,60)
  \put(15,40){\line(0,1){10}}  \put(17,38){$P_0$}
  \put(15,32){\oval(20,20)[tl]} 
  \put(15,42){\circle*{3}}    
  \put(40,30){\line(0,1){30}}  \put(42,38){$P_1$}
  \put(40,32){\oval(20,20)[tl]} 
  \put(40,42){\circle*{3}}     \put(40,30){\circle*{3}}
  \put(40,60){\circle*{3}}
  \put(65,30){\line(0,1){30}}  \put(67,38){$P_2$}
  \put(65,32){\oval(20,20)[tl]} 
  \put(65,42){\circle*{3}}     \put(65,30){\circle*{3}}
  \put(65,60){\circle*{3}}
  \put(90,40){\line(0,1){10}}  \put(92,38){$P_3$}
  \put( 5,32){\line(0,-1){12}} \put(15, 2){${\scriptstyle  2}$}
  \put(30,28){\line(0,-1){18}} \put(22,15){${\scriptstyle  2}$}
  \put(55,28){\line(0,-1){ 8}} \put(40, 2){${\scriptstyle  2}$}
  \put(30,20){\oval(50,20)[b]} \put(30,10){\circle*{3}}
  \put(40,40){\oval(50,20)[bl]}\put(65,40){\oval(50,20)[br]}
  \put(40,30){\line(1,0){20}}  \put(42,20){$i$} 
  \put(40,50){\oval(50,20)[tl]}\put(65,50){\oval(50,20)[tr]}
  \put(40,60){\line(1,0){20}}  \put(52,50){$j$}
\end{picture}\end{array}  
=\sum_k~ W^{(4)}_{[012]}{}_i^k ~~\delta^j_k ~~
  \frac{\theta(P_0,P_1,j)\theta(P_2,P_3,j)}{\Delta_j}. 
\label{eq:Vol4vertex1}
\end{eqnarray}
Using the relation
\begin{equation}
\begin{array}{c}
  \setlength{\unitlength}{1 pt}
  \begin{picture}(55,50)
       \put(25, 5){\line(0,1){10}}   \put(27,1){$i$}
       \put(25,15){\circle*{3}}
       \put(25,15){\line(-1, 1){10}} \put(12,14){${\scriptstyle  P_2}$}
       \put(25,15){\line( 1, 1){10}} \put(32,14){${\scriptstyle  P_3}$}
       \put(35,25){\circle*{3}}\put(15,25){\circle*{3}}
       \put(15,25){\line(-1, 1){10}} \put( 4,37){$2$}
       \put(35,25){\line( 1, 1){10}} \put(42,37){$j$}
       \put(15,25){\line(1,0){20}}   \put(25,29){${\scriptstyle  P_2}$}
\end{picture}\end{array} 
 = \frac{
     {Tet\left[\begin{array}{ccc} 
            i   &  j  & P_3 \\
            P_2 & P_2 & 2  
     \end{array}\right]}
   }{\theta(2,j,i)}
\begin{array}{c}\setlength{\unitlength}{1 pt}
\begin{picture}(40,40)
       \put(15,15){\line(-1, 1){10}} \put( 4,27){$2$}
       \put(15,15){\line( 1, 1){10}} \put(22,27){$j$}
       \put(15, 5){\line(0,1){10}}   \put(17,1){$i$}
       \put(15,15){\circle*{3}}
\end{picture}\end{array}
~,
\label{3vertexReduction}
\end{equation}
we obtain:
\begin{eqnarray}
W^{(4)}_{[012]}{}_i^j &=& 
  \frac{P_0 P_1 P_2 
        \left\{\begin{array}{ccc}
           {P_0} & {P_1} & {j} \\
           {P_0} & {P_1} & {i} \\
            {2}  &  {2}  & {2}
        \end{array}\right\}
        {Tet\left[\begin{array}{ccc} 
            i   &  j  & P_3 \\
            P_2 & P_2 & 2  
        \end{array}\right]}
      }{\theta(2,j,i)}
  \cdot
\frac{\Delta_j}{\theta(P_0,P_1,j)\theta(P_2,P_3,j)}
~~~.
\end{eqnarray}
We now prove that the matrix ${\rm i}\cdot
W^{(4)}_{[012]}{}_i^j$ its diagonalizable with real
eigenvalues and, as a consequence, that its absolute values
is well defined.  To this aim, let us define the notation:
\begin{eqnarray}
{A}_i^j &=& \frac{P_0 P_1 P_2 
  \left\{\begin{array}{ccc} 
         P_0 & P_1 & j \\
         P_0 & P_1 & i \\
         2   &  2  & 2  
  \end{array}\right\}
        {Tet\left[\begin{array}{ccc} 
            i   &  j  & P_3 \\
            P_2 & P_2 & 2  
        \end{array}\right]}
      }{\theta(2,j,i)}
\label{eq:Wtilda1}\\[2 mm]
M(i) &=& 
  \sqrt{\frac{\Delta_i}{\theta(P_0,P_1,i)\theta(P_2,P_3,i)}}
\\
\tilde W_i^j &=& M(i)~M(j) A_i^j  
\\
S_i^j  &=& \delta_i^j ~M(i) ~~.
\end{eqnarray} 
The matrix $S_i^j$ can be consider as a change of basis in
the space of the $4$-valent vertices and the matrix
${\rm i}\cdot W^{(4)}_{[012]}{}_i^j$ can be rewritten as:
\begin{equation}
  {\rm i} W^{(4)}_{[012]}{}_i^j 
  = (S^{-1}){}_i^k \cdot ({\rm i} \tilde W_k^l) \cdot S{}_l^j ,
\label{eq:Wtilda}
\end{equation}
where, because of the antisymmetry properties of the
$9J$-symbol under exchange of two rows and the symmetry
property of the $Tet$ symbol\footnote{For a 
discussion of the symmetry properties of the $9J$-symbol
and related quantities, see for instance \cite{Brink68}},
the matrix $\tilde W_k^l$ is antisymmetric. We have shown
that in the basis 
\begin{equation}
\bbox{n}_i =  \sqrt{\frac{
     \begin{array}{c}\setlength{\unitlength}{.5 pt}
     \begin{picture}(35,25)
        \put(15, 0){\line(0,1){10}}\put(20, 0){\line(0,1){10}}
        \put(15, 0){\line(1,0){ 5}}\put(15,10){\line(1,0){ 5}}
        \put(15,25){\line(1,0){5}}  \put(15,15){$\scriptstyle i$}
        \put(15,15){\oval(30,20)[l]}\put(20,15){\oval(30,20)[r]}
     \end{picture}\end{array}
   }{\begin{array}{c}\setlength{\unitlength}{.5 pt}\begin{picture}(40,40)
        \put(18,32){$\scriptstyle P_0$}
        \put(18,17){$\scriptstyle P_1$} 
        \put(18, 2){$\scriptstyle i$} 
        \put(20,15){\oval(40,30)} \put( 0,15){\line(1,0){40}} 
        \put( 0,15){\circle*{3}}  \put(40,15){\circle*{3}}
     \end{picture}\end{array}
     \begin{array}{c}\setlength{\unitlength}{.5 pt}\begin{picture}(40,40)
        \put(18,32){$\scriptstyle P_2$}
        \put(18,17){$\scriptstyle P_3$} 
        \put(18, 2){$\scriptstyle i$} 
        \put(20,15){\oval(40,30)} \put( 0,15){\line(1,0){40}} 
        \put( 0,15){\circle*{3}}  \put(40,15){\circle*{3}}
     \end{picture}\end{array}
   }} 
\begin{array}{c}\setlength{\unitlength}{1 pt}
\begin{picture}(50,40)
          \put( 0,0){$P_0$}\put( 0,30){$P_1$}
          \put(45,0){$P_3$}\put(45,30){$P_2$}
          \put(10,10){\line(1,1){10}}\put(10,30){\line(1,-1){10}}
          \put(30,20){\line(1,1){10}}\put(30,20){\line(1,-1){10}}
          \put(20,20){\line(1,0){10}}\put(22,25){$i$}
          \put(20,20){\circle*{3}}\put(30,20){\circle*{3}}
\end{picture}\end{array}
\label{eq:newbasis}
\end{equation}
the action, eq.\ (\ref{eq:ActOver4}), of the operator 
$\hat W^{(4)}_{[012]}{}$ is given by:
\begin{equation}
\hat W^{(4)}_{[012]}{}  \bbox{n}_i
= \sum_j~ \tilde{W}^{(4)}_{[012]}{}_i^j ~ \bbox{n}_j
\label{eq:ActOver4bis}
\end{equation}
where $\tilde{W}^{(4)}_{[012]}{}_i^j$ a {\it real
antisymmetric\ } matrix.  Moreover, from the admissibility
condition for the $3$-valent node of eq.\
(\ref{3vertexReduction}), we see that $\tilde W_k^l$ vanishes
unless $k=l$ or $k=l\pm2$. Thus, we have show that the operator
${\rm i}~ \hat W^{(4)}_{[012]}$ may be represented by a purely
imaginary antisymmetric matrix ${\rm i} \tilde W_k^l$ with
non-vanishing matrix elements only for $k=l\pm2$.  Such matrix
is diagonalizable and has real eigenvalues.

Furthermore, notice the following.  We write the dependence on
the coloring of the external edges explicitly; namely we write
$W^{(4)}_{[012]}{}_i^j(P_0,P_1,P_2,P_3)$.  Using eq.\
(\ref{eq:Vol4vertex1}), it is easy to see that the following
relations hold between the matrices
$W^{(4)}_{[i_1i_2i_3]}{}_i^j(P_0,P_1,P_2,P_3)$
\begin{eqnarray}
W^{(4)}_{[013]}{}_i^j(P_0,P_1,P_2,P_3)
  &=& W^{(4)}_{[012]}{}_i^j(P_0,P_1,P_3,P_2),
\label{eq:id4vertex}\\
W^{(4)}_{[023]}{}_i^j(P_0,P_1,P_2,P_3)
  &=& - W^{(4)}_{[123]}{}_i^j(P_3,P_2,P_1,P_0),
\nonumber\\
W^{(4)}_{[123]}{}_i^j(P_0,P_1,P_2,P_3)
  &=& - W^{(4)}_{[012]}{}_i^j(P_3,P_2,P_0,P_1).
\nonumber
\end{eqnarray}

We have shown that there exists a basis $\bbox{n}_i$ in which
the four operators ${\rm i} \hat W^{(4)}_{[i_1i_2i_3]}$ that
define the action of the volume on four valent vertices, are
purely imaginary antisymmetric matrices. The eigenvalues of the
four operators ${\rm i}\hat W^{(4)}_{[i_1i_2i_3]}$ are real
and, if $x$ is an eigenvalue, so is $-x$.  Therefore, the
absolute value of the matrices ${\rm i}\hat
W^{(4)}_{[i_1i_2i_3]}$ is well defined. It is given by a
non-negative (i.e., having real eigenvalues equal or greater
than zero) antisymmetric matrix.  But the sum of non-negative
matrices is a non-negative matrix.  Therefore the sum of the
the absolute values of the four matrices ${\rm i}~ \hat
W^{(4)}_{[i_1i_2i_3]}$ is a non-negative antisymmetric matrix
as well. Thus, the volume operator is diagonalizable on the
spin network basis, {\it with positive real eigenvalues}, if
all the vertices have valence $3$, $4$.  Below, we show that
these results extend to vertices of arbitrary valence.

\subsection{The case of an $n$-vertex}

We now shown that there exists a basis in which
all the operators ${\rm i} \hat
W^{(n)}_{[i_1i_2i_3]}(P_0,\ldots,P_{n-1})$ are represented
by a purely imaginary antisymmetric matrix.  Consider eq.\
(\ref{eq:recVol}).  By repeated application of
the recoupling theorem, eq.\ (\ref{eq:recVol}) can be
rewritten as 
\FL 
\begin{equation}
P_r P_t P_s
\begin{array}{c}\setlength{\unitlength}{1 pt}
\begin{picture}(120,60)
  \put( 5,40){\line(0,1){8}}    \put(1,50){$P_0$}
  \put(25,30){\line(0,1){18}}   \put(20,50){$P_r$}
  \put(25,32){\oval(20,20)[tl]} 
  \put(25,42){\circle*{3}}      \put(25,30){\circle*{3}}
  \put(50,30){\line(0,1){18}}   \put(45,50){$P_t$}
  \put(50,32){\oval(20,20)[tl]} 
  \put(50,42){\circle*{3}}      \put(50,30){\circle*{3}}
  \put(75,30){\line(0,1){18}}   \put(70,50){$P_s$}
  \put(75,32){\oval(20,20)[tl]} 
  \put(75,42){\circle*{3}}      \put(75,30){\circle*{3}}
  \put(15,32){\line(0,-1){12}} \put(25, 2){${\scriptstyle  2}$}
  \put(40,28){\line(0,-1){18}} \put(32,15){${\scriptstyle  2}$}
  \put(65,28){\line(0,-1){ 8}} \put(50, 2){${\scriptstyle  2}$}
  \put(40,20){\oval(50,20)[b]} \put(40,10){\circle*{3}}
  \put(60,40){\oval(110,20)[b]}
  \put(27,32){$\hat{i}_2$} 
  \put(52,32){$\hat{i}_3$}
  \put(77,32){$\hat{i}_4$}
  \put( 95,40){\line(0,1){8}}    \put(90,50){$\ldots$}
  \put( 95,30){\circle*{3}}
  \put(115,40){\line(0,1){8}}    \put(110,50){$P_{n-1}$}
\end{picture}\end{array} = 
\sum_{\hat{k}_{2}\ldots\hat{k}_{n-2}}
W^{(n)}_{[rst]}
{}_{\hat{i}_{2}\ldots\hat{i}_{n-2}
 }^{\hat{k}_{2}\ldots\hat{k}_{n-2}}
\cdot
\begin{array}{c}\setlength{\unitlength}{1 pt}
\begin{picture}(120,40)
  \put( 5,20){\line(0,1){8}}    \put(1,30){$P_0$}
  \put(25,10){\line(0,1){18}}   \put(20,30){$P_r$}
  \put(25,10){\circle*{3}}
  \put(50,10){\line(0,1){18}}   \put(45,30){$P_t$}
  \put(50,10){\circle*{3}}
  \put(75,10){\line(0,1){18}}   \put(70,30){$P_s$}
  \put(75,10){\circle*{3}}
  \put(60,20){\oval(110,20)[b]}
  \put(27,12){$\hat{k}_2$} 
  \put(52,12){$\hat{k}_3$}
  \put(77,12){$\hat{k}_4$}
  \put( 95,20){\line(0,1){8}}    \put(90,30){$\ldots$}
  \put( 95,10){\circle*{3}}
  \put(115,20){\line(0,1){8}}    \put(110,30){$P_{n-1}$}
\end{picture}\end{array} 
\end{equation}
(we have assumed, without loss of generality, 
that there is no grasp on
the $P_0$ or $P_{n-1}$ edge). 
Closing the vertex with itself and using the relation
(\ref{ThetaInALine}) and (\ref{TetInALine}), we find 
\begin{equation}
W^{(n)}_{[rst]}
{}_{\hat{i}_{2}\ldots\hat{i}_{n-2}
 }^{\hat{k}_{2}\ldots\hat{k}_{n-2}}
 = P_r P_t P_s 
  \left\{\begin{array}{ccc} 
       \hat{k}_2 & {P_t} & {\hat{k}_3} \\
       \hat{i}_2 & {P_t} & {\hat{i}_3} \\
         2   &  2  & 2  
  \end{array}\right\}  \cdot 
\frac{ -1 \lambda^{\hat{i}_2 2}_{\hat{k}_2}
\delta^{\hat{k}_4}_{\hat{i}_4}
\cdots\delta^{\hat{k}_{n-2}}_{\hat{i}_{n-2}}
\cdot {Tet\left[\begin{array}{ccc} 
            {P_r}      &{P_r}      &{P_0}\\
            {\hat{k}_2}&{\hat{i}_2}&{2} 
      \end{array}\right]}
     {Tet\left[\begin{array}{ccc} 
            {\hat{i}_3} & {\hat{k}_3} & {\hat{k}_4}\\
            {P_s}       &  {P_s}      &   {2}
     \end{array}\right]}
 \Delta_{\hat{k}_2} \Delta_{\hat{k}_3}
}{\theta(\hat{k}_2,2,\hat{i}_2)
  \theta(\hat{k}_3,2,\hat{i}_3)
  \theta(P_0,P_r,\hat{k}_2)
  \theta(\hat{k}_2,P_t,\hat{k}_3)
  \theta(\hat{k}_3,P_s,\hat{k}_4)
}
\label{eq:rootingVolN}
\end{equation}
We now change basis in the same fashion as we did for the
4-valent vertex, [see Eq.\ (\ref{eq:newbasis})]. We define a
new basis in which any edge (real or virtual) is multiplied by
$\sqrt{\Delta_i}$ ($i$ coloring of the edge) and any vertex is
divided by $\sqrt{\theta(a,b,c)}$ ($a$, $b$ and $c$ the
coloring of the edges adjacent to the vertex). It is then easy
to see that in this new basis the matrix on eq.\
(\ref{eq:rootingVolN}) becomes real antisymmetric. Indeed, we
have simply reduced the general problem to the case of four
valent vertices.  Now, the key result, that we shall prove in
the next section is that, in the basis we have defined, the
recoupling theorem is a {\it unitary\ } transformation.  A
unitary transformation preserves the property of a matrix of
being diagonalizable and having real eigenvalues.  It follows
that the results we have obtained for the four-valent vertices
hold in general.

We are now ready to find an explicit expression for the
recoupling matrix ${\rm i} W^{(n)}_{[rst]} {}^{i_{n-2}\ldots
i_3i_2}_{k_{n-2}\ldots k_3k_2} (P_{n-1},\ldots,P_0)$ of eq.\
(\ref{eq:recVol}) for a general valence $n$ of the vertex.  Let
us begin by sketching the procedure that we follow. First, the
recoupling theorem allows us to move one of the three grasps
from the external edge, say $P_r$, of eq.\ (\ref{eq:recVol}),
and bring it to a virtual vertex.  We denote this operation as
Move 1:
\begin{eqnarray}
\begin{array}{c}
  \setlength{\unitlength}{1 pt}
  \begin{picture}(40,45)
   \put( 2,12){${\scriptstyle  i_r}$}    \put( 0,20){\line(1,0){10}}
   \put( 8, 2){${\scriptstyle  2}$}      \put(10,10){\line(0,1){ 8}}
   \put(22,12){${\scriptstyle  i_{r+1}}$}\put(10,20){\line(1,0){20}}
   \put(20,20){\line(0,1){20}}\put(22,25){${\scriptstyle  P_r}$}
                                \put(22,35){${\scriptstyle  P_r}$}
   \put(20,22){\oval(20,20)[tl]}
   \put(20,32){\circle*{3}}\put(20,20){\circle*{3}}
\end{picture}\end{array}
&=& \sum_{k_r}  \left\{\begin{array}{ccc}
                      {i_{r+1}}&{i_{r}}&{k_r} \\
                         {2}   & {P_r} &{P_r} 
                \end{array}\right\}
\begin{array}{c}
  \setlength{\unitlength}{1 pt}
  \begin{picture}(40,45)
   \put( 2,12){${\scriptstyle  i_r}$}    \put( 0,20){\line(1,0){10}}
   \put(20,12){${\scriptstyle  k_r}$}    \put(10,20){\line(1,0){10}}
   \put( 8, 2){${\scriptstyle  2}$}      \put(10,10){\line(0,1){ 8}}
   \put(32,12){${\scriptstyle  i_{r+1}}$}\put(20,20){\line(1,0){20}}
   \put(30,20){\line(0,1){20}}\put(32,35){${\scriptstyle  P_r}$}
   \put(15,22){\oval(10,10)[t]}
   \put(20,20){\circle*{3}}\put(30,20){\circle*{3}}
\end{picture}\end{array} 
= \sum_{k_r} \left\{\begin{array}{ccc}
                  {i_{r+1}}&{i_{r}}&{k_r}\\
                    {2}    & {P_r} & {P_r} 
                \end{array}\right\}
    \left[ \lambda^{2i_r}_{k_r} \right]^{-1}
\begin{array}{c}
  \setlength{\unitlength}{1 pt}
  \begin{picture}(30,45)
   \put( 2,12){${\scriptstyle  i_r}$}     \put( 0,20){\line(1,0){10}}
   \put( 8, 2){${\scriptstyle  2}$}       \put(10,10){\line(0,1){10}}
   \put(12,22){${\scriptstyle  k_r}$}     \put(10,20){\line(1,0){10}}
   \put(22,12){${\scriptstyle  i_{r+1}}$} \put(20,20){\line(1,0){10}}
   \put(20,20){\line(0,1){20}} \put(22,35){${\scriptstyle  P_r}$}
   \put(10,20){\circle*{3}}\put(20,20){\circle*{3}}
\end{picture}\end{array}
~~. 
\end{eqnarray}
Second, we can use recoupling theorem repeatedly to move the
grasp all the way to the edge $P_0$. We denote this operation
as Move 2:
\begin{equation}
\begin{array}{c}
  \setlength{\unitlength}{1 pt}
  \begin{picture}(40,45)
     \put( 2,12){${\scriptstyle  i_{r-1}}$} \put( 0,20){\line(1,0){10}}
     \put(18, 2){${\scriptstyle  2}$}       \put(20,10){\line(0,1){10}}
     \put(12,23){${\scriptstyle  i_r}$}     \put(10,20){\line(1,0){10}}
     \put(32,19){${\scriptstyle  k_{r}}$}   \put(20,20){\line(1,0){10}}
     \put(10,20){\line(0,1){20}} \put(12,35){${\scriptstyle  P_{r-1}}$}
     \put(10,20){\circle*{3}}\put(20,20){\circle*{3}}
\end{picture}\end{array} 
=\sum_{k_r} \left\{\begin{array}{ccc}
                      {k_{r}}  &  {2}    &{k_{r-1}} \\
                      {i_{r-1}}&{P_{r-1}}&{i_r} 
            \end{array}\right\}
\begin{array}{c}
  \setlength{\unitlength}{1 pt}
  \begin{picture}(40,55)
     \put( 2,23){${\scriptstyle  i_{r-1}}$} \put( 0,20){\line(1,0){10}}
     \put( 8, 2){${\scriptstyle  2}$}       \put(10,10){\line(0,1){10}}
     \put(12,12){${\scriptstyle  k_{r-1}}$} \put(10,20){\line(1,0){10}}
     \put(32,19){${\scriptstyle  k_{r}}$}   \put(20,20){\line(1,0){10}}
     \put(20,20){\line(0,1){20}} \put(22,35){${\scriptstyle  P_{r-1}}$}
     \put(10,20){\circle*{3}}\put(20,20){\circle*{3}}
\end{picture}\end{array}
\ \ .
\end{equation}
In this way we can bring all three grasps to the edge
$P_0$. The final step is just given by recognizing that we have
Tet structure on the edge $P_0$.

Let us begin by applying Move 1 to the node $r$. We obtain
\begin{equation}
\bigg\langle{ 
\begin{array}{c}
  \setlength{\unitlength}{1 pt}
  \begin{picture}(155,60)
  \put( 0, 0){$P_0$}\put(15, 0){\line(1,1){10}}
  \put(10,30){\line(0,1){18}}  \put(7,50){$P_1$} 
  \put(25,10){\line(-3,4){15}} \put(27,10){${\scriptstyle  i_1}$}
  \put(10,30){\line(1,0){10}}  \put(40,22){${\scriptstyle   }$}
  \put(10,30){\circle*{3}}
  \put(25,30){\line(1,0){25}}   \put(28,22){${\scriptstyle  i_{r-1}}$}
  \put(40,30){\line(0,1){18}}   \put(37,50){$P_{r-1}$}
  \put(40,30){\circle*{3}}
  \put(50,30){\line(1,0){20}}   
  \put(47,22){${\scriptstyle  i_r}$}       \put(57,22){${\scriptstyle  k_r}$}
  \put(65,30){\line(0,1){18}}   \put(62,50){$P_r$}
  \put(55,30){\circle*{3}}      \put(65,30){\circle*{3}}
  \put(75,30){\line(1,0){20}}   \put(82,22){${\scriptstyle  i_t}$}
  \put(90,30){\line(0,1){18}}   \put(90,50){$P_t$}
  \put(90,32){\oval(20,20)[tl]} 
  \put(90,42){\circle*{3}}      \put(90,30){\circle*{3}}
  \put(115,32){\oval(20,20)[tl]}
  \put(115,42){\line(1,0){12}}  \put(127,42){\circle*{3}}
  \put(135,45){$P_{s}$}         \put(115,30){\line(1,1){15}}
  \put(115,30){\line(1,-1){10}} \put(107,22){${\scriptstyle  i_{s}}$}
  \put( 55,28){\line(0,-1){10}} \put(65, 2){${\scriptstyle  2}$}
  \put( 80,28){\line(0,-1){18}} \put(72,15){${\scriptstyle  2}$}
  \put(105,28){\line(0,-1){ 8}} \put(90, 2){${\scriptstyle  2}$}
  \put( 80,20){\oval(50,20)[b]} \put(80,10){\circle*{3}}
  \put(105,30){\line(1,0){10}}  \put( 93,20){$ $}
  \put(125,20){\line(0,-1){10}} \put(107,8){${\scriptstyle   }$}
  \put(125,20){\circle*{3}}     \put(115,30){\circle*{3}}
  \put(140,20){$ $}             \put(125,20){\line(1,0){10}}
  \put(140, 0){$P_{n-1}$}       \put(125,10){\line(1,-1){10}}
\end{picture}\end{array}
 ~~}\bigg|
\end{equation}
Then, using Move 2 we can move the $(i_r,k_r,2)$ node
to the left of the node $(i_{r-1},P_{r-1},i_t)$:
\begin{equation}
\bigg\langle{\begin{array}{c}
  \setlength{\unitlength}{1 pt}
  \begin{picture}(155,60)
  \put( 0, 0){$P_0$}\put(15, 0){\line(1,1){10}}
  \put(10,30){\line(0,1){18}}  \put(7,50){$P_1$} 
  \put(25,10){\line(-3,4){15}} \put(27,10){${\scriptstyle  i_1}$}
  \put(10,30){\line(1,0){10}}  \put(40,22){$ $}
  \put(10,30){\circle*{3}}
  \put(25,30){\line(1,0){25}}   \put(25,22){${\scriptstyle  i_{r-1}}$}
  \put(55,30){\line(0,1){18}}   \put(40,50){$P_{r-1}$}
  \put(55,30){\circle*{3}}
  \put(50,30){\line(1,0){20}}   
  \put(42,22){${\scriptstyle  k_{r-1}}$}   \put(60,22){${\scriptstyle  k_r}$}
  \put(65,30){\line(0,1){18}}   \put(65,50){$P_r$}
  \put(40,30){\circle*{3}}      \put(65,30){\circle*{3}}
  \put(75,30){\line(1,0){20}}   \put(82,22){${\scriptstyle  i_t}$}
  \put(90,30){\line(0,1){18}}   \put(90,50){$P_t$}
  \put(90,32){\oval(20,20)[tl]} 
  \put(90,42){\circle*{3}}      \put(90,30){\circle*{3}}
  \put(115,32){\oval(20,20)[tl]}
  \put(115,42){\line(1,0){12}}  \put(127,42){\circle*{3}}
  \put(135,45){$P_{s}$}         \put(115,30){\line(1,1){15}}
  \put(115,30){\line(1,-1){10}} \put(107,22){${\scriptstyle  i_{s}}$}
  \put( 40,28){\line(0,-1){10}} \put(65, 2){${\scriptstyle  2}$}
  \put( 80,28){\line(0,-1){18}} \put(72,15){${\scriptstyle  2}$}
  \put(105,28){\line(0,-1){ 8}} \put(90, 2){${\scriptstyle  2}$}
  \put( 80,20){\oval(80,20)[bl]}
  \put( 80,20){\oval(50,20)[br]}\put(80,10){\circle*{3}}
  \put(105,30){\line(1,0){10}}  \put( 93,20){$ $}
  \put(125,20){\line(0,-1){10}} \put(107,8){${\scriptstyle   }$}
  \put(125,20){\circle*{3}}     \put(115,30){\circle*{3}}
  \put(140,20){$ $}             \put(125,20){\line(1,0){10}}
  \put(140, 0){$P_{n-1}$}       \put(125,10){\line(1,-1){10}}
\end{picture}\end{array}
  ~~}\bigg|
\end{equation}
We repeat move 2 until the first node with the $2$ 
edge is coupled to the $P_0$ edge.  In this way, after 
a finite number of moves 2, we have transformed the 
original network to
\begin{equation}
\bigg\langle{ \begin{array}{c}
  \setlength{\unitlength}{1 pt}
  \begin{picture}(155,60)
  \put( 0, 0){$P_0$}\put(15, 0){\line(1,1){10}}
  \put(10,40){$P_1$}\put(25,40){\line(1,-1){10}}
  \put(25,10){\line(0,1){10}} \put(27,10){${\scriptstyle  i_1}$}
  \put(25,20){\line(1,1){10}} \put(32,20){${\scriptstyle  k_1}$}
  \put(40,22){${\scriptstyle  k_2}$}
  \put(35,30){\line(1,0){10}} \put(40,22){${\scriptstyle   }$}
  \put(25,20){\circle*{3}}\put(35,30){\circle*{3}}
  \put(50,30){\line(1,0){20}}   \put(57,22){${\scriptstyle  k_r}$}
  \put(65,22){${\scriptstyle  i_{r+1}}$}
  \put(65,30){\line(0,1){18}}   \put(62,50){$P_r$}
  \put(65,30){\circle*{3}}
  \put(75,30){\line(1,0){20}}   \put(82,22){${\scriptstyle  i_t}$}
  \put(90,30){\line(0,1){18}}   \put(90,50){$P_t$}
  \put(90,32){\oval(20,20)[tl]} 
  \put(90,42){\circle*{3}}      \put(90,30){\circle*{3}}
  \put(115,32){\oval(20,20)[tl]}
  \put(115,42){\line(1,0){12}}  \put(127,42){\circle*{3}}
  \put(135,45){$P_{s}$}         \put(115,30){\line(1,1){15}}
  \put(115,30){\line(1,-1){10}} \put(107,22){${\scriptstyle  i_{s}}$}
  \put( 25,20){\line(2,-1){20}}
  \put( 45,10){\line(1,0){35}}  \put(65, 2){${\scriptstyle  2}$}
  \put( 80,28){\line(0,-1){18}} \put(72,12){${\scriptstyle  2}$}
  \put(105,28){\line(0,-1){ 8}} \put(90, 2){${\scriptstyle  2}$}
  \put( 80,20){\oval(50,20)[br]}\put(80,10){\circle*{3}}
  \put(105,30){\line(1,0){10}}  \put( 93,20){$ $}
  \put(125,20){\line(0,-1){10}} \put(107,8){${\scriptstyle   }$}
  \put(125,20){\circle*{3}}     \put(115,30){\circle*{3}}
  \put(140,20){$ $}             \put(125,20){\line(1,0){10}}
  \put(140, 0){$P_{n-1}$}       \put(125,10){\line(1,-1){10}}
\end{picture}\end{array}
 ~~}\bigg|
\end{equation}
Before repeating this procedure for each of the three
grasps, it is convenient to rename the colors $k_a$ of the
virtual edges as $\bar{k}_a$ (and to replace 
the remaining $i_a$ by $k_a$ as well; this can be done by
inserting a sum over a $\bar{k}_a$ multiplied 
by a $\delta_{i_a}^{\bar{k}_a}$). 

Repeating the sequence of moves for the two grasps 
over the edges $r$ and $s$, we transform the grasped vertex to 
the following final form:
\begin{equation}
{\bigg\langle{ \begin{array}{c}
  \setlength{\unitlength}{1 pt}
  \begin{picture}(140,55)
  \put( 0, 0){$P_0$}\put(15,0){\line(1,1){10}}
  \put(12,16){${\scriptstyle  \bar{k}_1}$}   \put(25,10){\line(0,1){20}}
  \put(20,35){${\scriptstyle  \tilde{k}_1}$} \put(25,30){\line(1,0){20}} 
  \put(45,30){\line(1,0){20}} \put(50,22){${\scriptstyle  k_1}$}
  \put(25,10){\line(1,0 ){20}} \put(32, 2){${\scriptstyle  2}$}
  \put(25,30){\line(1,-1){20}} \put(32,12){${\scriptstyle  2}$}
  \put(45,30){\line(0,-1){20}} \put(36,22){${\scriptstyle  2}$}
  \put(25,10){\circle*{3}}\put(45,10){\circle*{3}}
  \put(25,30){\circle*{3}}\put(45,30){\circle*{3}}
  \put(55,30){\circle*{3}}
  \put( 40,40){$P_2$}           \put(55,40){\line(0,-1){10}}
  \put( 70,40){$\cdots$}        \put(70,30){$\ldots$}
  \put( 90,30){\line(1,0){10}}  \put(78,20){$ $}
  \put(100,30){\line(1,-1){10}} \put(85,20){${\scriptstyle  k_{n-2}}$}
  \put(110,20){\line(0,-1){10}} \put(92,8){${\scriptstyle  k_{n-1}}$}
  \put(120,20){\circle*{3}} \put(100,30){\circle*{3}}
  \put(115,40){$P_{n-3}$}   \put(100,30){\line(1,1){10}}
  \put(125,20){$P_{n-2}$}   \put(110,20){\line(1,0){10}}
  \put(125, 0){$P_{n-1}$}   \put(110,10){\line(1,-1){10}}
\end{picture}\end{array} ~~}\bigg|} .
\end{equation}
This it is equal to the original $n$-valent vertex with the
$i_a$ replaced by $k_a$ and multiplied by
$Tet[k_1,\bar{k}_1,\tilde{k}_1;2,2,2]$ (see eq.\
(\ref{TetInALine})).  Bringing all together, we have show that
the action of the volume operator is described by the sum
(\ref{eq:VolOpSN}) extended over all vertices of the spin
network, where the explicit form for the recoupling matrix
(\ref{eq:recVol}) is given by
\begin{eqnarray}
&&
W^{(n)}_{[rst]}{}_{i_{2}\ldots i_{n-2}}^{k_{2}\ldots k_{n-2}} 
  (P_0,\ldots,P_{n-1}) = \sum_{\bar{k}_1,\ldots,\bar{k}_{n-2}}
          \sum_{\tilde{k}_1,\ldots,\tilde{k}_{n-2}}
 P_t ~P_r ~P_s 
 ~\cdot~\frac{{Tet\left[\begin{array}{ccc} 
                 \bar{k}_1 & \tilde{k}_1 & k_1 \\
                  2  &   2         &   2  
               \end{array}\right]}
        }{\Delta_{P_0}}
\nonumber \\ &&~~~~~~ ~~\qquad\qquad\cdot
 \bigg[ \prod_{a=r+1}^{n-2} \delta_{i_a}^{\bar{k}_a} \bigg]
 \cdot { {\rm M} \left[\begin{array}{ccc} 
                    {i_{r+1}}&{i_r}&{\bar{k}_r}\\
                       {2}   & {r} & {P_r} 
                 \end{array}\right] }
 \cdot \bigg[ \prod_{a=1}^{r-1} 
             \left\{\begin{array}{ccc}
                   {\bar{k}_{a+1}}& {2} &{\bar{k}_{a}}\\
                      {i_a}       &{P_a}&{i_{a+1}}
              \end{array}\right\}
        \bigg]
\label{eq:recMatrix}\\ &&~~~~~~ ~~\qquad\qquad\cdot
  \bigg[ \prod_{b=t+1}^{n-2} \delta_{\bar{k}_b}^{\tilde{k}_b} 
  \bigg]
  \cdot { {\rm M} \left[\begin{array}{ccc} 
                   \bar{k}_{r+1} & \bar{k}_r & \tilde{k}_r \\
                      2         &   t       &    P_t 
                  \end{array}\right] }
  \cdot \bigg[ \prod_{b=1}^{t-1} 
              \left\{\begin{array}{ccc}
               {\tilde{k}_{b+1}} &  {2} & {\tilde{k}_{b}} \\
                 \bar{k}_b       & {P_a}& {\bar{k}_{b+1}} 
              \end{array}\right\}
  \bigg]
\nonumber \\ &&~~~~~~ ~~\qquad\qquad\cdot
  \bigg[ \prod_{c=s+1}^{n-2} \delta_{\tilde{k}_c}^{{k}_c} \bigg]
  \cdot { {\rm M}  \left\{\begin{array}{ccc}
                      a  & b & i \\
                      c  & d & j  
                   \end{array}\right\}
\left[\begin{array}{ccc} 
                   \tilde{k}_{s+1} & \tilde{k}_s & {k}_s \\
                      2         &   s       &    P_s 
\end{array}\right] }
  \cdot \bigg[ \prod_{c=1}^{s-1}  
         \left\{\begin{array}{ccc}
                  k_{c+1}  & 2    & k_c \\
              \tilde{k}_c  & {P_a}&{\tilde{k}_{c+1}}  
         \end{array}\right\}
  \bigg]
\nonumber
\end{eqnarray}
and 
\begin{equation}
{\rm M}\left[\begin{array}{ccc} 
                     {i}_{r+1} & {i}_r & {k}_r \\
                      2        &   r   &  P_r 
\end{array}\right]
= \left\{ 
\begin{array}{lcl} 
    1,  &&r=0; \\ 
    \left[\lambda^{2i_r}_{k_r}\right]^{-1} 
       \left\{\begin{array}{ccc}
                 {i_{r+1}} & {i_r} & {k_r}\\
                    {2}    & {P_r} & {P_r}
       \end{array}\right\}
    ,  &&0<r<n-1 ; \\
    \lambda^{2P_{n-1}}_{P_{n-1}} = - 1,  &&r=n-1 . 
          \end{array} \right. 
\end{equation}
where $i_1=k_1=P_0$ and
$i_{n-1}=k_{n-1}=\bar{k}_{n-1}=\tilde{k}_{n-1}=P_{-1}$. (We
have used the fact that for $A=-1$, $\lambda^{2a}_a=-1$.)

This formula can be specialized to the case of three-vertex
($n=3$) and four-vertex ($n=4$).  In the case of three-vertex
we have:
\begin{eqnarray}
W^{(3)}{}(P_0,P_1,P_2) 
   &=& \bigg| \sum_{\tilde{k}_1} P_0\,P_1\,P_2\, 
        \left[ \lambda^{2\tilde{k}_1}_{P_0}\right]^{-1}
              \left\{\begin{array}{ccc}
                      P_2      &  2  & P_0 \\
                    \tilde{k}_1& P_1 & P_2 
              \end{array}\right\} \,
              \left\{\begin{array}{ccc}
                      P_2  & P_0 & {\tilde{k}}_1 \\
                       2   & P_1 & P_1  
              \end{array}\right\} \,
    \frac{  \left[\begin{array}{ccc}
                      P_0 &{\tilde{k}}_1 & {P_0} \\
                       2  &     2        &  2  
             \end{array}\right]
         }{\Delta_{P_0}}
    \bigg|.
\end{eqnarray}
and a direct computation confirms that the volume 
of any three-vertex is 0.
For the case of four-valent vertex, we obtain the formula: 
\begin{eqnarray}
W^{(4)}_{[013]}{}_{i}^{k} &=&
 \sum_{\tilde{k}_1}\,P_0\,P_1\,P_3\,
      (-1) \left[ \lambda^{2\tilde{k}_1}_{P_0}\right]^{-1}
              \left\{\begin{array}{ccc}
                      i  & P_0 & {\tilde{k}}_1 \\
                      2  & P_1 &  P_1
              \end{array}\right\} \,
              \left\{\begin{array}{ccc}
                      P_3 & 2   & k \\ 
                       i  & P_2 & P_3
              \end{array}\right\}  \,
              \left\{\begin{array}{ccc}
                       k        & 2   & P_0 \\
                     \tilde{k}_1& P_1 & i
              \end{array}\right\}  \, ~.
    \frac{{Tet\left[\begin{array}{ccc} 
                 P_0 & \tilde{k}_1 & P_0 \\
                  2  &   2         &   2  
           \end{array}\right]}
         }{\Delta_{P_0}}
\nonumber 
\end{eqnarray}
and the other 3 matrix that appear in the definition of
the action of the volume operator are easily deduced
from the identities (\ref{eq:id4vertex}).


\subsection{Summary of the volume's action}

Finally, let us summarize the procedure for computing the
eigenvalues and eigenvectors of the volume.  Consider the
spin-network states ${\langle S |}$ with a fixed graph and a fixed
coloring of the real edges, but with arbitrary intersections.
The set of these spin networks forms a finite dimensional
subspace $V$ of the quantum state space.  The subspace $V$ is
invariant under the action of the volume operator.  We denote
the valence of the real vertex $i$ by $n_i$.  Fix a trivalent
decomposition of each vertex $i\in \{ S\cup{\cal
R}\}$. Consider all compatible colorings of the virtual edges.
For every vertex, the number of the compatible colorings
depends on the valence of the vertex, as well as on the
coloring of the external edges.  Let $N_i$ be the number of
compatible colorings of the vertex $n_i$. The dimension $N$ of
the subspace $V$ we are considering is $N=\prod_i N_i$.  Our
aim is to diagonalize the volume operator in $V$.

We indicate a basis in $V$ as follows.  Given a vertex $i$ with
valence $n_i$, we have previously denoted compatible colorings
of the internal edges by $(i_2,\ldots,i_{n_i-2})$. It is more
convenient here to simplify the notation by introducing a
single index $K_i=1,N_i$, which labels all compatible internal
colorings of the vertex $i$.

We now recall the basic expression we have obtained for the
volume, namely eq.\ (\ref{eq:VolOpSN}):
\begin{eqnarray}
 \hat V[{\cal V}] &=& {l_0^3}
 \sum_{i\in\{{S\cap{\cal V}}\}} \hat V_i, 
\nonumber \\
\hat V_i &=&  
\sqrt{  \sum_{\begin{array}{l} 
	      {\scriptstyle  r=0,\ldots,n-3}  \\[-1mm]
              {\scriptstyle  t=r+1,\ldots,n-2} \\[-1mm]
              {\scriptstyle  s=t+1,\ldots,n-1} 
       	\end{array}}
  \left| \frac{{\rm i}}{16}~ \hat W_{[rts]}^{(n_i)} ~\right|} ,
\label{eq:VolOpSN2}
\end{eqnarray}
where the first sum is over the vertices and the second sum is
over the triples of edges adjacent to the vertex.  We have
shown that the operators i$\hat W_{[rts]}^{(n_i)}$ are
diagonalizable matrices with real eigenvalues. These matrices
have components
\begin{equation}
\hat W_{[rts]}^{(n_i)}{}_{K_{I_i}}^{\bar{K}_{I_i}}
    = ~\left[\text{l.h.s. of eq.\ (\ref{eq:recMatrix})}\right] .
\label{def:MatOp}
\end{equation}
Since the matrices 
i$\hat W_{[rts]}^{(n_i)}{}_{K_{I_i}}^{\bar{K}_{I_i}}$
are diagonalizable with real  
eigenvalues, from the spectral theorem we can write them as:
\begin{equation}
 {\rm i}\hat W_{[rts]}^{(n_i)} 
= \sum_{\alpha} ~{}^{\scriptstyle  \alpha}\!\lambda^{(n_i)}_{[rts]} 
  ~{}^{\scriptstyle  \alpha}\!\hat P_{[rts]}^{(i)},
\end{equation} 
where ${}^{\scriptstyle  \alpha}\!\lambda^{(n_i)}_{[rts]}$ 
are real quantities and 
the $~{}^{\scriptscriptstyle \alpha}\!\hat P_{[rts]}^{(i)}$ 
are the spectral projectors of the finite dimensional
matrix operator $\hat W_{[rts]}^{(n_i)}$, acting on the
$i$-th vertex's basis. 

From (\ref{eq:VolOpSN2}), we have then 
\begin{equation}
\hat V_i^2 = 
 \sum_{\begin{array}{l} 
       {\scriptstyle  r=0,\ldots,n_i-3}  \\[-1mm]
       {\scriptstyle  t=r+1,\ldots,n_i-2} \\[-1mm]
       {\scriptstyle  s=t+1,\ldots,n_i-1} 
       \end{array}}
 \sum_\alpha
 \frac{ |~\lambda_{\alpha}^{[rts]} |}{16}~ 
 ~{}^{\scriptstyle  \alpha}\!\hat P_{[rts]}^{(n_i)}.
\label{def:matrixV}
\end{equation}
Being the sum of hermitian non-negative matrices, $\hat
V_i^2 $ as well is diagonalizable with real non-negative
eigenvalues, which we denote as $\lambda_{\beta_i}^2$, and
spectral projectors $P_{\beta_i}$:
\begin{equation}
\hat V_i^2 = 
 \sum_{\beta_i} \lambda^2_{\beta_i} \hat P_{\beta_i}.
\end{equation}
with $\lambda_{\beta_i} \geq 0$.  Therefore we have 
\begin{equation}
\hat V_i = \sum_{\beta_i} 
    \lambda_{\beta_i}  \hat P_{\beta_i}
\end{equation}
and the volume is given by 
\begin{equation}
 \hat V[{\cal V}] = {l_0^3}
 \sum_{i\in\{{S\cap{\cal V}}\}}
 \sum_{\beta_i} \lambda_{\beta_i} \hat P_{\beta_i}. 
\end{equation}
Now, the projectors acting on different vertices commute among
themselves: $\hat P_{\beta_i}\hat P_{\beta'_j}=\hat
P_{\beta'_j}\hat P_{\beta_i}$ if $i \neq j$. Therefore the
eigenvectors of $\hat V$ are the common eigenvectors of all
$\hat V_i$.  They are labeled by one $\beta_i$ for every vertex
$i$, namely by a multi-index $\vec \beta=(\beta_1...\beta_p)$,
where $p$ is the number of vertices in the region.  The
corresponding spectral projectors $\hat P_{\vec\beta}$ of $\hat
V$ are the products over the vertices of the spectral
projectors of the vertex volume operators $\hat V_i$
\begin{equation} 
\hat P_{\vec\beta} = \prod_i \hat P_{\beta_i}.
\end{equation}
It is immediate to conclude that
\begin{equation}
\hat V= {l_0^3} \sum_{\vec\beta} \lambda_{\vec\beta} \hat
P_{\vec\beta},
\end{equation}
where the eigenvalues of the volume are the sums of the
eigenvalues of the volume of each intersection:
\begin{equation}
 \lambda_{\vec\beta} = \sum_i \lambda_{\beta_i}.
\end{equation}

The problem of the determination of the spectrum of the
volume is reduced to a well defined calculation of the
eigenvalues $\lambda_{\beta_i}$, which depend on the valence
and coloring of adjacent vertices of the vertex $i$.  Let us
summarize the various steps of this computation.  Given an
arbitrary real vertex $i$ with coloring of adjacent edges
$P_0,\ldots,P_{n_i-1}$: {\bf (i)} determine the set of the
possible colorings of its virtual edges, and label them by
an index $K_i$; {\bf (ii)} using eq.\ (\ref{eq:recMatrix})
compute the matrix elements $\hat
W_{[rts]}^{(n_i)}{}_{K_{i}}^{\bar{K}_{i}}$; {\bf (iii)} for
each of this matrices, compute its spectral decomposition,
i.e. the eigenvalues $~{}^{\scriptscriptstyle
\alpha}\!\lambda^{(n_i)}_{[rts]}$ and the spectral
projectors $~{}^{\scriptscriptstyle \alpha}\!\hat
P_{[rts]}^{(n_i)}$; {\bf (iv)} compute the matrix $\hat V_i$
from eq.\ (\ref{def:matrixV}); {\bf (v)} compute the
eigenvalues of the matrix $\hat V_i$. The square root of
these give the $\lambda_{\beta_i}$'s.  All these steps can
be fully performed using an algebraic manipulation program
such as {\it Mathematica}.  We have written a {\it
Mathematica} program that performs these calculations, and
we will give free access to this program on line.  In
Appendix F we give the values of the quantities
$\lambda_{\beta_i}(P_0,\ldots,P_{n_i-1})$ for some
$4$-valent and $5$-valent vertex, computed using this
program.

\subsection{Complex Ashtekar connection}

Before closing this section, let us discuss the modifications
that are necessary in order to use the complex Ashtekar connection 
instead of the real one we have used here. On this subject, see also
\cite{Immirzi}.
The difference is
simply the appearance of a factor $i$ in the commutator
between the connection and the triad. This yields to an extra
$i$ in the factor associated to each grasp.  This additional
imaginary factor 
destroies the reality of the eigenvalues of area and volume, which 
is a main result here.  Probably this should be taken as an indication 
that spin networks constructed from the propagator of the {\it complex\/}  
Ashtekar connection are not physical states. We can
illustrate this by means of an analogy. Imagine 
that we study the eigenvalue equation for the momentum operator 
$i\hbar{\partial\over\partial x}$ in the quantum mechanics of a single 
particle. Formally, the functions $\psi(x)=\exp\{kx\}$ solve the eigenvalue 
equation for any real $k$. However, the corresponding eigenvalues are 
imaginary -- an indication that these states are not physical. Indeed, 
they are outside the relevant Hilbert space. The physical eigenstates of 
the momentum are of the form $\psi(x)=\exp\{ikx\}$, with an $i$; and 
these are 
correct (generalized) physical states.  Something similar happens here. 
In fact, one can check that if we insert an $i$ in the exponent of the 
holonomies, namely if we replace (\ref{p}) and (\ref{pp}) by 
\begin{equation}
  \frac{d}{d\tau} U_\gamma (\tau,\tau_0)
       + i \  \frac{d \gamma^a(\tau)}{d\tau}
       A_a(\gamma(\tau)) U_\gamma (\tau,\tau_0) = 0 
\label{pc}
\end{equation} 
and
\begin{equation} 
U_\gamma
(\tau,\tau_0) = {\cal P} e^{- i \int_{\tau_0}^\tau d\tau \
                         \dot{\gamma}^a A_a(\gamma(\tau))}, 
\label{ppc}
\end{equation} 
(where $A_a^i$ is now the {\it complex\/} Ashtekar connection)  then the 
eigenvalues of area and volume result to be real.  Using
(\ref{ppc}) as the definition of the holonomy implies that the 
spin network states correspond (in the connection representation) to 
combination of parallel propagators of $i$ times the Ashtekar connection. 
These seem therefore to be the correct physical states related to real 
geometries. However, this strategy (explored in a previous 
version of this paper) is not viable, at least in this form.  The reason is 
that (\ref{ppc}) is  not invariant under the internal gauge 
transformations generated by the Gauss constraint. Perhaps this 
difficulty can be circumvented by exploiting the complexity of the group 
and the nontriviality of the reality conditions, but for the moment we 
have not been able to find a construction viable for the Riemannian 
Ashtekar connection.  We leave this problem to future 
investigations.


\section{The scalar product}

The results above allow us to introduce a scalar product in
the loop representation.  The original definition of the
loop representation of quantum general relativity left the
problem of fixing the scalar product undetermined: the
scalar product had to be determined by requiring quantum
observables to be hermitian \cite{Rovelli90}.  The problem
was complicated by the fact that the loop ``basis'' is
overcomplete.  Later, the introduction of the
non-overcomplete spin network basis, and the realization
that spin network states (with suitable bases chosen on the
high-valent vertices) are eigenstates of the geometry, lead
to the natural suggestion that spin network states ought to
be orthogonal.  For no reason, however, these states ought
to be ortho-{\it normal}; namely the norm of the spin
network states remained undetermined.  The methods
introduced in this paper allow us to complete the process,
suggest a norm for the spin network states, and thus yield a
complete definition for a scalar product $\langle\ |\
\rangle$.  Here, we define a scalar product $\langle\ |\
\rangle$, and motivate the choice.  We have no compelling
argument for the uniqueness of this scalar product, but we
will show that it satisfies all consistency requirements so
far considered.  Therefore, it is reasonable to take it as a
first ansatz.

Let us begin by considering an $n$-valent vertex. This can
be arbitrarily expanded in trivalent vertices. Let $i_1
.... i_{n-3}$ be the colors of the internal edges, and let
us represent by $|i_1 .... i_{n-3}\rangle$ the $n$-valent
vertex expanded in trivalent vertices colored $i_1
.... i_{n-3}$.  We would like to determine an orthogonal
basis from the quantities $|i_1 .... i_{n-3}\rangle$.  We
have two highly non-trivial requirements.  First, that this
works independently from the way the $n$-valent vertex is
expanded in trivalent ones. Second, that the volume be
hermitian in this basis.  Rather remarkably, we believe,
both requirements can be satisfied.

Let us begin by considering a 4-valent vertex, for
simplicity. There are two ways in which we can expand it in
trivalent vertices. Thus, we have two distinct bases
$|i\rangle$ and $|i'\rangle$ for the 4-valent vertices. If
we wanted both of them to be orthonormal, the transformation
between the two had to be given by a unitary matrix.  Now,
the transformation matrix between the two bases is provided
by the recoupling theorem.  The matrix is given by a Six-J
symbol, seen as a matrix in its two rightmost entries.  It
is easy to see that this matrix is not unitary.  However, we
now show that we can rescale the length of the basis vectors
$|i\rangle$ in such a way that the transformation matrix
becomes unitary (indeed, orthogonal). Indeed, let

\begin{eqnarray}
\bbox{n}_j &=&  \sqrt{\frac{
     \begin{array}{c}\setlength{\unitlength}{.5 pt}
     \begin{picture}(35,25)
        \put(15, 0){\line(0,1){10}}\put(20, 0){\line(0,1){10}}
        \put(15, 0){\line(1,0){ 5}}\put(15,10){\line(1,0){ 5}}
        \put(15,25){\line(1,0){5}}  \put(15,15){$\scriptstyle j$}
        \put(15,15){\oval(30,20)[l]}\put(20,15){\oval(30,20)[r]}
     \end{picture}\end{array}
   }{\begin{array}{c}\setlength{\unitlength}{.5 pt}\begin{picture}(40,40)
        \put(18,32){$\scriptstyle a$}
        \put(18,17){$\scriptstyle b$} 
        \put(18, 2){$\scriptstyle j$} 
        \put(20,15){\oval(40,30)} \put( 0,15){\line(1,0){40}} 
        \put( 0,15){\circle*{3}}  \put(40,15){\circle*{3}}
     \end{picture}\end{array}
     \begin{array}{c}\setlength{\unitlength}{.5 pt}\begin{picture}(40,40)
        \put(18,32){$\scriptstyle c$}
        \put(18,17){$\scriptstyle d$} 
        \put(18, 2){$\scriptstyle j$} 
        \put(20,15){\oval(40,30)} \put( 0,15){\line(1,0){40}} 
        \put( 0,15){\circle*{3}}  \put(40,15){\circle*{3}}
     \end{picture}\end{array}
   }} 
\begin{array}{c}\setlength{\unitlength}{1 pt}
\begin{picture}(50,40)
          \put( 0,0){$a$}\put( 0,30){$b$}
          \put(45,0){$d$}\put(45,30){$c$}
          \put(10,10){\line(1,1){10}}\put(10,30){\line(1,-1){10}}
          \put(30,20){\line(1,1){10}}\put(30,20){\line(1,-1){10}}
          \put(20,20){\line(1,0){10}}\put(22,25){$j$}
          \put(20,20){\circle*{3}}\put(30,20){\circle*{3}}
\end{picture}\end{array}
\\
\bbox{\tilde{n}}_i &=&  \sqrt{\frac{
     \begin{array}{c}\setlength{\unitlength}{.5 pt}
     \begin{picture}(35,25)
        \put(15, 0){\line(0,1){10}}\put(20, 0){\line(0,1){10}}
        \put(15, 0){\line(1,0){ 5}}\put(15,10){\line(1,0){ 5}}
        \put(15,25){\line(1,0){5}}  \put(15,15){$\scriptstyle i$}
        \put(15,15){\oval(30,20)[l]}\put(20,15){\oval(30,20)[r]}
     \end{picture}\end{array}
   }{\begin{array}{c}\setlength{\unitlength}{.5 pt}\begin{picture}(40,40)
        \put(18,32){$\scriptstyle a$}
        \put(18,17){$\scriptstyle d$} 
        \put(18, 2){$\scriptstyle i$} 
        \put(20,15){\oval(40,30)} \put( 0,15){\line(1,0){40}} 
        \put( 0,15){\circle*{3}}  \put(40,15){\circle*{3}}
     \end{picture}\end{array}
     \begin{array}{c}\setlength{\unitlength}{.5 pt}\begin{picture}(40,40)
        \put(18,32){$\scriptstyle c$}
        \put(18,17){$\scriptstyle b$} 
        \put(18, 2){$\scriptstyle i$} 
        \put(20,15){\oval(40,30)} \put( 0,15){\line(1,0){40}} 
        \put( 0,15){\circle*{3}}  \put(40,15){\circle*{3}}
     \end{picture}\end{array}
   }} 
\begin{array}{c}\setlength{\unitlength}{1 pt}
\begin{picture}(40,40)
      \put( 0,0){$a$}\put( 0,40){$b$}
      \put(35,0){$d$}\put(35,40){$c$}
      \put(10,10){\line(1,1){10}}\put(10,40){\line(1,-1){10}}
      \put(20,30){\line(1,1){10}}\put(20,20){\line(1,-1){10}}
      \put(20,20){\line(0,1){10}}\put(22,22){$i$}
      \put(20,20){\circle*{3}}\put(20,30){\circle*{3}}
\end{picture}\end{array}
\end{eqnarray}
In this basis, the recoupling theorem becomes 
\begin{eqnarray}
\bbox{n}_j &=& \sum_i \sqrt{\frac{
     \begin{array}{c}\setlength{\unitlength}{.5 pt}
     \begin{picture}(35,25)
        \put(15, 0){\line(0,1){10}}\put(20, 0){\line(0,1){10}}
        \put(15, 0){\line(1,0){ 5}}\put(15,10){\line(1,0){ 5}}
        \put(15,25){\line(1,0){5}}  \put(15,15){$\scriptstyle j$}
        \put(15,15){\oval(30,20)[l]}\put(20,15){\oval(30,20)[r]}
     \end{picture}\end{array}
     \begin{array}{c}\setlength{\unitlength}{.5 pt}\begin{picture}(40,40)
        \put(18,32){$\scriptstyle a$}
        \put(18,17){$\scriptstyle d$} 
        \put(18, 2){$\scriptstyle i$} 
        \put(20,15){\oval(40,30)} \put( 0,15){\line(1,0){40}} 
        \put( 0,15){\circle*{3}}  \put(40,15){\circle*{3}}
     \end{picture}\end{array}
     \begin{array}{c}\setlength{\unitlength}{.5 pt}\begin{picture}(40,40)
        \put(18,32){$\scriptstyle c$}
        \put(18,17){$\scriptstyle b$} 
        \put(18, 2){$\scriptstyle i$} 
        \put(20,15){\oval(40,30)} \put( 0,15){\line(1,0){40}} 
        \put( 0,15){\circle*{3}}  \put(40,15){\circle*{3}}
     \end{picture}\end{array}
   }{
     \begin{array}{c}\setlength{\unitlength}{.5 pt}\begin{picture}(40,40)
        \put(18,32){$\scriptstyle a$}
        \put(18,17){$\scriptstyle b$} 
        \put(18, 2){$\scriptstyle j$} 
        \put(20,15){\oval(40,30)} \put( 0,15){\line(1,0){40}} 
        \put( 0,15){\circle*{3}}  \put(40,15){\circle*{3}}
    \end{picture}\end{array}
    \begin{array}{c}\setlength{\unitlength}{.5 pt}\begin{picture}(40,40)
        \put(18,32){$\scriptstyle c$}
        \put(18,17){$\scriptstyle d$} 
        \put(18, 2){$\scriptstyle j$} 
        \put(20,15){\oval(40,30)} \put( 0,15){\line(1,0){40}} 
        \put( 0,15){\circle*{3}}  \put(40,15){\circle*{3}}
    \end{picture}\end{array}
    \begin{array}{c}\setlength{\unitlength}{.5 pt}
    \begin{picture}(35,25)
        \put(15, 0){\line(0,1){10}}\put(20, 0){\line(0,1){10}}
        \put(15, 0){\line(1,0){ 5}}\put(15,10){\line(1,0){ 5}}
        \put(15,25){\line(1,0){5}}  \put(15,15){$\scriptstyle i$}
        \put(15,15){\oval(30,20)[l]}\put(20,15){\oval(30,20)[r]}
    \end{picture}\end{array}
   }
}  \left\{\begin{array}{ccc}
          a  & b & i \\
          c  & d & j  
   \end{array}\right\}
~~\bbox{\tilde{n}}_i
\nonumber \\
&=& \sum_i \sqrt{\frac{
     \begin{array}{c}\setlength{\unitlength}{.5 pt}
     \begin{picture}(35,25)
        \put(15, 0){\line(0,1){10}}\put(20, 0){\line(0,1){10}}
        \put(15, 0){\line(1,0){ 5}}\put(15,10){\line(1,0){ 5}}
        \put(15,25){\line(1,0){5}}  \put(15,15){$\scriptstyle j$}
        \put(15,15){\oval(30,20)[l]}\put(20,15){\oval(30,20)[r]}
     \end{picture}\end{array}
     \begin{array}{c}\setlength{\unitlength}{.5 pt}
     \begin{picture}(35,25)
        \put(15, 0){\line(0,1){10}}\put(20, 0){\line(0,1){10}}
        \put(15, 0){\line(1,0){ 5}}\put(15,10){\line(1,0){ 5}}
        \put(15,25){\line(1,0){5}}  \put(15,15){$\scriptstyle i$}
        \put(15,15){\oval(30,20)[l]}\put(20,15){\oval(30,20)[r]}
     \end{picture}\end{array}
   }{
     \begin{array}{c}\setlength{\unitlength}{.5 pt}\begin{picture}(40,40)
        \put(18,32){$\scriptstyle a$}
        \put(18,17){$\scriptstyle d$} 
        \put(18, 2){$\scriptstyle i$} 
        \put(20,15){\oval(40,30)} \put( 0,15){\line(1,0){40}} 
        \put( 0,15){\circle*{3}}  \put(40,15){\circle*{3}}
     \end{picture}\end{array}
     \begin{array}{c}\setlength{\unitlength}{.5 pt}\begin{picture}(40,40)
        \put(18,32){$\scriptstyle c$}
        \put(18,17){$\scriptstyle b$} 
        \put(18, 2){$\scriptstyle i$} 
        \put(20,15){\oval(40,30)} \put( 0,15){\line(1,0){40}} 
        \put( 0,15){\circle*{3}}  \put(40,15){\circle*{3}}
     \end{picture}\end{array}
     \begin{array}{c}\setlength{\unitlength}{.5 pt}\begin{picture}(40,40)
        \put(18,32){$\scriptstyle a$}
        \put(18,17){$\scriptstyle b$} 
        \put(18, 2){$\scriptstyle j$} 
        \put(20,15){\oval(40,30)} \put( 0,15){\line(1,0){40}} 
        \put( 0,15){\circle*{3}}  \put(40,15){\circle*{3}}
    \end{picture}\end{array}
    \begin{array}{c}\setlength{\unitlength}{.5 pt}\begin{picture}(40,40)
        \put(18,32){$\scriptstyle c$}
        \put(18,17){$\scriptstyle d$} 
        \put(18, 2){$\scriptstyle j$} 
        \put(20,15){\oval(40,30)} \put( 0,15){\line(1,0){40}} 
        \put( 0,15){\circle*{3}}  \put(40,15){\circle*{3}}
    \end{picture}\end{array}
   }} 
\begin{array}{c}\setlength{\unitlength}{1 pt}\begin{picture}(55,40)
        \put( 0,15){\line(1,-1){15}} \put(0,22){${\scriptstyle b}$}
        \put( 0,15){\line(1, 1){15}} \put(0, 0){${\scriptstyle a}$}
        \put( 0,15){\circle*{3}}
        \put(30,15){\line(-1, 1){15}} \put(28,20){${\scriptstyle c}$}
        \put(30,15){\line(-1,-1){15}} \put(28, 2){${\scriptstyle d}$}
        \put(30,15){\circle*{3}}
        \put( 0,15){\line(1,0){30}} \put(12,16){${\scriptstyle j}$}
        \put(15,30){\line(1,0){25}} \put(15,30){\circle*{3}}
        \put(15, 0){\line(1,0){25}} \put(15, 0){\circle*{3}}
        \put(40, 0){\line(0,1){30}} \put(42,12){${\scriptstyle i}$}
\end{picture}\end{array}
\!\!\!\!\!\cdot \bbox{\tilde{n}}_i
\nonumber \\
&=& \sum_i ~U(a,b,c,d){}^{~i}_j ~~\bbox{\tilde{n}}_i
\label{eq:RecOrto}
\end{eqnarray}
We now prove that the matrix $U(a,b,c,d){}^{~i}_j$ is real
orthogonal.  The inverse transformation matrix from the
$\bbox{\tilde{n}}_i$ basis to the $\bbox{{n}}_j$ basis is given
by the same expression (\ref{eq:RecOrto}), with a reordering of
the external edges' colorings, i.e:
\begin{equation}
\bbox{\tilde{n}}_k =\sum_k ~U(d,a,b,c){}^{~k}_i ~~\bbox{{n}}_k. 
\end{equation}
Therefore we have the relation 
\begin{equation}
 \sum_i U(a,b,c,d){}^{~i}_j ~U(d,a,b,c){}^{~k}_i = \delta_j^k
\label{eq:OrtRel}
\end{equation}
From direct inspection of eq.\ (\ref{eq:RecOrto}) it
is easy to see that $U(a,b,c,d){}^{~i}_k
=~U(d,a,b,c){}^{~k}_i$. As an immediate consequence of eq.\
(\ref{eq:OrtRel}) we have orthogonality.  Looking at eq.\
(\ref{eq:valSIM}) and (\ref{eq:valTheta}) we can easily compute
the sign of the argument of the square root, which is:
\begin{eqnarray}
\text{sign}(\sqrt{}) &=& \frac{ (-1)^i (-1)^j}{
    (-1)^{(a+b+j)/2 + (c+d+j)/2 + (a+d+i)/2 +(b+c+i)/2}}
\nonumber\\
  &=& (-1)^{a+b+c+d} = +1
\end{eqnarray}
We have thus shown that there exists a basis in which the
recoupling theorem yields a unitary transformation.  For
higher valence vertices, the transformation from one
trivalent expansion to another can be obtained by a repeated
applications of the recoupling theorem transformation, and
therefore by a product of orthogonal matrices. Thus, the
argument above extends immediately to higher valence.

Now, the normalization we have found is exactly the one in
which the volume operator is represented by a real
antisymmetric matrix, as shown by eqs.\ (\ref{eq:Wtilda1})
and (\ref{eq:Wtilda}). Therefore we have found a basis that
satisfies all our requirements.

We thus define the normalized spin-network states by the
following normalization: given an arbitrary spin-network state
${\langle S |}$, we label with an index $i\in{V}$ all the $3$-valent
vertices of the expanded state (virtual and real)) and with a
index $e\in {\cal E}$ all its edges (virtual and real).  We
denote the color of the edge $e$ by with $p_e$ and the color of
the three edges adjacent to the vertex $i$ by $a_i$, $b_i$ and
$c_i$. We define the {\it normalized\ } spin network-state
${\langle S |}_{{\textstyle N}}$ by
\begin{equation}
{\langle S |}_{{\textstyle N}} ~=  
   \sqrt{\prod_{i\in{V}} \prod_{e\in {\cal E}}
        \frac{ \Delta_{p_e}}{\theta(a_i,b_i,c_i)} } 
  ~~ {\langle S |}
\nonumber
\end{equation}
And we define a scalar product on $\cal V$ by requiring that
these states are orthonormal.  We have immediately from the
discussion above that the definition does not depend on the
trivalent expansion chosen, and that the volume and area
operators are symmetric with respect to this scalar product.

We think that the scalar product defined in this way is
precisely the one defined on the loop representation by the
loop transform \cite{Rovelli90,Isham96} of the
Ashtekar-Lewandowski measure \cite{Lewandowski}, namely the
conventional Haar measure lattice gauge theory scalar product
for each graph.  The precise relation is discussed by
Reisenberger \cite{Reisenberger96} and in \cite{DePietri96}.  
In turn, we expect that
(the norm derived from) the scalar product we have defined is
equivalent to the evaluation of the Kauffman bracket of the
state, and to the trace of the Temperley-Lieb algebra,
discussed in Appendix B.


\section{Conclusions and future directions}

We have reviewed the kinematics of the 
loop representation of quantum gravity,
and presented a number of novel results.  We have 
modified the definition of the theory by 
inserted a minus sign in the definition of the loop
observables.  With this convention, the spinor identity is
transformed into the binor identity, allowing immediately a
local graphical calculus for the grasping operation and the use
of recoupling theory.  We have shown that the loop states obey
the axioms of recoupling theory, and the corresponding
graphical formalism provides a powerful tool for computing the
action of geometrical operators. We have discussed in detail
the way in which recoupling theory can be used in this context.

Using recoupling theory, we have re-derived known results on
the eigenstates of the area, and the volume of trivalent and
4-valent vertices. We have given a general expression for the
volume of higher valence vertices.  We have proven that the
square root in the volume operator is well defined, because the
relevant operator is hermitian.  We have defined a scalar
product by a suitable normalization of the trivalent spin
networks. We have shown that that the scalar product is well
defined and independent from the trivalent expansion chosen,
and that the volume is symmetric with respect to this scalar
product.

Notice that the area and volume operators $\hat A$ and $\hat V$
do {\it not\/} correspond to physical observables: they are not
gauge invariant and do not commute with GR's constraints. The
areas and volumes that we routinely measure are associated to
spatial regions determined by matter.  Indeed, the area and
volume of regions determined by physical matter {\it are\ }
represented on the phase space of the coupled gravity-matter
theory by observables which are gauge invariant (see for
instance \cite{Rovelli91b}).  However, it was suggested in
Ref.\ \cite{Rovelli93a} that it is reasonable to expect that
these physical areas and volumes (of spatial regions determined
by matter) be {\it still\ } expressed by (operators unitary
equivalent to) $\hat A$ and $\hat V$.  See Refs.\
\cite{Rovelli91b} and \cite{Rovelli93a} for the details of the
argument.  If this suggestion is correct, the spectra computed
here can be taken as physical predictions on short scale
geometry, following from the loop representation of quantum
gravity \cite{Rovelli93a}.  These predictions are testable in
principle, and could perhaps lead to indirect observable 
consequences.

We consider the following open problems particularly important
for the development of the theory.
\begin{itemize}
\item We have not explored the degenerate cases in the action
     of the area operator (But see \cite{AshtekarLewand,Frittelli}).
\item We believe that the formalism is now well established for
     a precise discussion of the Hamiltonian and for computing
     transition amplitudes \cite{Rovelli94b,Rovelli95b}.
\item Can a weave \cite{Ashtekar92a} be found for which not
     just the area but the volume as well approximates smooth
     geometries?  Can a weave related to a four dimensional
     geometry \cite{Iwasaki} be constructed?
\item A way of implementing the Lorentzian reality conditions
     is, to our knowledge, still lacking (For an attempt to
     address this problem, see \cite{Thiemann}).
\item Under the optimistic assumption that the above technical
       problems could be addressed, a possible first task for
       the theory could be the following.  Compute the clock
       time evolution of a weave representing a black hole;
       show that Hawking's radiation \cite{Hawking} is emitted,
       and determine the final stage of the black hole after
       evaporation.
\item Supposing that area and volume eigenvalues computed here
	describe an actual physical discreteness (in the
	quantum sense) of Planck scale geometry, could there be
	any low energy observable consequence of such
	discreteness?\footnote{
	Note added: Two applications of these
	results have been studied after the appearance of the 
	preprint of this paper: one on black hole's emission spectrum
	\cite{Barreira}, and one on black hole entropy.\cite{Rovelli96}
	}
\end{itemize} 


\acknowledgments

We thank Mike Reisenberger and Lee Smolin for teaching us the
relevance of recoupling theory and of its tangle theoretical
version; Roumen Borissov, Simonetta Frittelli, 
Viqar Husain, Giorgio Immirzi, Luis Lehner, Renate Loll 
and the anonymous referee for a careful reading 
of the manuscript and many important suggestions; Seth Major 
and Abhay Ashtekar for valuable criticisms and insights.  One of us (RDP) 
thanks the members of the Relativity group of Pittsburgh 
--where this work begun-- for their warm hospitality, as well
as Massimo Pauri and Luca Lusanna for their continuous support
and encouragement.  This work has been
partially supported by the NSF grant PHY-90-12099 (USA), by the
INFN grant ``Iniziativa specifica FI-41'' (Italy), and by the
Human Capital and Mobility Program ``Constrained Dynamical
Systems'' (European Union).


\appendix


\section{Pauli matrices identities}

Defining $\tau_i = - \frac{{\rm i}}{2} ~\sigma_i$, where
$\sigma_i$ are the Pauli matrices, we have the following
identities:
\begin{eqnarray}
 {\rm Tr}[ \tau_i \tau_j ]
 &=& - \frac{1}{2} \delta_{ij}, 
\label{tr2tau}\\
 {\rm Tr}[ \tau_i \tau_j \tau_k ]
 &=& - \frac{1}{4} \epsilon_{ijk} ,
\label{tr3tau}\\
 \delta^{ij} \tau_i{}_A^{~B} \tau_j{}_C^{~D}
    &=& -\frac{1}{4}\left( 
       \delta_A^{~D}\delta_B^{~C} - \epsilon^{BD} \epsilon_{AC}
       \right), \\
 \delta^{ij} {\rm Tr}[ A \tau_i ]{\rm Tr}[ B \tau_j ]
   &=& -\frac{1}{4}\left\{ {\rm Tr}[ A B] - {\rm Tr}[ A B^{-1}] 
                 \right\}, \\
 A^{-1}{}_A^{~B} &=& \epsilon^{BD} \epsilon_{AC} A{}_D^{~C}, \\
 \delta_A^{~B} \delta_D^{~C} 
  &=& \delta_A^{~C} \delta_D^{~B} + \epsilon^{BC}\epsilon_{AD}, 
\\ 
{\rm Tr}[A]{\rm Tr}[B] &=& {\rm Tr}[AB]+{\rm Tr}[AB^{-1}],
\label{spinorID}
\end{eqnarray}
where $A$ and $B$ are $SL(2,C)$ matrices.


\section{Kauffman Brackets and Temperley-Lieb Recoupling theory}

In the context of Knot theory \cite{Kauffman94}, the appearance 
of recoupling theory is based on the observation that the 
Kauffman bracket satisfies the properties (and is completely
determined by the properties) 
\begin{equation}
 {\big\langle} \begin{array}{c}{\setlength{\unitlength}{1 pt}
\begin{picture}(25,15) 
        \put( 0, 0){\line( 5, 3){10}}
        \put(25,15){\line(-5,-3){10}}
        \put(25, 0){\line(-5, 3){25}}
\end{picture}}\end{array}
~{\big\rangle} = A
     {\big\langle} \begin{array}{c}{\setlength{\unitlength}{1 pt}
\begin{picture}(10,15) 
\put(5, 0){\oval(10,10)[t]}
\put(5,15){\oval(10,10)[b]} 
\end{picture}}\end{array}
 ~{\big\rangle} + A^{-1}
     {\big\langle}~\begin{array}{c}{\setlength{\unitlength}{1 pt}
\begin{picture}(10,15) 
\put( 0,0){\line(0,1){15}}
\put(10,0){\line(0,1){15}}
\end{picture}}\end{array}
 ~{\big\rangle}
\label{eq:KB1}
\label{SkeinId}
\end{equation}
and
\begin{equation}
{\big\langle} \begin{array}{c}{\setlength{\unitlength}{1 pt}
\begin{picture}(10,10) \put(5,5){\oval(10,10)}
\end{picture}}\end{array}
  ~\cup~ \bbox{K}~{\big\rangle} = 
   d~ {\big\langle} ~\bbox{K}~ {\big\rangle}
\label{eq:KB2}
\end{equation}
where $\langle ~ \rangle$ denotes the Kauffman bracket, where
             $d = - A^2 - A^{-2}$ and $\bbox{K}$ is any diagram
             that does not intersect the added loop.  These
             properties of the Kauffman bracket are sufficient
             to generate the entire formalism of recoupling
             theory.  In particular, they generate a ``tangle
             theoretic'' interpretation of the Temperley-Lieb
             algebra as follows.

A planar tangle is a set of lines on a plane.  It is possible
to write an arbitrary tangle inside the Kauffman brackets as
the sum of non-intersecting tangles by applying eq.\
(\ref{eq:KB1}) to all crossings.  In \cite{Kauffman90b} it is
shown that every planar non intersecting $n$-tangle with $n$
inputs and $n$ outputs is equivalent to the product of
elementary tangles $\openone_{n}$, $U_1$, $\ldots$,$U_{n-1}$,
given by
\begin{eqnarray*}
\begin{array}{c}\setlength{\unitlength}{1 pt}
\begin{picture}(20,20)
       \multiput(5,0)(2,0){6}{\line(0,1){5}}
       \put(1,5){\framebox(18,10){$\openone_{n}$}}
       \multiput(5,15)(2,0){6}{\line(0,1){5}}
\end{picture}\end{array}
&=& \begin{array}{c}\setlength{\unitlength}{1 pt}
    \begin{picture}(20,20)
       \multiput(2,0)(2,0){6}{\line(0,1){20}}
    \end{picture}\end{array}
\\
\begin{array}{c}\setlength{\unitlength}{1 pt}
\begin{picture}(20,20)
       \multiput(5,0)(2,0){6}{\line(0,1){5}}
       \put(1,5){\framebox(18,10){$U_1$}}
       \multiput(5,15)(2,0){6}{\line(0,1){5}}
\end{picture}\end{array}
&=& \begin{array}{c}\setlength{\unitlength}{1 pt}
    \begin{picture}(20,15)
       \multiput(10,0)(2,0){4}{\line(0,1){15}}
       \put(5,0){\oval(6,10)[t]}
       \put(5,15){\oval(6,10)[b]}
    \end{picture}\end{array} 
\\
 \vdots &\vdots&\vdots\\
\begin{array}{c}\setlength{\unitlength}{1 pt}
\begin{picture}(20,20)
       \multiput(5,0)(2,0){6}{\line(0,1){5}}
       \put(1,5){\framebox(18,10){$U\!\!{}_{n\!{\scriptscriptstyle -\!1}}$}}
       \multiput(5,15)(2,0){6}{\line(0,1){5}}
\end{picture}\end{array}
&=& \begin{array}{c}\setlength{\unitlength}{1 pt}
    \begin{picture}(20,15)
       \multiput(2,0)(2,0){4}{\line(0,1){15}}
       \put(14,0){\oval(6,10)[t]}
       \put(14,15){\oval(6,10)[b]}
    \end{picture}\end{array}
\end{eqnarray*}
where the product is interpreted as a stacking of two
diagrams.  Two such products represent tangles equivalent
under the Kauffman brackets if and only they can be
transformed into each other by the relations
\begin{eqnarray}
 && U_i^2 = d~U_i \label{eq:C1}\\
 && U_i U_{i\pm1} U_i = U_i \\
 && U_i U_j = U_j U_i \qquad, ~|i-j|>1 ~~,
\end{eqnarray}
which are at the basis of the Temperley-Lieb algebra. 
For example (\ref{eq:C1}), means:
\begin{equation}
{\begin{array}{c}\setlength{\unitlength}{1 pt}
       \begin{picture}(20,40)
       \multiput(5,0)(2,0){6}{\line(0,1){5}}
       \put(1,5){\framebox(18,10){$U_1$}}
       \multiput(5,15)(2,0){6}{\line(0,1){5}}
       \multiput(5,20)(2,0){6}{\line(0,1){5}}
       \put(1,25){\framebox(18,10){$U_1$}}
       \multiput(5,35)(2,0){6}{\line(0,1){5}}
\end{picture}\end{array}}
= \begin{array}{c}\setlength{\unitlength}{1 pt}
  \begin{picture}(20,40)
       \multiput(10,0)(2,0){4}{\line(0,1){40}}
       \put(5,0){\oval(6,10)[t]}
       \put(5,20){\oval(6,20)}
       \put(5,40){\oval(6,10)[b]}
  \end{picture}\end{array}
= \begin{array}{c}\setlength{\unitlength}{1 pt}
  \begin{picture}(25,40)
       \multiput(15,0)(2,0){4}{\line(0,1){40}}
       \put(10,0){\oval(6,20)[t]}
       \put( 3,20){\oval(6,10)}
       \put(10,40){\oval(6,20)[b]}
  \end{picture}\end{array}
= d~ 
  \begin{array}{c}\setlength{\unitlength}{1 pt}
  \begin{picture}(20,20)
       \multiput(5,0)(2,0){6}{\line(0,1){5}}
       \put(1,5){\framebox(18,10){$U_1$}}
       \multiput(5,15)(2,0){6}{\line(0,1){5}}
  \end{picture}\end{array} .
\end{equation}

Given an $n$-tangle ${\bbox{x}}$, let $\bar{{\bbox{x}}}$ denote
the standard closure of ${\bbox{x}}$, obtained by 
attaching the $k^{th}$ input to the $k^{th}$
output.
$$
\begin{array}{c}\setlength{\unitlength}{1 pt}
       \begin{picture}(20,20)
       \multiput(5,0)(2,0){6}{\line(0,1){5}}
       \put(1,5){\framebox(18,10){$x$}}
       \multiput(5,15)(2,0){6}{\line(0,1){5}}
\end{picture}\end{array}
\Rightarrow 
 \bar{x} = \begin{array}{c}
   \setlength{\unitlength}{1 pt}
   \begin{picture}(50,50)
       \put(20,15){\oval(10,10)[b]}
       \put(20,15){\oval(14,14)[b]}
       \put(20,15){\oval(18,18)[b]}
       \put(20,15){\oval(22,22)[b]}
       \put(20,15){\oval(26,26)[b]}
       \put(20,15){\oval(30,30)[b]}
       \multiput(5,15)(2,0){6}{\line(0,1){5}}
       \put(1,20){\framebox(18,10){$x$}}
       \multiput(5,30)(2,0){6}{\line(0,1){5}}
       \multiput(25,15)(2,0){6}{\line(0,1){20}}
       \put(20,35){\oval(10,10)[t]}
       \put(20,35){\oval(14,14)[t]}
       \put(20,35){\oval(18,18)[t]}
       \put(20,35){\oval(22,22)[t]}
       \put(20,35){\oval(26,26)[t]}
       \put(20,35){\oval(30,30)[t]}
\end{picture}\end{array} . 
$$
The Temperley-Lieb algebra $T_n$ is the free additive
algebra over $\tilde{Z}[A,A^{-1}]$ with multiplicative
generators $\openone_n$, $U_1$, $\ldots$,$U_{n-1}$. The
trace on the algebra $T_n$ is defined by:
\begin{description}
\item {(i)} If ${\bbox{x}}$ is an $n$-tangle then $tr({\bbox{x}}) =
     \langle \bar{{\bbox{x}}} \rangle$ where $\langle ~ \rangle$
     denotes the Kauffman bracket, or, which is the same,
     the recursive evaluation of $\bar{x}$ using
     (\ref{eq:KB1}) and (\ref{eq:KB2}) .
\item {(ii)}
     $tr(\bbox{x}+\bbox{y})=tr(\bbox{x})+tr(\bbox{y})$.
\end{description}

\subsection{The Jones-Wenzel projector}

It can be shown \cite{Kauffman94} that in the Temperley-Lieb
algebra $T_n$ there exist one (and only one) element $\Pi_n
\in T_n$ such that $\Pi_n^2=\Pi_n$ and $\Pi_n U_i = U_i
\Pi_n$, $i=1,\ldots,n-1$. This unique element it is called
the Jones-Wenzel projector of $T_n$.  Its explicit
expression is given by:
\begin{equation}
 \Pi_n = \begin{array}{c}\setlength{\unitlength}{1 pt}
       \begin{picture}(10,25)
       \put(5, 0){\line(0,1){10}}
       \put(0,10){\framebox(10,5){}}
       \put(5,15){\line(0,1){10}}\put(7,17){n}
\end{picture}\end{array}
= \frac{1}{n!} \sum_p
    (A^{-3})^{|p|}\ P^{(p)}_{n}.
\label{eq:defJW}
\end{equation}
$P{(p)}, p=1...n!$ is the $n$-tangle obtained by all possible
permutations $p$ in the way the $n$ lines entering $e$ are
connected to the $n$ outgoing lines, $P^{(p)}_{n}$ its a minimal
representation of the permutation $(p)$, and $|p|$ its the
parity of the representation. Since any $n$-tangle 
can be expanded, using eq.\ (\ref{eq:KB1}), in 
a sum of {\it non-intersecting} tangles, the expression
(\ref{eq:defJW}) is an elements of $T_n$. As an example we give
the definition of $\Pi_2$:
\begin{eqnarray*}
\Pi_2 &=& 
\begin{array}{c}\setlength{\unitlength}{1 pt}
\begin{picture}(10,25)
     \put(5, 0){\line(0,1){10}}
     \put(0,10){\framebox(10,5){}}
     \put(5,15){\line(0,1){10}}\put(7,17){2}
\end{picture}\end{array}
=\frac{1}{\{2\}!}\left[
\begin{array}{c}\setlength{\unitlength}{1 pt}
     \begin{picture}(10,15)\put(0,0){\line(0,1){15}}
     \put(10,0){\line(0,1){15}}
\end{picture}\end{array} 
 +A^{-3} 
\begin{array}{c}{\setlength{\unitlength}{1 pt}
\begin{picture}(25,15) 
        \put( 0, 0){\line( 5, 3){10}}
        \put(25,15){\line(-5,-3){10}}
        \put(25, 0){\line(-5, 3){25}}
\end{picture}}\end{array}
\right]  \\
&=& \frac{1}{1+A^{-4}}\left[
\begin{array}{c}\setlength{\unitlength}{1 pt}
     \begin{picture}(10,15)   
     \put(0,0){\line(0,1){15}}\put(10,0){\line(0,1){15}}
     \end{picture}
\end{array}
 +A^{-4}\begin{array}{c}\setlength{\unitlength}{1 pt}
     \begin{picture}(10,15) 
     \put(0,0){\line(0,1){15}}\put(10,0){\line(0,1){15}}
     \end{picture}
     \end{array} 
 +A^{-2} \begin{array}{c}\setlength{\unitlength}{1 pt}
     \begin{picture}(10,15)
     \put(5,0){\oval(10,10)[t]}\put(5,15){\oval(10,10)[b]}
     \end{picture}\end{array} 
\right]  \\
&=&  \begin{array}{c}\setlength{\unitlength}{1 pt}
     \begin{picture}(10,15)
     \put(0,0){\line(0,1){15}}\put(10,0){\line(0,1){15}}
     \end{picture}\end{array} 
  + \frac{1}{A^{2}+A^{-2}} \begin{array}{c}\setlength{\unitlength}{1 pt}
     \begin{picture}(10,15)
     \put(5,0){\oval(10,10)[t]}\put(5,15){\oval(10,10)[b]}
     \end{picture}\end{array}  \\
&=&  \begin{array}{c}\setlength{\unitlength}{1 pt}
     \begin{picture}(10,15)
     \put(0,0){\line(0,1){15}}\put(10,0){\line(0,1){15}}
     \end{picture}\end{array}  
  - \frac{1}{d}  \begin{array}{c}\setlength{\unitlength}{1 pt}
     \begin{picture}(10,15)
     \put(5,0){\oval(10,10)[t]}\put(5,15){\oval(10,10)[b]}
     \end{picture}\end{array}   
 = \openone_2 - \frac{1}{d} ~U_1.
\end{eqnarray*}
In the $A=-1$ case the projectors reduce to antisymmetrizers.

\subsection{A special sum of tangles: the three vertex}

A special sum of tangles is indicated by a $3$-vertex. 
Each line of the vertex is labeled
with a positive integer $a$, $b$ or $c$ as shown below
$$
\begin{array}{c}\setlength{\unitlength}{1 pt}
\begin{picture}(40,40)
       \put(15,15){\line(-1, 1){10}} \put( 4,27){$a$}
       \put(15,15){\line( 1, 1){10}} \put(22,27){$b$}
       \put(15, 5){\line(0,1){10}}   \put(17,1){$c$}
       \put(15,15){\circle*{3}}
\end{picture}\end{array}
$$
and it is assumed that $m=(a+b-c)/2$, $n=(b+c-a)/2$ and  
$p=(c+a-b)/2$ are positive integer.  This last condition
is called the {\it admissibility condition} for the $3$-vertex
$(a,b,c)$. A line labeled by a positive integer $a$ 
is interpreted as the non-intersecting $n$-tangle $\openone_a$. 
The $3$-vertex is then defined as:
\begin{equation}
\begin{array}{c}\setlength{\unitlength}{1 pt}
\begin{picture}(40,40)
       \put(15,15){\line(-1, 1){10}} \put( 4,27){$a$}
       \put(15,15){\line( 1, 1){10}} \put(22,27){$b$}
       \put(15, 5){\line(0,1){10}}   \put(17,1){$c$}
       \put(15,15){\circle*{3}}
\end{picture}\end{array}
\stackrel{def}{=}
\begin{array}{c}\setlength{\unitlength}{1 pt}
\begin{picture}(70,40)
       \put(20,30){\line( 1, 0){30}} \put(28,32){$m$}
       \put(35,15){\line(-1, 1){15}} \put(20,17){$p$}
       \put(35,15){\line( 1, 1){15}} \put(44,17){$n$}
       \put(35, 1){\line(0,1){10}}   \put(42,1){$c$}
       \put(27,11){\framebox(16,4){}}
       \put(16,22){\framebox(4,16){}} 
       \put(6 ,30){\line(1,0){10}}\put(0,30){$a$}
       \put(50,22){\framebox(4,16){}}
       \put(54,30){\line(1,0){10}}\put(65,30){$b$}
\end{picture}\end{array}
\end{equation}
Here, it is understood that each Temperley-Lieb projector
is fully expanded. For instance
\begin{eqnarray*}
\begin{array}{c}\setlength{\unitlength}{1 pt}
\begin{picture}(40,40)
       \put(15,15){\line(-1, 1){10}} \put( 4,27){$1$}
       \put(15,15){\line( 1, 1){10}} \put(22,27){$1$}
       \put(15, 5){\line(0,1){10}}   \put(17,1){$2$}
       \put(15,15){\circle*{3}}
\end{picture}\end{array}
&=&
\begin{array}{c}\setlength{\unitlength}{1 pt}
\begin{picture}(25,35)
       \put(10,20){\line(-1, 1){15}} 
       \put(14,20){\line( 1, 1){15}} 
       \put(10, 0){\line(0,1){20}}   
       \put(14, 0){\line(0,1){20}}   
\end{picture}\end{array}
 -\frac{1}{d} 
\begin{array}{c}\setlength{\unitlength}{1 pt}
\begin{picture}(25,35)
       \put(10,20){\line(-1, 1){15}} 
       \put(14,20){\line( 1, 1){15}} 
       \put(12,20){\oval(4,16)[b]}   
       \put(12, 0){\oval(4,16)[t]}   
\end{picture}\end{array} \\
\begin{array}{c}\setlength{\unitlength}{1 pt}
\begin{picture}(40,40)
       \put(15,15){\line(-1, 1){10}} \put( 4,27){$2$}
       \put(15,15){\line( 1, 1){10}} \put(22,27){$2$}
       \put(15, 5){\line(0,1){10}}   \put(17,1){$2$}
       \put(15,15){\circle*{3}}
\end{picture}\end{array}
&=&
\begin{array}{c}\setlength{\unitlength}{1 pt}
\begin{picture}(25,35)
       \put(10,20){\line(-1, 1){15}} 
       \put(12,22){\line(-1, 1){13}} 
       \put(12,22){\line( 1, 1){13}} 
       \put(14,20){\line( 1, 1){15}} 
       \put(10, 0){\line(0,1){20}}   
       \put(14, 0){\line(0,1){20}}   
\end{picture}\end{array}
+ \frac{2}{d^2} 
\begin{array}{c}\setlength{\unitlength}{1 pt}
\begin{picture}(35,35)
       \put(13,21){\line(-1, 1){14}} 
       \put(13,25){\line(-1, 1){10}}
       \put(13,21){\line(0,1){4}} 
       \put(21,21){\line(0,1){4}} 
       \put(21,25){\line( 1, 1){10}} 
       \put(21,21){\line( 1, 1){14}} 
       \put(17, 0){\oval(4,26)[t]}   
\end{picture}\end{array} \\
&& \!\!\!\!\!\!\!\!\!\!\!\!\!\!
-\frac{1}{d} \left[ 
\begin{array}{c}\setlength{\unitlength}{1 pt}
\begin{picture}(35,35)
       \put(13,21){\line(-1, 1){14}} 
       \put(13,25){\line(-1, 1){10}}
       \put(13,21){\line(0,1){4}} 
       \put(15,20){\line( 1, 1){15}} 
       \put(19,20){\line( 1, 1){15}} 
       \put(15, 0){\line(0,1){20}}   
       \put(19, 0){\line(0,1){20}}  
\end{picture}\end{array} 
 + 
\begin{array}{c}\setlength{\unitlength}{1 pt}
\begin{picture}(25,35)
       \put(10,20){\line(-1, 1){15}} 
       \put(12,22){\line(-1, 1){13}} 
       \put(12,22){\line( 1, 1){13}} 
       \put(14,20){\line( 1, 1){15}} 
       \put(12,20){\oval(4,16)[b]}   
       \put(12, 0){\oval(4,16)[t]} 
\end{picture}\end{array} 
+
\begin{array}{c}\setlength{\unitlength}{1 pt}
\begin{picture}(35,35)
       \put(15,20){\line(-1, 1){15}} 
       \put(19,20){\line(-1, 1){15}} 
       \put(21,21){\line(0,1){4}} 
       \put(21,25){\line( 1, 1){10}} 
       \put(21,21){\line( 1, 1){14}} 
       \put(15, 0){\line(0,1){20}}   
       \put(19, 0){\line(0,1){20}}  
\end{picture}\end{array} 
\right]
\end{eqnarray*}

\subsection{Chromatic evaluation}

If we joining trivalent vertices by their edges, we obtain
trivalent networks.  Thus, in the present context a trivalent
spin network is defined as a trivalent graph with an admissible
coloring.  Notice that in this context spin networks are not
embedded in a three dimensional space.  An edge of color $n$
represents $n$ parallel lines and a Jones-Wenzel projector, and
a vertex is understood as completed expanded in terms of
non-intersecting tangles, as above.  Thus, a trivalent spin
network determines a closed tangle. We can compute the Kauffman
bracket, or the Temperley-Lieb trace, of such tangle.  This is
also called the chromatic evaluation, or network evaluation.
The explicit calculation of the trace is generally based on a
generalization of the chromatic method of spin-network
evaluation \cite{Moussoris79}. In ref.\ \cite{Moussoris79} this
method is used in order to compute the Clebsh-Gordon
coefficients for the group $SU(2)$.

Chromatic evaluations of simple networks are given in Appendix
E. We refer to \cite{Kauffman94} for the details of the
computations. Here, we perform one such computation explicitly,
as an example.  Let us consider the spin network formed by two
trivalent vertices joined to each other. This is called the
$\theta$ network. Consider the case with edges of color
$2,1,1$:
\begin{eqnarray} 
\begin{array}{c}\setlength{\unitlength}{.5 pt}\begin{picture}(40,40)
        \put(18,32){$\scriptstyle 1$}
        \put(18,17){$\scriptstyle 2$} 
        \put(18, 2){$\scriptstyle 1$} 
        \put(20,15){\oval(40,30)} \put( 0,15){\line(1,0){40}} 
        \put( 0,15){\circle*{3}}  \put(40,15){\circle*{3}}
\end{picture}\end{array}
&=&
\begin{array}{c}\setlength{\unitlength}{1 pt}
\begin{picture}(40,40)
      \put(18,34){$1$} \put(18, 0){$1$} \put(24,19){$2$}
      \put( 0,13){\line(1,0){18}} \put(22,13){\line(1,0){18}} 
      \put( 0,17){\line(1,0){18}} \put(22,17){\line(1,0){18}} 
      \put(20,17){\oval(40,30)[t]}\put(20,13){\oval(40,30)[b]} 
      \put(18,10){\framebox(4,10){}}
\end{picture}\end{array}
= \begin{array}{c}\setlength{\unitlength}{1 pt}
  \begin{picture}(40,40)
        \put(18,34){$1$} \put(18, 0){$1$} \put(24,19){$2$}
        \put( 0,13){\line(1,0){40}}
        \put( 0,17){\line(1,0){40}} 
        \put(20,17){\oval(40,30)[t]}   
        \put(20,13){\oval(40,30)[b]}   
\end{picture}\end{array}
- \frac{1}{d}
\begin{array}{c}\setlength{\unitlength}{1 pt}
\begin{picture}(40,40)
    \put(18,34){$1$} \put(18, 0){$1$} \put(24,19){$2$}
    \put( 0,13){\line(1,0){16}} \put(24,13){\line(1,0){16}} 
    \put( 0,17){\line(1,0){16}} \put(24,17){\line(1,0){16}} 
    \put(20,17){\oval(40,30)[t]} \put(20,13){\oval(40,30)[b]}   
    \put(16,15){\oval(4,4)[r]}  \put(24,15){\oval(4,4)[l]} 
\end{picture}\end{array} =
\nonumber
\\&=& d^2 - \frac{1}{d} d = d^2 - 1 
   = \left( -(A^2+A^{-2}) \right)^2 -1 = \nonumber\\
&=& [3] 
= \frac{ (-1)^{0+1+1} [3]! [1]! [1]! [0]! }{[2]![1]![1]!} 
~~.
\end{eqnarray} 
We have: (1) expanded the trivalent vertices explicitly; (2)
computed the trace using (\ref{eq:KB2}); (3) written the
expression in terms of quantum integer (\ref{eq:quantumInt});
(4) compared the result with the general formula of the
chromatic evaluation of the $\theta$ net
(\ref{eq:valTheta}). In the $A=-1$ case, the above gives
\begin{equation} 
\begin{array}{c}\setlength{\unitlength}{.5 pt}\begin{picture}(40,40)
        \put(18,32){$\scriptstyle 1$}
        \put(18,17){$\scriptstyle 2$} 
        \put(18, 2){$\scriptstyle 1$} 
        \put(20,15){\oval(40,30)} \put( 0,15){\line(1,0){40}} 
        \put( 0,15){\circle*{3}}  \put(40,15){\circle*{3}}
\end{picture}\end{array} = 3 
~~. 
\end{equation}
 

\section{Penrose theory of spin-network}

In this appendix we discuss the relation between
the Penrose theory of spin networks and the Kauffman bracket
and Temperley-Lieb recoupling theory. This appendix is
based essentially on Penrose's original formulation
\cite{Penrose71} and on an article by Kauffman
\cite{Kauffman90c}. A basic idea used by Penrose
(in his doctoral thesis)  
it is to rewrite any tensor expression in which there 
are sums of indices in a graphical way \cite{Penrose71a}.
Consider the calculus of spinors.  Penrose represents 
the basic element of spinor calculus as 
\begin{mathletters}
\begin{eqnarray}
 \delta_C^{~A} &=& \begin{array}{c}\setlength{\unitlength}{1 pt}
              \begin{picture}(10,15)
               \put(5,5){\line(0,1){10}}
               \put(5,5){\circle*{3}}\put(5,15){\circle*{3}}
               \put(6,0){${\scriptstyle  C}$}
               \put(6,14){${\scriptstyle  A}$}
               \end{picture}\end{array} \\
 \epsilon_{AC} &=& \begin{array}{c}\setlength{\unitlength}{1 pt}
              \begin{picture}(20,15)
               \put(5,5){\circle*{3}}\put(15,5){\circle*{3}}
               \put(10,5){\oval(10,20)[t]}
               \put(3,0){${\scriptstyle  A}$}
               \put(12,0){${\scriptstyle  C}$}
               \end{picture}\end{array} \qquad\qquad
 \epsilon^{AC} = \begin{array}{c}\setlength{\unitlength}{1 pt}
              \begin{picture}(20,15)
              \put(10,10){\oval(10,20)[b]}
              \put(5,10){\circle*{3}}\put(15,10){\circle*{3}}
              \put(3,11){${\scriptstyle  A}$}
              \put(12,11){${\scriptstyle  C}$}
              \end{picture}\end{array} \\
 \eta_{A} &=& \begin{array}{c}\setlength{\unitlength}{1 pt}
              \begin{picture}(20,15)
               \put(0,5){\framebox(10,10){$\eta$}}
               \put(6,-2){${\scriptstyle  A}$}
               \put(5,0){\line(0,1){5}}\put(5,0){\circle*{3}}
               \end{picture}\end{array} \qquad\qquad
 \eta^{A} = \begin{array}{c}\setlength{\unitlength}{1 pt}
              \begin{picture}(20,15)
              \put(0,0){\framebox(10,10){$\eta$}}
              \put(6,13){${\scriptstyle  A}$}
              \put(5,10){\line(0,1){5}}\put(5,15){\circle*{3}}
             \end{picture}\end{array}
\end{eqnarray}
and generally to any tensor object
\begin{equation}
  X_{AB}^{C} = \begin{array}{c}\setlength{\unitlength}{1 pt}
          \begin{picture}(20,25)
          \put(0,7){\framebox(20,10){${\scriptstyle  X}$}}
          \put(6,0){${\scriptstyle  A}$}\put(5,2){\line(0,1){5}}
          \put(5,2){\circle*{3}}
          \put(16,0){${\scriptstyle  B}$}\put(15,2){\line(0,1){5}}
          \put(15,2){\circle*{3}}
          \put(11,20){${\scriptstyle  C}$}\put(10,17){\line(0,1){5}}
          \put(10,22){\circle*{3}}
          \end{picture}\end{array}
~~.
\end{equation}
\end{mathletters}
This convention provides the possibility of writing the product
of any two tensors in a graphical way. For example:
\begin{eqnarray}
   \epsilon_{AB} \eta^A \eta^B &=&
\begin{array}{c}\setlength{\unitlength}{1 pt}
  \begin{picture}(30,20)
  \put(0,0){\framebox(10,10){$\eta$}}
  \put(6,13){${\scriptstyle  A}$}\put(5,10){\line(0,1){5}}
  \put(5,15){\circle*{3}}
  \put(20,0){\framebox(10,10){$\eta$}}
  \put(26,13){${\scriptstyle  B}$}\put(25,10){\line(0,1){5}}
  \put(25,15){\circle*{3}}
  \put(15,15){\oval(20,10)[t]}
\end{picture}\end{array}
\\ =- \epsilon_{AD} \epsilon_{BC} \epsilon^{CD} \eta^A \eta^B 
&=&-
\begin{array}{c}\setlength{\unitlength}{1 pt}
\begin{picture}(60,20)
  \put(0,0){\framebox(10,10){$\eta$}}
  \put(6,13){${\scriptstyle  A}$}\put(5,10){\line(0,1){5}}
  \put(5,15){\circle*{3}}
  \put(20,0){\framebox(10,10){$\eta$}}
  \put(26,13){${\scriptstyle  B}$}\put(25,10){\line(0,1){5}}
  \put(25,15){\circle*{3}}
  \put(30,15){\oval(10,10)[t]} \put(25,15){\oval(40,15)[t]}
  \put(36,13){${\scriptstyle  C}$}\put(46,13){${\scriptstyle  D}$}
  \put(40,15){\oval(10,30)[b]}
  \put(35,15){\circle*{3}}\put(45,15){\circle*{3}}
\end{picture}\end{array}
\\  =- \epsilon_{CD}\delta_A^D \delta_B^C \eta^A \eta^B &=&-
\begin{array}{c}\setlength{\unitlength}{1 pt}
\begin{picture}(30,30)
  \put(0,0){\framebox(10,10){$\eta$}}
  \put(20,0){\framebox(10,10){$\eta$}}
  \put(6,13){${\scriptstyle  A}$}\put(5,10){\line(0,1){5}}
  \put(5,15){\circle*{3}}
  \put(26,13){${\scriptstyle  B}$}\put(25,10){\line(0,1){5}}
  \put(25,15){\circle*{3}}
  \put(6,23){${\scriptstyle  C}$}\put(5,15){\line(2,1){20}}
  \put(5,25){\circle*{3}}
  \put(26,23){${\scriptstyle  D}$}\put(25,15){\line(-2,1){20}}
  \put(25,25){\circle*{3}}
  \put(15,25){\oval(20,10)[t]}
\end{picture}\end{array} 
~~~.
\end{eqnarray}

In the light of the example above, Penrose considered a
modification of the spinor calculus, which he denoted as binor
calculus. The binor calculus is obtained by adding two
conventions to the calculus above:
\begin{enumerate}
\item Assign a minus sign to each minimum.
\item Assign a minus sign to each crossing
\item Maxima and minima are taken with respect to a fixed
      direction in the plane. (This direction is conventionally
      taken to be the vertical direction on the written page)
\item A segment with transversal intersection with all
      horizontal direction is taken to be a Kronecker delta.
\end{enumerate}
The advantage of these additional rules is that they make the
calculus topological invariant, namely one can arbitrarily
smoothly deform a graphical expression without changing its
meaning.

The other way around, any curve can now be decomposed in a
product of $\delta$'s and $\epsilon$'s and any two curves that
are ambient isotopic, i.e. that can be transformed one in the
other by a sequence of Reidemeister moves, represent the
tensorial expression as product of epsilons and deltas.
  
A closed loop (with this convention) has value $(-2)$, because
\begin{equation}
 \begin{array}{c}{\setlength{\unitlength}{1 pt}
\begin{picture}(10,10) \put(5,5){\oval(10,10)}
\end{picture}}\end{array}
 = - \epsilon_{AB}\ \epsilon^{AB} = - 2
\end{equation}
and we have the basic binor identity, which reads:
\begin{eqnarray}
&&\begin{array}{c}{\setlength{\unitlength}{1 pt}
\begin{picture}(10,15) 
\put( 0,0){\line( 2,3){10}}
\put(10,0){\line(-2,3){10}}
\end{picture}}\end{array}
 + \begin{array}{c}{\setlength{\unitlength}{1 pt}
\begin{picture}(10,15) 
\put( 0,0){\line(0,1){15}}
\put(10,0){\line(0,1){15}}
\end{picture}}\end{array}
 + \begin{array}{c}{\setlength{\unitlength}{1 pt}
\begin{picture}(10,15) 
\put(5, 0){\oval(10,10)[t]}
\put(5,15){\oval(10,10)[b]} 
\end{picture}}\end{array}
 =
(-1)   ~\delta^{\cdot C}_{B}\ \delta^{\cdot D}_{A}  
   +      ~\delta^{\cdot C}_{A}\ \delta^{\cdot D}_{B}
   + (-1) ~\epsilon_{AB}\ \epsilon^{CD} = 0  
\end{eqnarray}

It is easy to see that these relations are exactly the same as
the properties (\ref{eq:KB1}) and (\ref{eq:KB2}) of the
Kauffman bracket with $A=-1$ and $d=-2$.  Notice from equation
(\ref{eq:KB1}) that if $A=-1$ undercrossing and overcrossing
are equivalent: indeed they give the same expansion.  Clearly
in Penrose's binor calculus there is no meaning of the
distinction between over and under crossing.  The theory can
then be developed as the recoupling theory of appendix B with
the special value $A=-1$.  Thus, the $A=-1$ Kauffman bracket of
a spin network is the same at the Penrose's spin network
evaluation.

For more detail on the exact relation between {\it
tangle-theoretic} recoupling theory and spin-networks see, for
example, \cite{Kauffman94,Kauffman91,Kauffman90c}.  An
important point that emerges from this brief discussion is the
possibility of using a topological invariant calculus for
writing generic $SL(2,C)$ invariant tensor expressions. (This
was one of the original motivations of Penrose for introducing
binors.)  It is possible to write any $SL(2,C)$ Mandelstam
identities (\ref{eq:Mandelstam}) in a graphical way and in
particular we can express these identities in spin-network-like
graphical relations, in which each edge it is the
antisymmetrization of the holonomies along the edge.


\section{Graphical calculus of angular momentum and its
relation with the tangle-theoretical recoupling theory}

Finally, the $A=-1$ case of recoupling theory is equivalent to
the graphical calculus of the algebra of the $SU(2)$
representations.  In the literature there is a great number of
results on the Wigner $3nJ$ symbols and a well developed theory
of graphical calculus for angular momentum.  To our knowledge
the most used graphical method of computation in the
representation theory of $su(2)$ are the one due to
Levinson\cite{Levinson57} and developed by Yutsis, Levinson and
Vanagas\cite{Yutsin62} and the slightly modified version of
Brink and Satcheler\cite{Brink68}.  We discuss here the
connection between the tangle-theoretical recoupling theory (in
the case $A=1$)\footnote{The correspondence between the case
$A=-1$ and $A=1$ and their equivalence is discussed by R.\
Penrose in \cite{Penrose71a}.} and the graphical method of
Brink and Satcheler\cite{Brink68}. We indicate a diagram in the
Brink convention with a subscript $B$, and the $3nJ$-Symbol (in
the standard normalization\footnote{We recall the fact that we
use color and not spin to denote the $su(2)$ representation
associated to an edge. In the angular momentum literature, the
spin notation is prevalent.  As a consequence, numbers in Brink
diagrams, or in $3nJ$-Symbols in standard normalization, must
be understood as the spin of the edge; or, equivalently, the
color divided by two.})  with a subscript $W$.  The two methods
are identical up to a different normalization of the $3$-valent
vertex and the fact the the orientations of any vertex are
explicit denoted with a $+$ for a counter-clockwise orientation
and $-$ for a clockwise one. (In this appendix we are imprecise
about this overall sign.)  Following Kauffman, we have chosen
to denote the recoupling matrix of a $4$-valent node by curl
brackets, while curl brackets are used in the angular momentum
literature to indicate Wigner's $6J$-symbols, which are the
evaluation of the tetragonal net.  In other words, the Wigner
$3J$ and $6J$-symbol are defined as the evaluation
\begin{eqnarray}
  \{ a,b,c \}_W         
&=& \text{norm} \cdot \theta(a,b,c), \\
  { \left\{\begin{array}{ccc}
                      a  & b & c \\
                      d  & e & f  
    \end{array}\right\} }_W 
&=& \text{norm} \cdot {       
        {Tet\left[\begin{array}{ccc} 
            a   &  b  & c\\
            d   &  e  & f  
        \end{array}\right]}}, 
\end{eqnarray}
where the normalization factor ``norm'' of \cite{Brink68}
corresponds to the choice
\begin{equation}
\{ a,b,c \}_W   = 
\left( \begin{array}{c}\setlength{\unitlength}{1 pt}\begin{picture}(40,40)
        \put(18,32){$a$}
        \put(18,17){$b$} 
        \put(18, 2){$c$} 
        \put(20,15){\oval(40,30)} \put( 0,15){\line(1,0){40}} 
        \put( 0,15){\circle*{3}}  \put(40,15){\circle*{3}}
\end{picture}\end{array} \right)_B = +1
\label{eq:NormCleb}
\end{equation}
This is also the standard normalization of the Clebsh-Gordon
coefficient that gives the usual normalization of the Wigner
$3nJ$-Symbol. With this normalization the recoupling theorem
(eq.\ (\ref{eq:recTheorem})) becomes: 
\begin{eqnarray}
&& \left(
\begin{array}{c}\setlength{\unitlength}{1 pt}
\begin{picture}(50,40)
          \put( 0,0){$a$}\put( 0,30){$b$}
          \put(45,0){$d$}\put(45,30){$c$}
          \put(10,10){\line(1,1){10}}\put(10,30){\line(1,-1){10}}
          \put(30,20){\line(1,1){10}}\put(30,20){\line(1,-1){10}}
          \put(20,20){\line(1,0){10}}\put(22,25){$j$}
          \put(20,20){\circle*{3}}\put(30,20){\circle*{3}}
\end{picture}\end{array}
\right)_B = 
\sum_i ~\Delta_i ~  \left\{\begin{array}{ccc}
                                 a  & b & i \\
                                 c  & d & j  
                               \end{array}\right\}_W 
         \left(
\begin{array}{c}\setlength{\unitlength}{1 pt}
\begin{picture}(40,40)
      \put( 0,0){$a$}\put( 0,40){$b$}
      \put(35,0){$d$}\put(35,40){$c$}
      \put(10,10){\line(1,1){10}}\put(10,40){\line(1,-1){10}}
      \put(20,30){\line(1,1){10}}\put(20,20){\line(1,-1){10}}
      \put(20,20){\line(0,1){10}}\put(22,22){$i$}
      \put(20,20){\circle*{3}}\put(20,30){\circle*{3}}
\end{picture}\end{array}
\right)_B 
~~,
\label{eq:recTheoremBrink} 
\end{eqnarray}
where $\Delta_i$ is interpreted as the dimension of the 
representation of spin $i/2$.
From eqs.\ (\ref{eq:NormCleb}) and (\ref{eq:recTheoremBrink})
we have the correspondence between a Bring diagram 
and one of ours: one has to divide any $3$-valent node
by $\sqrt{\theta(a,b,c)}$. 
As an example, let us consider the relation between the
tetrahedron evaluation ($Tet$) and the Wigner $6J$ 
symbol. 
\begin{eqnarray}
&&  \left\{\begin{array}{ccc}
            A  & B & E \\
            C  & D & F  
    \end{array}\right\}_W =
  \left(
  \begin{array}{c}\setlength{\unitlength}{1 pt}\begin{picture}(55,40)
        \put( 0,15){\line(1,-1){15}} \put(0,22){${\scriptstyle B}$}
        \put( 0,15){\line(1, 1){15}} \put(0, 0){${\scriptstyle A}$}
        \put( 0,15){\circle*{3}}
        \put(30,15){\line(-1, 1){15}} \put(28,20){${\scriptstyle C}$}
        \put(30,15){\line(-1,-1){15}} \put(28, 2){${\scriptstyle D}$}
        \put(30,15){\circle*{3}}
        \put( 0,15){\line(1,0){30}} \put(12,16){${\scriptstyle F}$}
        \put(15,30){\line(1,0){25}} \put(15,30){\circle*{3}}
        \put(15, 0){\line(1,0){25}} \put(15, 0){\circle*{3}}
        \put(40, 0){\line(0,1){30}} \put(42,12){${\scriptstyle E}$}
  \end{picture}\end{array} \right)_B \\
&&\qquad = 
\frac{\begin{array}{c}\setlength{\unitlength}{1 pt}\begin{picture}(55,40)
        \put( 0,15){\line(1,-1){15}} \put(0,22){${\scriptstyle B}$}
        \put( 0,15){\line(1, 1){15}} \put(0, 0){${\scriptstyle A}$}
        \put( 0,15){\circle*{3}}
        \put(30,15){\line(-1, 1){15}} \put(28,20){${\scriptstyle C}$}
        \put(30,15){\line(-1,-1){15}} \put(28, 2){${\scriptstyle D}$}
        \put(30,15){\circle*{3}}
        \put( 0,15){\line(1,0){30}} \put(12,16){${\scriptstyle F}$}
        \put(15,30){\line(1,0){25}} \put(15,30){\circle*{3}}
        \put(15, 0){\line(1,0){25}} \put(15, 0){\circle*{3}}
        \put(40, 0){\line(0,1){30}} \put(42,12){${\scriptstyle E}$}
      \end{picture}\end{array}
 }{\sqrt{
    \begin{array}{c}\setlength{\unitlength}{.5 pt}\begin{picture}(40,40)
        \put(18,32){$\scriptstyle A$}
        \put(18,17){$\scriptstyle B$} 
        \put(18, 2){$\scriptstyle F$} 
        \put(20,15){\oval(40,30)} \put( 0,15){\line(1,0){40}} 
        \put( 0,15){\circle*{3}}  \put(40,15){\circle*{3}}
    \end{picture}\end{array}
    \begin{array}{c}\setlength{\unitlength}{.5 pt}\begin{picture}(40,40)
        \put(18,32){$\scriptstyle C$}
        \put(18,17){$\scriptstyle D$} 
        \put(18, 2){$\scriptstyle E$} 
        \put(20,15){\oval(40,30)} \put( 0,15){\line(1,0){40}} 
        \put( 0,15){\circle*{3}}  \put(40,15){\circle*{3}}
    \end{picture}\end{array}
    \begin{array}{c}\setlength{\unitlength}{.5 pt}\begin{picture}(40,40)
        \put(18,32){$\scriptstyle A$}
        \put(18,17){$\scriptstyle D$} 
        \put(18, 2){$\scriptstyle E$} 
        \put(20,15){\oval(40,30)} \put( 0,15){\line(1,0){40}} 
        \put( 0,15){\circle*{3}}  \put(40,15){\circle*{3}}
    \end{picture}\end{array}
    \begin{array}{c}\setlength{\unitlength}{.5 pt}\begin{picture}(40,40)
        \put(18,32){$\scriptstyle B$}
        \put(18,17){$\scriptstyle C$} 
        \put(18, 2){$\scriptstyle E$} 
        \put(20,15){\oval(40,30)} \put( 0,15){\line(1,0){40}} 
        \put( 0,15){\circle*{3}}  \put(40,15){\circle*{3}}
    \end{picture}\end{array}
    }}
\nonumber \\[1 mm]
&&\qquad = \frac{{Tet\left[\begin{array}{ccc} 
                        A & B & E \\
                        C & D & F  
                  \end{array}\right]}
    }{\sqrt{\theta(A,B,F)\theta(C,D,F)
      \theta(A,D,E)\theta(B,C,E)
    }}. 
\nonumber
\end{eqnarray}


\section{Basic Formulas of Recoupling theory}

We collect here the basic formulas 
of recoupling theory in the case $A=-1$ and $d=-2$.
Using the ``quantum'' integer
\begin{eqnarray}
~[n]     &=& \frac{A^{2n}-A^{-2n}}{A^2-A^{-2}}
          = (-1)^{n-1} \Delta_{n-1} = n  
\nonumber\\
\{ n \}  &=& \frac{1-A^{-4n}}{1-A^{-4}}
          = A^{2n-2} ~[n] = n  
\label{eq:quantumInt}\\
\{ n \} !&=& \{1\}\cdot\{2\} \cdots \{n\} 
          = n! ~~
\nonumber
\end{eqnarray}
we define

\noindent {(1)} The symmetrizer
\begin{equation}
\Delta_n = \begin{array}{c}\setlength{\unitlength}{1 pt}
           \begin{picture}(35,25)
           \put(15, 0){\line(0,1){10}}\put(20, 0){\line(0,1){10}}
           \put(15, 0){\line(1,0){ 5}}\put(15,10){\line(1,0){ 5}}
           \put(15,25){\line(1,0){5}}  \put(15,15){$n$}
           \put(15,15){\oval(30,20)[l]}\put(20,15){\oval(30,20)[r]}
           \end{picture}\end{array} 
   = (-1)^n [n+1]  =  (-1)^n (n+1). 
\label{eq:valSIM}
\end{equation}

\noindent {(2)} The exchange of line in a 3-Vertex
      \begin{equation}
\begin{array}{c}
\setlength{\unitlength}{1 pt}
\begin{picture}(40,50)
       \put(30,25){\line(-3, 2){30}} \put( 0,35){$a$}
       \put( 0,25){\line( 3, 2){13}} 
       \put(30,45){\line(-3,-2){13}} \put(30,35){$b$}
       \put(15,25){\oval(30,20)[b]}  \put(15,15){\circle*{3}}
       \put(15, 5){\line(0,1){10}}   \put(17,1){$c$}
\end{picture}\end{array} 
      = \lambda^{ab}_c 
\begin{array}{c}\setlength{\unitlength}{1 pt}
\begin{picture}(40,40)
       \put(15,15){\line(-1, 1){10}} \put( 4,27){$a$}
       \put(15,15){\line( 1, 1){10}} \put(22,27){$b$}
       \put(15, 5){\line(0,1){10}}   \put(17,1){$c$}
       \put(15,15){\circle*{3}}
\end{picture}\end{array}
      \end{equation}
Where $\lambda^{ab}_c = (-1)^{(a+b-c)/2} ~A^{(a'+b'-c')/2}$,
      and $x'=x(x+2)$.

\noindent {(3)} The $\theta$ evaluation
\begin{eqnarray}
&&\theta(a,b,c) = 
\begin{array}{c}\setlength{\unitlength}{1 pt}\begin{picture}(40,40)
        \put(18,32){$a$}
        \put(18,17){$b$} 
        \put(18, 2){$c$} 
        \put(20,15){\oval(40,30)} \put( 0,15){\line(1,0){40}} 
        \put( 0,15){\circle*{3}}  \put(40,15){\circle*{3}}
\end{picture}\end{array}
= \frac{ (-1)^{m+n+p} [m+n+p+1]! ~[m]!~[p]!~[p]!}{
                         [a]! ~[b]! ~[c]!}
\label{eq:valTheta}
\end{eqnarray}
where $m=(a+b-c)/2$, $n=(b+c-a)/2$, $p=(c+a-b)/2$. 


\noindent {(4)} The recoupling theorem:
\begin{eqnarray}
\begin{array}{c}\setlength{\unitlength}{1 pt}
\begin{picture}(50,40)
          \put( 0,0){$a$}\put( 0,30){$b$}
          \put(45,0){$d$}\put(45,30){$c$}
          \put(10,10){\line(1,1){10}}\put(10,30){\line(1,-1){10}}
          \put(30,20){\line(1,1){10}}\put(30,20){\line(1,-1){10}}
          \put(20,20){\line(1,0){10}}\put(22,25){$j$}
          \put(20,20){\circle*{3}}\put(30,20){\circle*{3}}
\end{picture}\end{array}
    &=& \sum_i  \left\{\begin{array}{ccc}
                      a  & b & i \\
                      c  & d & j  
                \end{array}\right\}
\begin{array}{c}\setlength{\unitlength}{1 pt}
\begin{picture}(40,40)
      \put( 0,0){$a$}\put( 0,40){$b$}
      \put(35,0){$d$}\put(35,40){$c$}
      \put(10,10){\line(1,1){10}}\put(10,40){\line(1,-1){10}}
      \put(20,30){\line(1,1){10}}\put(20,20){\line(1,-1){10}}
      \put(20,20){\line(0,1){10}}\put(22,22){$i$}
      \put(20,20){\circle*{3}}\put(20,30){\circle*{3}}
\end{picture}\end{array}
\label{eq:recTheorem}\\
 \left\{\begin{array}{ccc}
      a  & b & i \\
      c  & d & j  
 \end{array}\right\}
&=& \frac{ {\displaystyle \Delta_i ~ \left[\begin{array}{ccc}
                      a  & b & i \\
                      c  & d & j  
              \end{array}\right] }
}{\theta(a,d,i) \theta(b,c,i) }
~. 
\end{eqnarray}

\noindent {(5)} The Tetrahedral net
\begin{eqnarray}
&&{Tet\left[\begin{array}{ccc} 
            A & B & E \\
            C & D & F  
  \end{array}\right]}
= \begin{array}{c}\setlength{\unitlength}{1 pt}\begin{picture}(55,40)
        \put( 0,15){\line(1,-1){15}} \put(0,22){${\scriptstyle B}$}
        \put( 0,15){\line(1, 1){15}} \put(0, 0){${\scriptstyle A}$}
        \put( 0,15){\circle*{3}}
        \put(30,15){\line(-1, 1){15}} \put(28,20){${\scriptstyle C}$}
        \put(30,15){\line(-1,-1){15}} \put(28, 2){${\scriptstyle D}$}
        \put(30,15){\circle*{3}}
        \put( 0,15){\line(1,0){30}} \put(12,16){${\scriptstyle F}$}
        \put(15,30){\line(1,0){25}} \put(15,30){\circle*{3}}
        \put(15, 0){\line(1,0){25}} \put(15, 0){\circle*{3}}
        \put(40, 0){\line(0,1){30}} \put(42,12){${\scriptstyle E}$}
  \end{picture}\end{array} 
= \frac{{\cal I}}{{\cal E}} \sum_{m\leq S \leq M}
 \frac{ (-1)^{S} [S+1]!}{\prod_i  ~[S-a_i]!~\prod_j ~[b_j-S]! }
~~,
\end{eqnarray}
where
$$
{\displaystyle\begin{array}{rclcrcl} 
   a_1&=&{\displaystyle\frac{A+D+E}{2}},
   &\qquad&b_1&=&{\displaystyle\frac{B+D+E+F}{2}}, 
\\[2 mm]
   a_2&=&{\displaystyle \frac{B+C+E}{2}},
   &\qquad&b_2&=&{\displaystyle \frac{A+C+E+F}{2}}, 
\\[2 mm]
   a_3&=&{\displaystyle \frac{A+B+F}{2}},
   &\qquad&b_3&=&{\displaystyle \frac{A+B+C+D}{2}}, 
\\[2 mm]
   a_4&=&{\displaystyle \frac{C+D+F}{2}},&\qquad&     & &      
\end{array}}
$$
$$
{\displaystyle \begin{array}{rclcrcl} 
   m &=&{\rm max}\{ a_i \},   &&      
   M &=&{\rm min}\{ b_j \}, \\[2 mm]
   {\cal E} &=& [A]! [B]! [C]! [D]! [E]! [F]!,
 &&{\cal I} &=& \prod_{ij} [b_j-a_i]! 
~. 
\end{array}}
$$

\noindent {(6)} The reduction formula
\begin{eqnarray}
\begin{array}{c}\setlength{\unitlength}{1 pt}
\begin{picture}(40,50)
        \put(8,25){${b}$}
        \put(20,15){\line(0,-1){15}} \put(22,50){${a}$}
        \put(20,45){\line(0,1){15}}  \put(22,2){${a}$}
        \put(20,30){\oval(30,30)}   \put(37,32){${c}$}
        \put(20,15){\circle*{3}}\put(20,45){\circle*{3}}
\end{picture}\end{array}
&=& \frac{\begin{array}{c}\setlength{\unitlength}{.5 pt}\begin{picture}(40,40)
            \put(18,32){$\scriptstyle a$}
            \put(18,17){$\scriptstyle b$} 
            \put(18, 2){$\scriptstyle c$} 
            \put(20,15){\oval(40,30)} \put( 0,15){\line(1,0){40}} 
            \put( 0,15){\circle*{3}}  \put(40,15){\circle*{3}}
           \end{picture}\end{array}
         }{\begin{array}{c}\setlength{\unitlength}{.5 pt}
           \begin{picture}(35,25)
           \put(15, 0){\line(0,1){10}}\put(20, 0){\line(0,1){10}}
           \put(15, 0){\line(1,0){ 5}}\put(15,10){\line(1,0){ 5}}
           \put(15,25){\line(1,0){5}}  \put(15,15){$\scriptstyle a$}
           \put(15,15){\oval(30,20)[l]}\put(20,15){\oval(30,20)[r]}
           \end{picture}\end{array}
          } 
~\cdot~
\begin{array}{c}\setlength{\unitlength}{1 pt}
\begin{picture}(15,40)
        \put(2,0){\line(0,1){40}}
        \put(5,18){$a$}
\end{picture}\end{array}
\label{ThetaInALine}
\\
\begin{array}{c}\setlength{\unitlength}{1 pt}
\begin{picture}(50,40)
        \put( 0,25){\line(1,-1){15}} \put(0,32){${b}$}
        \put( 0,25){\line(1, 1){15}} \put(0,10){${c}$}
        \put( 0,25){\circle*{3}}
        \put(30,25){\line(-1, 1){15}} \put(28,30){${d}$}
        \put(30,25){\line(-1,-1){15}} \put(28,12){${e}$}
        \put(30,25){\circle*{3}}
        \put( 0,25){\line(1,0){30}} \put(12,26){${f}$}
        \put(15,40){\line(0,1){10}} \put(15,40){\circle*{3}}
        \put(15,10){\line(0,-1){10}} \put(15,10){\circle*{3}}
        \put(17,2){${a}$}\put(17,42){${a}$}
\end{picture}\end{array}
&=& \frac{\begin{array}{c}\setlength{\unitlength}{.8 pt}
        \begin{picture}(55,40)
        \put( 0,15){\line(1,-1){15}} \put(0,22){${\scriptstyle c}$}
        \put( 0,15){\line(1, 1){15}} \put(0, 0){${\scriptstyle b}$}
        \put( 0,15){\circle*{3}}
        \put(30,15){\line(-1, 1){15}} \put(28,20){${\scriptstyle d}$}
        \put(30,15){\line(-1,-1){15}} \put(28, 2){${\scriptstyle e}$}
        \put(30,15){\circle*{3}}
        \put( 0,15){\line(1,0){30}} \put(12,16){${\scriptstyle f}$}
        \put(15,30){\line(1,0){25}} \put(15,30){\circle*{3}}
        \put(15, 0){\line(1,0){25}} \put(15, 0){\circle*{3}}
        \put(40, 0){\line(0,1){30}} \put(42,12){${\scriptstyle a}$}
        \end{picture}\end{array}
    }{ \begin{array}{c}\setlength{\unitlength}{.5 pt}
       \begin{picture}(35,25)
           \put(15, 0){\line(0,1){10}}\put(20, 0){\line(0,1){10}}
           \put(15, 0){\line(1,0){ 5}}\put(15,10){\line(1,0){ 5}}
           \put(15,25){\line(1,0){5}}  \put(15,15){$\scriptstyle a$}
           \put(15,15){\oval(30,20)[l]}\put(20,15){\oval(30,20)[r]}
        \end{picture}\end{array}
     } 
~\cdot~
\begin{array}{c}\setlength{\unitlength}{1 pt}
\begin{picture}(15,40)
        \put(2,0){\line(0,1){40}}
        \put(5,18){$a$}
\end{picture}\end{array}
~~.\label{TetInALine}
\end{eqnarray}
These formulas are sufficient for the computations performed in
the paper. For details on their derivation, see
\cite{Kauffman94}.

\twocolumn
\section{Some volume eigenvalues}

Finally, we present here some volume eigenvalues of four- and 
five-valent vertices. The tables \ref{EVvolume5} and \ref{EVvolume}, 
give the colors of the external edges, the
dimension of the vertex (number of independent compatible
colorings), and the eigenvalues. The number in parenthesis
indicates the multiplicity of the eigenvalues.

\begin{table}
\begin{tabular}{cccccl}
$P_0$&$P_1$&$P_2$&$P_3$& Dim. & 
$\lambda_{\beta_i}=\lambda_{\beta_i}(P_0,\ldots,P_3)$ \\
\tableline
1&1&1&1& $2$ & $(2)~ \sqrt{\frac{1}{8}\sqrt{3}}$ \\
2&2&2&2& $3$ & $(1)~0$, $(2)~ \sqrt{\frac{1}{2}\sqrt{3}}$ \\
3&3&3&3& $4$ & $(2)~\sqrt{\frac{3}{8}\sqrt{3}}$,\\
 & & & &&      $(2)~\sqrt{\frac{3}{8} \sqrt{35}}$ \\
4&4&4&4& $5$ & $(1)~0$ \\,
 & & & &&$(2)~\sqrt{\frac{3}{4} \sqrt{22 - \sqrt{114}}}$,\\
 & & & &&$(2)~\sqrt{\frac{3}{4} \sqrt{22 + \sqrt{114}}}$ \\
5&5&5&5& $6$ & $(2)~\sqrt{\frac{1}{8} \sqrt{1155}}$,\\
 & & & &&$(2)~\sqrt{\frac{1}{8}\sqrt{2211 - 96\sqrt{481}}}$ \\
 & & & &&$(2)~\sqrt{\frac{1}{8}\sqrt{2211 + 96\sqrt{481}}}$ \\
6&6&6&6& $7$ & $(1)~0$,\\   
 & & & &   &   $(2)~ \sqrt{2 \sqrt{3}}$,\\
 & & & &   &   $(2)~\sqrt{\frac{9}{2} \sqrt{3}}$,\\
 & & & &   &   $(2)~\sqrt{\frac{1}{2} \sqrt{723}}$ \\
\hline
1&1&1&1& $2$ & $(2)~ \sqrt{\frac{1}{8} \sqrt{3}}$  \\
2&2&1&1& $2$ & $(2)~ \sqrt{\frac{1}{4} \sqrt{2}}$  \\
3&2&2&1& $2$ & $(2)~ \sqrt{\frac{1}{4} \sqrt{5}}$  \\
3&3&1&1& $2$ & $(2)~ \sqrt{\frac{1}{8} \sqrt{15}}$  \\
3&3&3&1& $2$ & $(2)~ \sqrt{\frac{1}{2} \sqrt{3}}$  \\
4&2&2&2& $2$ & $(2)~ \sqrt{\frac{1}{2} \sqrt{2}}$  \\
4&3&2&1& $2$ & $(2)~ \sqrt{\frac{3}{4}} $ \\
4&4&1&1& $2$ & $(2)~ \sqrt{\frac{1}{4 } \sqrt{6 }}$  \\
4&4&3&1& $2$ & $(2)~ \sqrt{\frac{1}{4 } \sqrt{21}}$  \\
5&3&2&2& $2$ & $(2)~ \sqrt{\frac{1}{4 } \sqrt{21}}$  \\
5&3&3&1& $2$ & $(2)~ \sqrt{\frac{3}{8 } \sqrt{7 }}$  \\
5&4&2&1& $2$ & $(2)~ \sqrt{\frac{1}{4 } \sqrt{14}}$  \\
5&4&4&1& $2$ & $(2)~ \sqrt{\frac{3}{2}}$  \\
5&5&1&1& $2$ & $(2)~ \sqrt{\frac{1}{8 } \sqrt{35}}$  \\
5&5&3&1& $2$ & $(2)~ \sqrt{\sqrt{2  }}  $ \\
5&5&5&1& $2$ & $(2)~ \sqrt{\frac{9}{8 } \sqrt{3 }}$  \\
6&4&2&2& $2$ & $(2)~ \sqrt{2 }$   \\
6&4&3&1& $2$ & $(2)~ \sqrt{\frac{1}{2 } \sqrt{6 }}$  \\
6&5&2&1& $2$ & $(2)~ \sqrt{\frac{1}{2 } \sqrt{5 }}$  \\
6&5&4&1& $2$ & $(2)~ \sqrt{\frac{3}{4 } \sqrt{6 }}$  \\
6&6&1&1& $2$ & $(2)~ \sqrt{\frac{1}{2 } \sqrt{3 }}$  \\
6&6&3&1& $2$ & $(2)~ \sqrt{\frac{3}{4 } \sqrt{5 }}$  \\
6&6&5&1& $2$ & $(2)~ \sqrt{\frac{3}{4 } \sqrt{10}}$  \\
7&3&3&3& $2$ & $(2)~ \sqrt{\frac{9}{8 } \sqrt{3 }}$  \\
7&4&3&2& $2$ & $(2)~ \sqrt{\frac{3}{4 } \sqrt{6 }}$  \\
7&4&4&1& $2$ & $(2)~ \sqrt{\frac{3}{2 }}$ \\
7&5&2&2& $2$ & $(2)~ \sqrt{\frac{3}{2 } \sqrt{5 }}$  \\
7&5&3&1& $2$ & $(2)~ \sqrt{\frac{3}{8 } \sqrt{15}}$  \\
7&5&5&1& $2$ & $(2)~ \sqrt{             \sqrt{5 }}$  \\
7&6&2&1& $2$ & $(2)~ \sqrt{\frac{3}{4 } \sqrt{3 }}$  \\
7&6&4&1& $2$ & $(2)~ \sqrt{\frac{5}{4 } \sqrt{3 }}$  \\
7&6&6&1& $2$ & $(2)~ \sqrt{\frac{1}{2 } \sqrt{33}}$  \\
7&7&1&1& $2$ & $(2)~ \sqrt{\frac{3}{8 } \sqrt{7 }}$  \\
7&7&3&1& $2$ & $(2)~ \sqrt{\frac{1}{2 } \sqrt{15}}$  \\
7&7&5&1& $2$ & $(2)~ \sqrt{\frac{3}{8 } \sqrt{55}}$  \\
7&7&7&1& $2$ & $(2)~ \sqrt{2            \sqrt{3 }}$  \\
8&4&3&3& $2$ & $(2)~ \sqrt{\frac{3}{4 } \sqrt{10}}$  \\
8&4&4&2& $2$ & $(2)~ \sqrt{             \sqrt{5 }}$  \\
\hline
\hline
2&2&2&2& $3$ & $(1)~0$, $(2)~\sqrt{\frac{1}{2} \sqrt{3}}$ \\
3&3&2&2& $3$ & $(1)~0$, $(2)~\sqrt{\frac{1}{4} \sqrt{26}}$  \\
4&3&3&2& $3$ & $(1)~0$, $(2)~\sqrt{\frac{3}{4} \sqrt{6}}$   \\
4&4&2&2& $3$ & $(1)~0$, $(2)~\sqrt{\frac{1}{2} \sqrt{11}}$  \\
4&4&4&2& $3$ & $(1)~0$, $(2)~\sqrt{\frac{3}{2} \sqrt{3}}$   \\
5&3&3&3& $3$ & $(1)~0$, $(2)~\sqrt{\frac{3}{2} \sqrt{3}}$   \\
5&4&3&2& $3$ & $(1)~0$, $(2)~\sqrt{\frac{1}{4}\sqrt{89 }}$   \\
5&5&2&2& $3$ & $(1)~0$, $(2)~\sqrt{\frac{1}{4}\sqrt{66 }}$   \\
5&5&4&2& $3$ & $(1)~0$, $(2)~\sqrt{\frac{1}{4}\sqrt{174}}$   \\
6&4&3&3& $3$ & $(1)~0$, $(2)~\sqrt{\frac{1}{4}\sqrt{174}}$   \\
6&4&4&2& $3$ & $(1)~0$, $(2)~\sqrt{ 3                   }$   \\
6&5&3&2& $3$ & $(1)~0$, $(2)~\sqrt{\frac{1}{4}\sqrt{131}}$   \\
6&5&5&2& $3$ & $(1)~0$, $(2)~\sqrt{\frac{1}{2}\sqrt{69 }}$   \\
6&6&2&2& $3$ & $(1)~0$, $(2)~\sqrt{\frac{1}{2}\sqrt{23 }}$   \\
6&6&4&2& $3$ & $(1)~0$, $(2)~\sqrt{\frac{3}{2}\sqrt{7  }}$   \\
6&6&6&2& $3$ & $(1)~0$, $(2)~\sqrt{3          \sqrt{3  }}$   \\
7&4&4&3& $3$ & $(1)~0$, $(2)~\sqrt{\frac{1}{2}\sqrt{69 }}$   \\
7&5&3&3& $3$ & $(1)~0$, $(2)~\sqrt{\frac{3}{2}\sqrt{7  }}$   \\
7&5&4&2& $3$ & $(1)~0$, $(2)~\sqrt{\frac{1}{4}\sqrt{209}}$   \\
7&6&3&2& $3$ & $(1)~0$, $(2)~\sqrt{\frac{3}{2}\sqrt{5  }}$   \\
7&6&5&2& $3$ & $(1)~0$, $(2)~\sqrt{\frac{1}{4}\sqrt{395}}$   \\
7&7&2&2& $3$ & $(1)~0$, $(2)~\sqrt{\frac{1}{4}\sqrt{122}}$   \\
7&7&4&2& $3$ & $(1)~0$, $(2)~\sqrt{\frac{3}{4}\sqrt{38 }}$   \\
7&7&6&2& $3$ & $(1)~0$, $(2)~\sqrt{\frac{3}{2}\sqrt{17 }}$   \\
\hline
\hline
3&3&3&3& $4$ & $(2)~\sqrt{\frac{3 }{8 } \sqrt{3  }}$, \\ 
 & & & &&      $(2)~\sqrt{\frac{3 }{8 } \sqrt{35 }}$  \\
4&4&3&3& $4$ & 
         $(2)~ \sqrt{\frac{3 }{4 } \sqrt{9 - \sqrt{57}}}$,\\ 
 & & & &&$(2)~ \sqrt{\frac{3 }{4 } \sqrt{9 + \sqrt{57}}}$ \\
5&4&4&3& $4$ & 
         $(2)~\sqrt{\frac{3 }{8 } \sqrt{66 - 2\sqrt{753}}}$,\\  
 & & & &&$(2)~\sqrt{\frac{3 }{8 } \sqrt{66 + 2\sqrt{753}}}$ \\
5&5&3&3& $4$ & 
         $(2)~\sqrt{\frac{1}{8}\sqrt{511 - 16 \sqrt{721}}}$,\\  
 & & & &&$(2)~\sqrt{\frac{1}{8}\sqrt{511 + 16 \sqrt{721}}}$ \\
5&5&5&3& $4$ & $(2)~\sqrt{\frac{1}{4 } \sqrt{3 }}$,  \\
 & & & &   &   $(2)~\sqrt{\frac{1}{4 } \sqrt{30}}$   \\
6&4&4&4& $4$ & $(2)~\sqrt{\frac{1}{4 } \sqrt{3 }}$,  \\
 & & & &   &   $(2)~\sqrt{\frac{1}{4 } \sqrt{30}}$   \\
6&5&4&3& $4$ & 
         $(2)~\sqrt{\frac{1}{8}\sqrt{918-18\sqrt{1801}}}$,\\  
 & & & &&$(2)~\sqrt{\frac{1}{8}\sqrt{918+18\sqrt{1801}}}$\\
6&6&3&3& $4$ & 
         $(2)~\sqrt{\frac{1}{4}\sqrt{183-3\sqrt{2641}}}$,\\  
 & & & &&$(2)~\sqrt{\frac{1}{4}\sqrt{183+3\sqrt{2641}}}$ \\
6&6&5&3& $4$ & 
         $(2)~\sqrt{\frac{1}{8}\sqrt{1602-18\sqrt{5281}}}$,\\  
 & & & &&$(2)~\sqrt{\frac{1}{8}\sqrt{1602 + 18 \sqrt{5281}}}$\\
7&6&4&3& $4$ & $(2)~ \sqrt{\frac{3 }{4 } \sqrt{6 }}$,\\
 & & & &   &   $(2)~ \sqrt{\frac{3 }{4 } \sqrt{66}}$   \\
7&7&7&3& $4$ & $(2)~ \sqrt{\frac{15}{8 } \sqrt{3 }}$,  \\
 & & & &   &   $(2)~ \sqrt{\frac{15}{8 } \sqrt{715}}$   \\
\end{tabular}
\caption{The Eigenvalues of the Volume for some 4-valent 
vertices}
\label{EVvolume}
\end{table}

\begin{table}
\begin{tabular}{ccccccl}
$P_0$&$P_1$&$P_2$&$P_3$&$P_4$& Dim. & 
$\lambda_{\beta_i}=\lambda_{\beta_i}(P_0,\ldots,P_{4})$ \\
\tableline
2&1&1&1&1& $3$ & $(3)~\sqrt{\frac{3\sqrt{2}+\sqrt{3}}{12}}$ \\
2&2&2&1&1& $4$ &
   $(2)~\sqrt{\frac{29\sqrt{2}+12\sqrt{3}+16\sqrt{5}}{96}}$,\\
 & & & & &&
   $(1)~\sqrt{\frac{5 \sqrt{2}+4 \sqrt{5}}{16}}$,\\
 & & & & &&
   $(1)~\sqrt{\frac{5 \sqrt{2}+4 \sqrt{5}}{24}}$ \\
2&2&2&2&2& $6$ & $(6)~\sqrt{\frac{5}{3}\sqrt{3}}$ \\
3&2&1&1&1& $3$ & 
   $(2)~\sqrt{\frac{21 \sqrt{2}+6 \sqrt{3}
               +18 \sqrt{5} + 14 \sqrt{15} }{192}}$,\\
 & & & & &&
   $(1)~\sqrt{\frac{15 \sqrt{2}+18 \sqrt{5}+4\sqrt{15}}{96}}$ \\
4&2&2&1&1& $3$ & 
           $(1)~\sqrt{\frac{40+10\sqrt{2}
               +4\sqrt{5}+5\sqrt{6}}{80}}$ \\
&&&&&&     $(1)~\sqrt{\frac{60+15\sqrt{2}+20\sqrt{3}
               +12\sqrt{5}+10\sqrt{6}}{80}}$ \\
&&&&&&     $(1)~\sqrt{\frac{20+\sqrt{2}
               +4\sqrt{3}+4\sqrt{5}}{80}}$ \\
\end{tabular}
\caption{The Eigenvalues of the Volume for some 
5-valent vertices.}
\label{EVvolume5}
\end{table}



\begin{thebibliography}{10}

\bibitem{Penrose72} R. Penrose, in {\em Magic Without Magic:
    John Archibald Wheeler}, edited by J.~R. Klauder
    (W. H. Freeman and Company, San Francisco, 1972).

\bibitem{Rovelli88} C. Rovelli and L. Smolin, Phys. Rev.  Lett.
    {\bf 61}, 1155 (1988).

\bibitem{Rovelli90} C. Rovelli and L. Smolin, Nucl.  Phys.
   {\bf B331}, 80 (1990).

\bibitem{Isham96} C. Isham, {\em Structural Issues in Quantum
Gravity}, to appear in the proceedings of the GR14 (Florence
1995).

\bibitem{jmp95} {\em Diffeomorphism invariant quantum field
theory and quantum geometry}.  Special Issue of Journal of
Mathematical Physics, J. Math. Phys. {\bf 36}, (1995).

\bibitem{Baez94b} J. Baez, {\em Knots and Quantum Gravity}
(Oxford University Press, Oxford, 1994).

\bibitem{Ehlers94} J. Ehlers and H. Friedrich, {\em Canonical
    Gravity: from Classical to Quantum} (Springer-Verlag,
    Berlin, 1994).

\bibitem{Rovelli91} C. Rovelli, Class. and Quantum Grav {\bf
8}, 1613 (1991).

\bibitem{Ashtekar92} A. Asthekar, in {\em Gravitation and
    Quantization, Les Houches, Session $LVII$, 1992}, edited by
    B. Julia and J. Zinn-Justin (Elseiver Science, 1995).

\bibitem{Smolin93} L. Smolin, in {\em Brill Feschrift
Proceedings}, edited by B. Hu and T.  Jacobson (Cambridge
University Press, Cambridge, 1993); in {\em Quantum Gravity and
Cosmology} (World Scientific, Singapore, 1993).

\bibitem{Gambini} R. Gambini, J. Pullin {\it Loops, knots, gauge theories and quantum gravity}, Cambridge University Press, in press.

\bibitem{Bruegmann} B. Br\"ugmann, {\it Loop Representations}, in 
\cite{Ehlers94}. 

\bibitem{Ezawa} K. Ezawa, ``Nonperturbative solutions for canonical quantum gravity: an overview'', gr-qc/9601050.

\bibitem{Isham} A. Ashtekar and C. Isham, Class. and Quantum
	Grav. {\em 9}, 1433 (1992).

\bibitem{Ashtekar95} A. Ashtekar, J. Lewandowsky, D. Marolf,
    J. Mourao and T. Thiemann, J. Math.  Phys. {\bf 36}, 6456
    (1995).

\bibitem{AshtekarLewand} A. Ashtekar, J. Lewandowski, ``Quantum
Theory of Geometry: Area Operator'', gr-qc/9602046.

\bibitem{Lewandowski96} J. Lewandoski, ``Volume and Quantization'',
preprint, March 1996.

\bibitem{DePietri96} R. De Pietri, ``On the relation between the 
     connection and the loop representation of quantum gravity'', 
     University of Parma preprint UPRF-96-466, May 1996.

\bibitem{Rovelli95a} C. Rovelli and L. Smolin, Phys. Rev. {\bf
    D53}, 5743 (1995).

\bibitem{Baez95a} J.~C.\ Baez, Advances in Mathematics, 
    {\bf 117}, 253 (1996), gr-qc/9411007; 
    {\em Spin Networks in Nonperturbative Quantum Gravity}, 
    in {\em The Interface of Knots and Physics}, 
    edited by L.H.\ Kauffman (American
    Mathematical Society, Providence, Rhode Island, 1996),
    gr-qc/9504036. 

\bibitem{Foxon} T. Foxon, Class. and Quant. Grav {\bf 12}, 951
(1995). 

\bibitem{Ashtekar92a} A. Ashtekar, C. Rovelli, and L. Smolin,
    Phys. Rev. Lett. {\bf 69}, 237 (1992).

\bibitem{Rovelli95} C. Rovelli and L. Smolin, Nucl. Phys. {\bf
    B442}, 593 (1995).

\bibitem{Loll95} R. Loll, Phys. Rev. Lett. {\bf 75}, 3048
    (1995). A. Ashtekar and J. Lewandowski, J. Geo. and Phys. {\bf 17}, 
191 (1995). J. Lewandowski lecture given at the {\it Workshop
on Canonical and Quantum Gravity\/}, Warsaw, 1995.

\bibitem{Rovelli93a} C. Rovelli, Nucl. Phys. {\bf B405}, 797
    (1993).

\bibitem{Rovelli94b} C. Rovelli and L. Smolin,
Phys. Rev. Lett. {\bf 72}, 446 (1994). R. Borissov, ``Graphical Evolution of
Spin Network States'', in preparation.  R. Gambini, J. Pullin, ``A rigorous
solution of the quantum Einstein equations, gr-qc/9511042; ``The 
general solution of the quantum Einstein equations?'', 
gr-qc/9603019.

\bibitem{Rovelli95b} C. Rovelli, J. Math. Phys. {\bf 36}, 6529
(1995).

\bibitem{Rovelli94} C. Rovelli and H. Morales-Tecotl,
    Phys. Rev. Lett. {\bf 72}, 3642 (1995); Nucl. Phys.  {\bf
    B451}, 325 (1995).

\bibitem{Krasnov95} K. Krasnov, Phys.\ Rev.\ {\bf D53}, 1874 (1995).

\bibitem{Pullin} R. Gambini and J. Pullim, Phys.\ Rev.\ {\bf D47}, 
    R5214 (1993).

\bibitem{Penrose71a} R. Penrose, in {\em Combinatorial
    Mathematics ant its Application}, edited by D. Welsh
    (Academic Press, New York, 1971).

\bibitem{Kauffman94} L.~H. Kauffman and S.~L. Lins, {\em
    {T}emperley-{L}ieb {R}ecoupling {T}heory and {I}nvariant of
    3-{M}anifolds} (Princeton University Press, Princeton,
    1994).

\bibitem{Reisenberger96} M. Reisenberger, {\em Spin-network
states and operators}, in preparation (1996).

\bibitem{Smolin95a} L. Smolin and S. Major, {\it Quantum
    deformation of quantum gravity\ } CGPG-95/12-3 (1995),
    gr-qc/9512020. L. Smolin, S. Major and R. Borissov, {\it
    The geometry of quantum spin networks\ }, CGPG-95/12-4
    (1995), gr-qc/9512043.

\bibitem{Borissov} R. Borissov, {\em Eigenvalue spectrum of the
       Volume Operator in Quantum Gravity}, in preparation
       (1996).

\bibitem{Loll95a} R.\ Loll, Nucl.\ Phys.\ {\bf B460}, 143 (1996).

\bibitem{Barbero} J.F. Barbero, Phys. Rev. {\bf D51} (1995) 5507.

\bibitem{Ashtekar91} A. Ashtekar, Phys. Rev. Lett. 57, 2244
   (1986); Phys. Rev. {\bf D36}, 1587 (1987).  For an
   introduction to the Ashtekar formalism, see
   Ref.~\cite{Rovelli91}, or: A. Ashtekar, {\em
   {N}onperturbative {C}anonical {G}ravity} (World Scientific,
   Singapore, 1991).

\bibitem{Thiemann}
T. Thiemann, ``Reality conditions inducing transforms for
quantum gauge field theory and quantum gravity'', gr-qc/9511057.
A. Ashtekar, Phys. Rev. {\bf D53} (1996) 2865.

\bibitem{Loll96} R. Loll, ``A real Alternative to Quantum Gravity in 
Loop Space'', gr-qc/9602041. 

\bibitem{Immirzi} G. Immirzi, ``Quantizing Regge Calculus'', 
gr-qc/9512040. 

\bibitem{major} S. Major, private communication.

\bibitem{Penrose71} R. Penrose, in {\em Quantum Theory and
    Beyond}, edited by T. Bastin (Cambridge University Press,
    Cambridge, 1971).

\bibitem{Rovelli93} C. Rovelli, E.~T. Newman, and
   J. Lewandowski, J. Math. Phys. {\bf 34}, 4646 (1993).

\bibitem{Gambini81} R. Gambini and A. Trias, Nucl. Phys.  {\bf
B278}, 436 (1986); Phys. Rev. {\bf D23}, 553 (1981).

\bibitem{Kauffman91} L.~H. Kauffman, in {\em {K}nots,
   {T}opology and {Q}uantum {F}ield {T}heories}, edited by
   L. Lusanna (World Scientific, Singapore, 1991).

\bibitem{Graph} M. Behzad and G. Chartrand, {\em Introduction
   to the Theory of Graphs} (Allyn and Bacon, Boston, 1971).

\bibitem{Reidemeister} K. Reidemeister, {\em Knotentheorie}
   (Julius Springer, Berlin, 1932).

\bibitem{FalseIntersec} B. Br\"ugmann, R. Gambini, and
   J. Pullin, Phys. Rev. Lett {\bf 68}, 431 (1992).
   B. Br\"ugmann, R. Gambini, and J. Pullin, Nucl. Phys {\bf
   B385}, 587 (1992).  B. Br\"ugmann, R. Gambini, and
   J. Pullin, Gen. Rel. Grav. {\bf 25}, 1 (1993).
   B. Br\"{u}gmann, in {\em Canonical Gravity: From Classical
   to Quantum}, edited by J. Ehlers and H. Friedrich
   (Springer-Verlag, Berlin, 1993).  J. Pullin, in {\em
   Proceedings of the Vth Mexican School of Particles and
   Fields}, edited by J. Lucio (World Scientific, Singapore,
   1993).

\bibitem{Rovelli93b} C. Rovelli, Phys. Rev. {\bf D47}, 1703
(1993).

\bibitem{Frittelli} S. Frittelli, L Lehner, C Rovelli, ``The Complete Spectrum
of the Area from Recoupling Theory in Loop Quantum Gravity'', 
preprint.

\bibitem{Brink68} D.~M. Brink and G.~R. Satchler, {\em Angular
    Momentum} (Claredon Press, Oxford, 1968).

\bibitem{Lewandowski} A. Ashtekar, J. Lewandowski, {\em
Representation theory of analytic holonomy $C^*$ algebras}, in
\cite{Baez94b}.

\bibitem{Rovelli91b} C. Rovelli, Class. and Quantum Grav. {\bf
         8}, 297 (1991); Class. and Quantum Grav. {\bf 8}, 317
         (1991).

\bibitem{Iwasaki} J. Iwasaki and C. Rovelli: Int. J. of
	Mod. Phys.  {\bf D1}, 533 (1993); Class. and Quantum
	Grav. {\bf 11}, 1653 (1994).

\bibitem{Hawking} S. Hawking, Comm. Math. Phys., {\bf 43}, 199
    (1975).


\bibitem{Barreira} M. Barreira, M. Carfora, C. Rovelli, ``Physics with 
nonperturbative quantum gravity: radiation from a quantum black hole'',
gr-qc/9603064. 

\bibitem{Rovelli96} C. Rovelli, ``Blak Hole Entropy from Loop Quantum 
Gravity'', gr-qc/9603063.

\bibitem{Kauffman90b} L.~H. Kauffman,
Trans. Amer. Math. Soc. {\bf 318}, 417 (1990).

\bibitem{Moussoris79} Moussoris, in {\em Advances in Twistor
Theory}, {\em Research Notes in Mathematics}, edited by Huston
and Ward (Pitman, 1979), pp.\ 308--312.

\bibitem{Kauffman90c} L.~H. Kauffman, Inter. Jour. of Modern
Physics A. {\bf 5}, 93 (1990).

\bibitem{Levinson57} I. Levinson, Liet. TSR Mokslu Acad. Darbai
B Ser. {\bf 2}, 17 (1956).

\bibitem{Yutsin62} A.~P. Yutsin, J.~B. Levinson, and
V.~V. Vanagas, {\em Mathematical Apparatus of the Theory of
angular momentum} (Israel program for Scientific Translation,
Jerusalem, 1962).

\end{thebibliography}
\onecolumn






\end{document}